\documentclass[5p, 9pt]{elsarticle}
\pdfoutput=1
\usepackage{pdflscape}
\usepackage{fullpage}
\usepackage{extsizes}
\usepackage{amsmath}
\usepackage{amssymb}
\usepackage{caption}
\usepackage{subcaption}
\usepackage{comment}
\usepackage[hidelinks]{hyperref}

\usepackage{pgfplots}
\pgfplotsset{compat=1.18}
\usepackage{tikz-3dplot}
\usetikzlibrary{shapes}	

\newcommand{\K}{\frac{\rho g}{\sigma}}
\newcommand{\T}{\mathbf{T}}

\newlength{\fthick}
\definecolor{straightline}{RGB}{135, 129, 39}
\definecolor{curveline}{RGB}{173, 117, 26}
\definecolor{lumped}{rgb}{0.8, 0 ,0}

\definecolor{groundStruct}{rgb}{0.18,0.529,0.721}
\definecolor{ndes}{RGB}{102,102,162}

\begin{document}


	\title{Topology and geometry optimization of grid-shells under self-weight loading}
	\author[shef]{Helen E. Fairclough}
    \author[warsaw,paris]{Karol Bo{\l}botowski}
    \author[shef]{Linwei He}    
    \author[uct]{Andrew Liew}
    \author[shef]{Matthew Gilbert}

    \affiliation[shef]{organization={School of Mechanical, Aerospace and Civil Engineering, The University of Sheffield},
    addressline={Mappin Street},
    city={Sheffield},
    postcode={S1 3JD},
    country={UK}}
    
    \affiliation[uct]{organization={Unipart Construction Technologies Ltd},
    addressline={Advanced Manufacturing Park},
    city={Sheffield},
    postcode={S60 5WG},
    country={UK}}
    
    \affiliation[warsaw]{organization={Department of Structural Mechanics and Computer Aided Engineering, Faculty of Civil Engineering, Warsaw University of Technology},
    addressline={16 Armii Ludowej Street},
    city={Warsaw},
    postcode={00-637},
    country={Poland}}

    \affiliation[paris]{organization={Lagrange Mathematics and Computing Research Center},
    addressline={103 rue de Grenelle},
    city={Paris},
    postcode={75007},
    country={France}}

    \begin{abstract}
        This manuscript presents an approach for simultaneously optimizing the connectivity and elevation of grid-shell structures acting in pure compression (or pure tension) under the combined effects of a prescribed external loading and the design-dependent self-weight of the structure itself. The method derived herein involves solving a second-order cone optimization problem, thereby ensuring convexity and obtaining globally optimal results for a given discretization of the design domain.  Several numerical examples are presented, illustrating characteristics of this class of optimal structures. It is found that, as self-weight becomes more significant, both the optimal topology and the optimal elevation profile of the structure change, highlighting the importance of optimizing both topology and geometry simultaneously from the earliest stages of design. It is shown that this approach can obtain solutions with greater accuracy and several orders of magnitude more quickly than a standard 3D layout/truss topology optimization approach.
    \end{abstract}

    \maketitle
    
    \section{Introduction}
	
	Grid-shells can be used as lightweight and elegant structures, which are especially suited to long-span roofs and similar structures. The term grid-shell was originally applied to timber structures constructed by deforming an initially flat network of thin elements \cite{chilton2016timber}. However, it has since come to be used for any shell structure constructed from discrete elements \cite{adriaenssens2014shell}; it is the latter definition which is adopted herein.   
	
	To ensure an axially loaded design, a wide variety of form-finding approaches have been developed. These typically correspond to problems which would, in other structural optimization communities, be referred to as size and/or shape optimization. This means that the connectivity or topology of the design is fixed, usually by specifying an initial structure. Such methods include particle-spring approaches, dynamic relaxation and force-density methods, as well as physical modeling approaches \cite{adriaenssens2014shell}. 
	
	A key issue in applying any of these approaches to long-span structures is the ability to model the weight of the structure itself. The simplest and most common approximation is to size the (straight) element based only on the axial load, and then apply a force equal to half the element's weight at each end \cite{linkwitz2014force, jiang2018form}. This approximation ignores the bending effects of the weight, which may be substantial, especially in longer elements. A refinement of this approach is to divide each element into multiple shorter sub-elements \cite{huttner2015efficiency}. A more realistic approach is to distribute the loading along the element, typically resulting in parabolic forms, based on the assumption of constant cross-sections \cite{deng2005shape}.
	
	Regardless of the self-weight model used, these form-finding approaches typically do not alter the initial connectivity of the problem. Within the few studies that attempt to take this into account, genetic algorithms are typically employed \cite{richardson2013coupled, richardson2014discrete}, these are very flexible in what may be considered, but there is no guarantee that even a locally optimal solution will be obtained. In \cite{gythiel2022gradient}, gradient-based optimization is used, ensuring that the results are at least locally optimal, however the problem was still non-convex and non-linear, so global optimality could not be obtained, and a only a restricted set of possible elements was employed.

    Within the broader field of axially loaded structures (e.g. trusses), the optimization of topology is typically achieved through the use of the ground structure method \cite{dorn1964automatic}, also known as truss topology optimization or layout optimization. Continuum-based topology optimization methods (e.g. based on finite element type meshes) are also available \cite{bendsoe2003topology}, however, these are less suited to structures containing slender elements such as grid-shells, and so will not be discussed here. 
	
	Layout optimization requires as input only the geometry of the permissible design domain, the location and nature of the support conditions and the forces to be supported. The design domain is then populated with nodes, and every possible pair of nodes is connected with lines representing the bars of the truss. This network of bar elements forms the ground structure. A mathematical optimization problem is then solved to find the minimum material structure contained within the ground structure that satisfies the loading and support conditions. In the basic case of a static loading regime, this becomes a linear programming problem which can be solved to global optimality rapidly, even for millions of potential elements.
	
	There are no additional challenges involved with applying this method in 3D space, although the number of nodes required to obtain the same nodal spacing will be much larger. The computational demands (in 2D or 3D problems) can be reduced by using the member adding method \cite{gilbert2003layout}, without impacting the resulting optimal solution. Using the member adding approach, problems containing billions of potential elements can be tackled. 
				
	One of the key challenges in applying truss topology optimization approaches in the design of grid-shells or other roof structures, is that the applied loading should move to track the structure as the topology optimization progresses. This can be implemented through the use of transmissible loads \cite{fuchs2000optimal}, where the representative point loads of the grid-shell are transmitted through or attached to vertical lines passing through the nodes. Two formulations of transmissible loads are commonly used, although if only single layer structures are permitted then they are equivalent \cite{jiang2018form, lu2021transmissible}. However, to obtain accurate results with either approach, a very dense distribution of nodes must be adopted in the ground structure/domain, particularly in the vertical direction. This can result in computationally expensive problems.  \citet{jiang2018form} suggest an iterative approach of refining the design domain to allow results to be obtained with fewer nodes, however this was still found to be 3 to 4 orders of magnitude slower than fixed-topology approaches such as the force density method. 
	
	For roofs of particularly long spans, a large portion of the loading is generated by the self-weight of the structure itself, as other imposed loads are restricted to e.g. wind and snow loading. The classical method of addressing this within the layout optimization framework is again to add a point load at the end of each element, equal to one half of its weight \cite{bendsoe1994optimization}. As with other form-finding methods, this produces acceptable results in problems of moderate span; but at long spans this erroneously favors solutions containing long elements, maximizing the `free' bending capacity. To address this, \citet{fairclough2018theoretically} developed an approach where each element is curved and variable in cross-section such that, at all points, it experiences a uniform, purely axial stress under the combined effect of the applied loads and the self-weight.
	
	In contrast to the numerical methods described above, exact analytical results for minimum material structures (a.k.a. Michell structures) are challenging to obtain, and as such only a few are known \cite{lewinski2018michell}. In particular, 3D examples of exact optimal structures have been obtained only for two classes of problems. Firstly, axisymmetric problems \cite[][Chapter 5]{lewinski2018michell}, most notably the torsion sphere identified in the seminal work of \citet{michell1904lviii}. Secondly, funicular structures, also known as Prager-structures, which are compression-only (or tension-only) structures, where the point of application of the load is to be optimized alongside the structural form. 
	
	For these Prager-structures, significant initial work focused on cases where the elements were restricted to lie along Cartesian directions \cite{rozvany1979new}, referred to as archgrids. The methodology involved assuming the two sets of arches could be optimized independently, and then noting that the optimal elevations of each layer were coincident, leading to a single-layer structure (the \textit{equal elevation} condition). This work also identified that for an optimal planar arch under external loading (i.e. without self-weight), the average of the slope value squared must be 1 (the \textit{unit mean square slope} condition). 
	
	Later, \cite{rozvany1980optimal} extended the work on arch-grids to cases where self-weight of the structure was to be considered. It was found that the unit mean square slope condition was no longer valid and must be replaced by a much more complicated equation \cite[Eqn 19]{rozvany1980optimal}, which results in a greater rise for a given problem when self-weight is included. Meanwhile, the equal elevation condition continued to hold, demonstrating that a single layer structure is still suitable when self-weight is considered.
	
	Extending the archgrid concept to problems where the layout of elements is also to be considered was achieved by \cite{wang1982prager} for certain classes of problems. It was also noted that the vertical displacement at any point in such a structure was proportional to the elevation at that point, and that the layouts obtained achieve the minimum material usage for under either a limit on maximum stress or on minimum stiffness. It is notable that consideration of the effects of the structure's self-weight loading was not included in \cite{wang1982prager}. 
	
	Recently, there has been a revival in interest in optimal design of vault and grid-shell structures, aligning with ambitions for efficient and sustainable construction. The most widely applicable results for the problem of combined topology and geometry optimization have been obtained by \cite{bolbotowski2022optimal}. For problems where self-weight is not considered, \cite{bolbotowski2022optimal} shows that the minimum material layout and elevation at each point can be obtained as the solution of a convex optimization problem. An analytical approach was presented, alongside a ground-structure based numerical approach which allowed the use of second-order cone programming to solve the problem. This allows for standard solvers to be used, significantly easing implementation, and as the optimization problems are convex, globally optimal results can be found. However, the self-weight of the structures was not considered, and so the applicability of the problems is limited to shorter spans and/or more lightweight materials. Structures designed for longer spans using the method of \cite{bolbotowski2022optimal} would become overloaded due to the extra forces generated by their own weight and would be likely to fail.

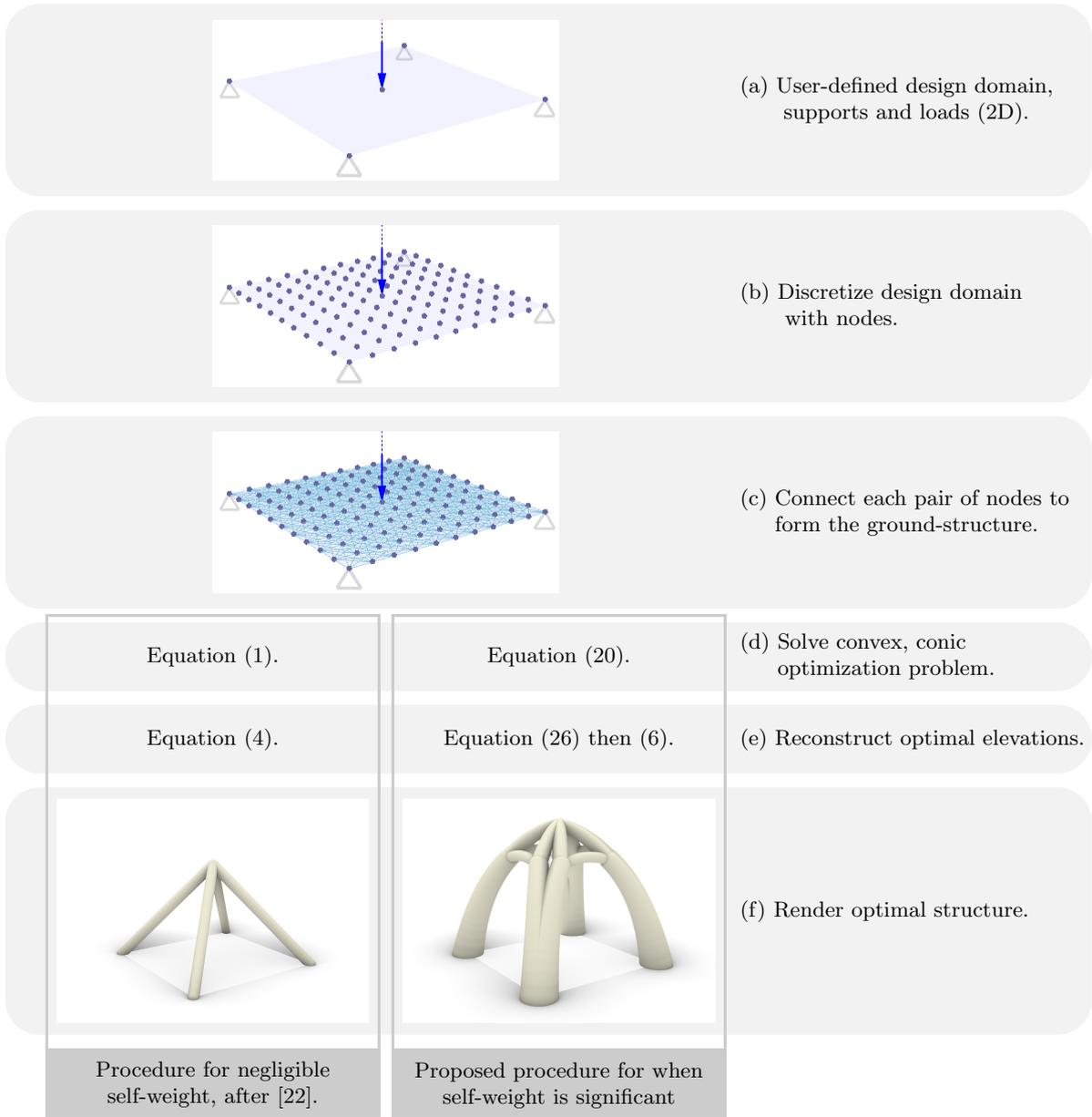
\begin{figure*}
	\centering    
    
        \begin{tikzpicture}
            \definecolor{bg}{rgb}{0.95, 0.95, 0.95}
            \definecolor{type}{rgb}{0.8, 0.8, 0.8}
            \begin{scope}
                \fill[rounded corners=0.5cm, bg] (-10.5, -1.4) rectangle (5.2, 1.4);
                \node [] at (-5cm,0) {\includegraphics[width = 5cm, trim = 7.5cm 0.7cm 7.2cm 7.7cm, clip]{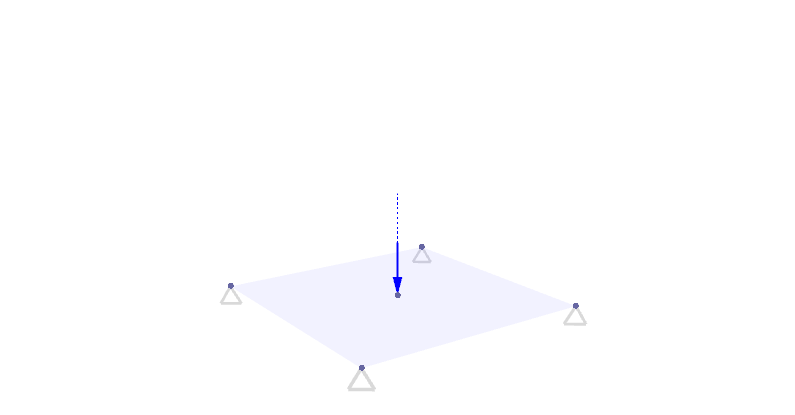}};
                \node [right, align=left, text width = 5cm] at (0,0) {(a) User-defined design domain, \\ \phantom{(a) } supports and loads (2D).};
            \end{scope}

            \begin{scope}[yshift = -3cm]
                \fill[rounded corners=0.5cm, bg] (-10.5, -1.4) rectangle (5.2, 1.4);
                 \node [] at (-5,0) {\includegraphics[width = 5cm, trim = 7.5cm 0.7cm 7.2cm 7.7cm, clip]{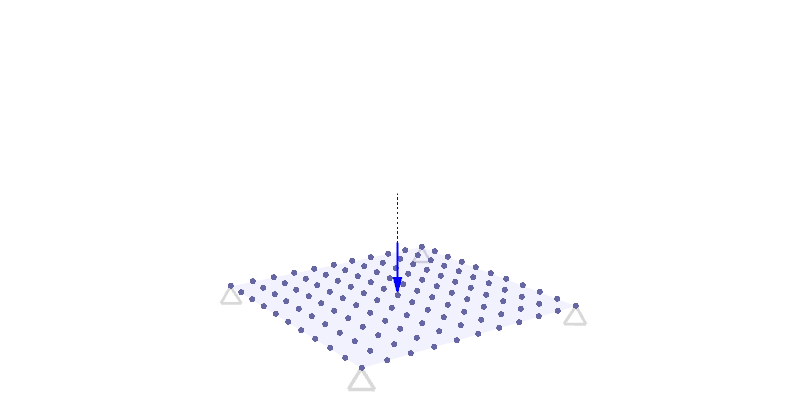}};
                \node [right, align=left, text width = 5cm] at (0,0) {(b) Discretize design domain \\ \phantom{(b) } with nodes.};
            \end{scope}

             \begin{scope}[yshift = -6cm]
                \fill[rounded corners=0.5cm, bg] (-10.5, -1.4) rectangle (5.2, 1.4);
                 \node [] at (-5,0) {\includegraphics[width = 5cm, trim = 7.5cm 0.7cm 7.2cm 7.7cm, clip]{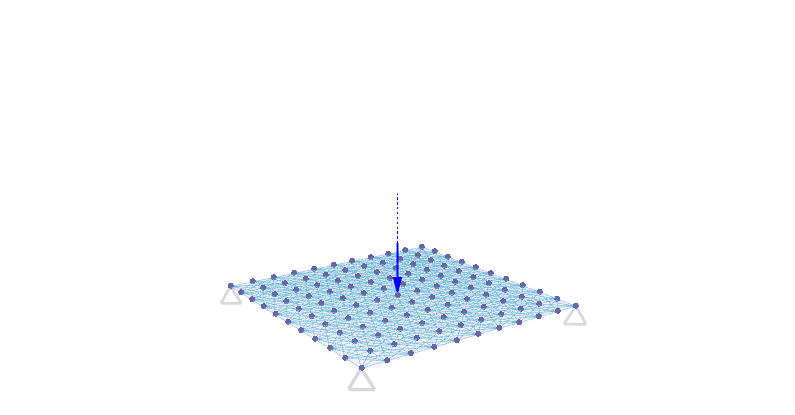}};
                \node [right, align=left, text width = 5cm] at (0,0) {(c) Connect each pair of nodes to \\ \phantom{(c) }form the ground-structure.};
            \end{scope}

            \begin{scope}[yshift = -8.1cm]
                \fill[rounded corners=0.5cm, bg] (-10.5, -0.5) rectangle (5.2, 0.5);
                \node [right, align=left, text width = 5cm] at (0,0) {(d) Solve convex, conic \\ \phantom{(d) }optimization problem.};
                \node at (-7.5,0) {Equation (\ref{eqn:vault}).};
                \node at (-2.5,0) {Equation (\ref{eqn:primal}).};
            \end{scope}

            \begin{scope}[yshift = -9.3cm]
                \fill[rounded corners=0.5cm, bg] (-10.5, -0.5) rectangle (5.2, 0.5);
                \node [right, align=left, text width = 5cm] at (0,0) {(e) Reconstruct optimal elevations.};
                \node at (-7.5,0) {Equation (\ref{eqn:lightVaultZ}).};
                \node at (-2.5,0) {Equation (\ref{eqn:zNodes}) then (\ref{eqn:centerline}).};
            \end{scope}

            \begin{scope}[yshift = -11.8cm]
                 \fill[rounded corners=0.5cm, bg] (-10.5, -1.8) rectangle (5.2, 1.8);
                 \node [] at (-7.5,0) {\includegraphics[width = 4.5cm, trim = 7.8cm 0 7.8cm 0, clip]{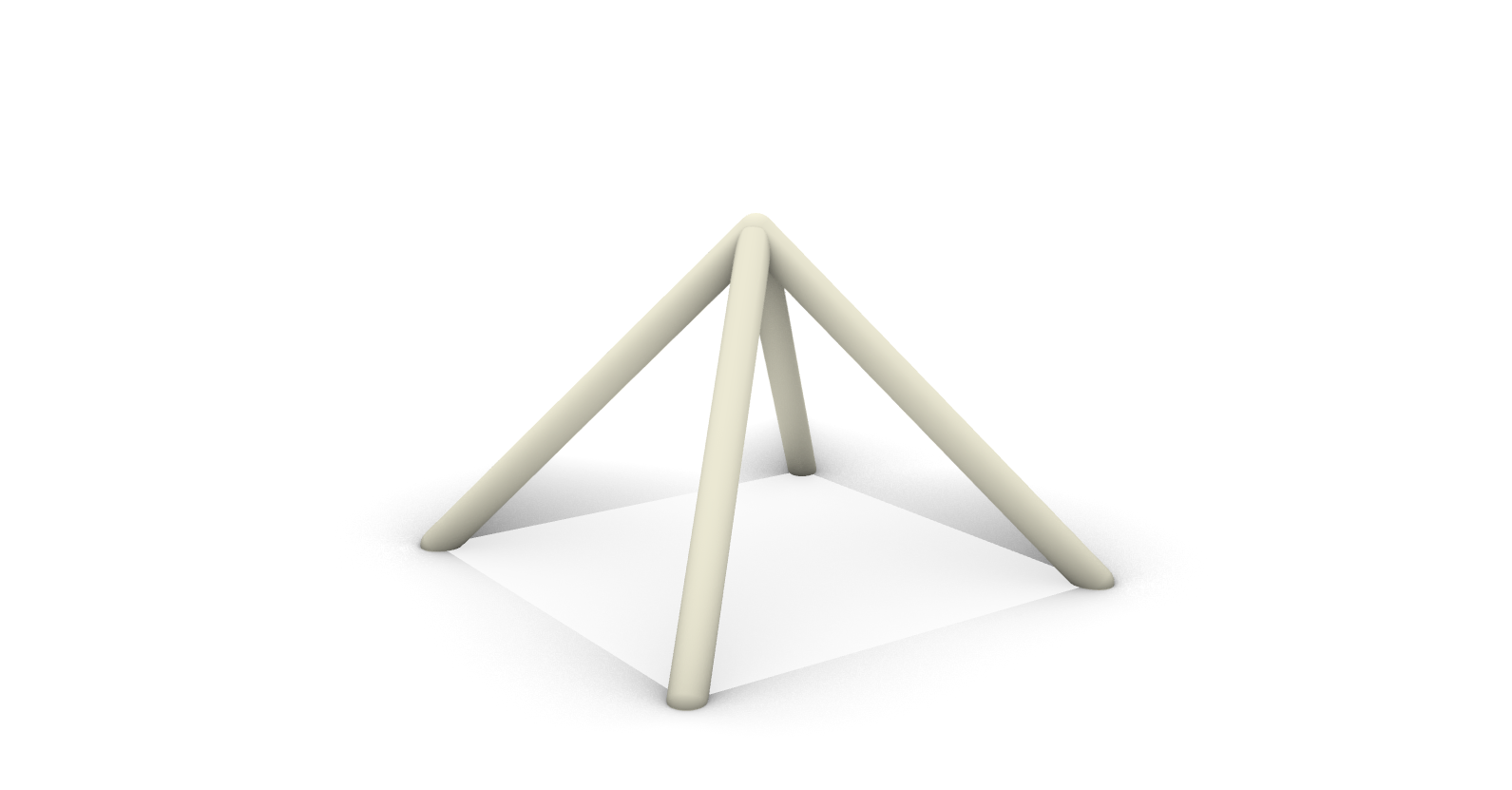}};
                 \node [] at (-2.5,0) {\includegraphics[width = 4.5cm, trim = 7.8cm 0 7.8cm 0, clip]{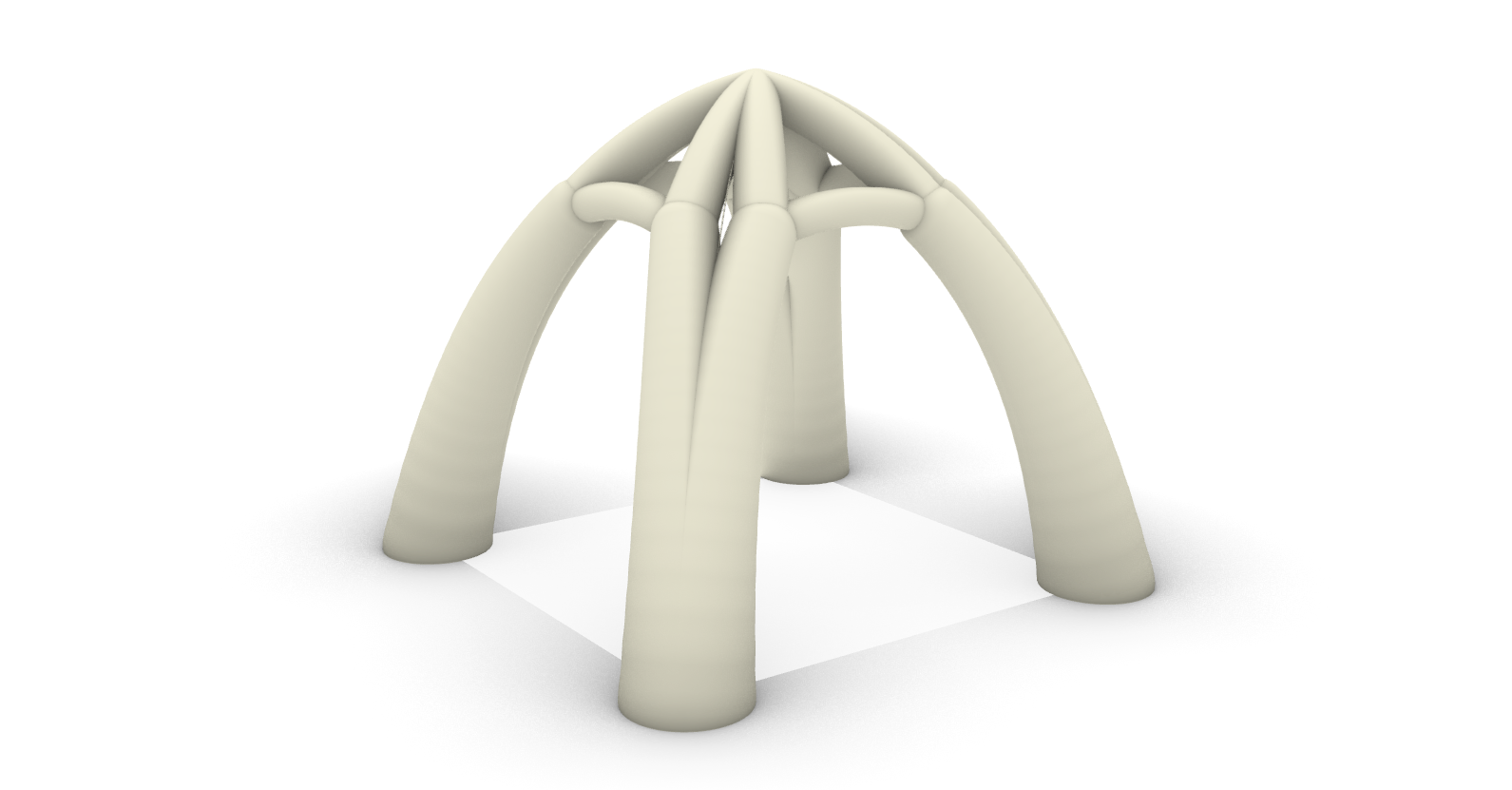}};
                \node [right] at (0,0) {(f) Render optimal structure.};
            \end{scope}

            \begin{scope}[type, very thick]
                \draw (-9.9,-14.8) -| (-5.1,-7.5) -| cycle;
                \draw (-4.9,-14.8) -| (-0.1,-7.5) -| cycle;
                \fill (-9.9,-14.8) rectangle (-5.1, -13.8);
                \fill (-4.9,-14.8) rectangle (-0.1, -13.8);
    
                \node at (-7.5, -14.8) [above, align = center, text width = 4cm, black] {Procedure for negligible self-weight, after \cite{bolbotowski2022optimal}.};
                \node at (-2.5, -14.8) [above, align = center, text width = 4.8cm,black] {Proposed procedure for when  self-weight is significant};
            \end{scope}
        \end{tikzpicture}
		\caption[Summary of methodology developed herein.]{Summary of methodology developed herein, compared to existing approach \cite{bolbotowski2022optimal} for cases where self-weight can be neglected. Note that the initial setup process is common to both cases, with the key differences being the formulation of the optimization problem and the reconstruction of the optimal elevations. The procedures for when self-weight is negligible are recapped in Section \ref{sec:vault}, whilst the new procedure including self-weight is derived in Section \ref{sec:vaultCat}.}
		\label{fig:summary}
	\end{figure*}

	In this paper, the framework of \citet{bolbotowski2022optimal} will be adapted to incorporate the self-weight of the structural grid-shell bars, as summarized in Figure \ref{fig:summary}. The self-weight formulation is based on the catenary approach of \citet{fairclough2018theoretically}, and \ref{sec:lumped} demonstrates why the simpler lumped model cannot be used. The structure of this paper is as follows, Section \ref{sec:formulations} presents existing formulations from \cite{bolbotowski2022optimal} and \cite{fairclough2018theoretically}, which underpin the approach developed here. Section \ref{sec:vaultCat} then derives the new procedure for grid-shell optimization under self-weight loading. The new formulation is then tested on a range of examples in Section \ref{sec:example} and concluding remarks are given in Section \ref{sec:conclusions}.
	
	\begin{figure}
	\tdplotsetmaincoords{60}{105}

	\begin{tikzpicture}[tdplot_main_coords,scale =2.5, >=latex]
		\coordinate (A1) at (2,0,0.4);
		\coordinate (B1) at (1,2,1.2);
		\draw[fill=blue!10, draw=none] 
		(0,0,0) -- (2,0,0) -- (2,2,0) -- (0,2,0) -- cycle;

		\foreach \xA in {0,1,2}{\foreach \yA in {0,1,2}{
				\foreach \xB in {0,1,2}{\foreach \yB in {0,1,2}{
						\draw[groundStruct!70] (\xA, \yA, 0) -- (\xB, \yB, 0);
				}}
		}}	
		\draw[ultra thick, groundStruct] (2,0,0) -- (1,2,0);
		\begin{scope}[canvas is xy plane at z=0]
			\fill[groundStruct] (2,0) -- ++(0,0.5) arc[start angle=90, end angle =117, radius = 0.5] 
			node[right, pos = 0.4,fill=blue!10, circle, inner sep = 0pt, outer sep = 0.5pt, fill opacity = 0.9] {$\phi$} -- cycle;
		\end{scope}
		
		\draw[thick, dashed] (A1) -- (2,0,0);	
		\draw[thick, dashed] (B1) -- (1,2,0);	
		
		\foreach \x in {0,1,2}{	\foreach \y in {0,1,2}{
				\node [circle, fill=ndes, draw =ndes, thick, inner sep=1.5pt] at (\x, \y, 0) {};	
		}}
		
		\draw[ultra thick, straightline, shorten >= 10pt, shorten <= 10pt] (A1) -- (B1);
		\node [left, xshift = -0pt, yshift = -2pt] at (A1) {$\hat{\mathrm{A}}$};
		\node [right, yshift =1pt] at (B1) {$\hat{\mathrm{B}}$};
        
		\begin{scope}[plane origin = {(2,0,0)}, plane y = {(2,0,1)}, plane x = {(2-1/2.23, 2/2.23, 0)}, canvas is plane]
			

			\draw[fill=straightline, straightline] (0, 0.4) -- (0.5,0.4)  arc[start angle = 0 , end angle = 19.7, radius = 0.5] 
			node [right, pos=0.5, fill=blue!10, circle, inner sep = 1pt, outer sep = 0.2pt,fill opacity = 1] {$\theta$} ;
            \draw[straightline, thick] (1, 0.4) -- (0,0.4);
                
			\draw [dashed, thick] (0.5,0.4) -- (0.5, 0);

            \begin{scope}[>=to, thick]
				\draw[<->, black!50!groundStruct, yshift=3pt] (0,0) -- (2.23, 0) node [pos=0.5, inner sep = 1pt, fill=blue!10] {$l$};
				
				\draw[<->, straightline, yshift=7pt, xshift = -2.3pt] (0,0.4) -- (2.23, 1.2) node [pos=0.5, fill=white, inner sep =2pt] {$\hat{l}$};
                \draw [thin, straightline, shorten >=-3pt] (0,0.4) --++(-2.3pt,7pt); 
                \draw [thin, straightline, shorten >=-3pt] (2.23, 1.2) --++(-2.3pt,7pt); 
			\end{scope}

			\begin{scope}[straightline!60]
				\draw [thick] (0,0.4) |- ++(19.7:0.7) node [pos=0.7, above, inner sep = 0pt, outer sep = 1pt] {$s$} node[pos=0.25, left,  fill=white, fill opacity =0.9, text opacity = 1, inner sep = 0.2pt, outer sep = 1pt] {$q$};
                \draw [<-, line width = \fthick, shorten <=2pt, straightline!45] (0,0.4) -- ++(19.7:0.7) node[below, pos=0.12, inner sep = 0pt, outer sep = 3pt] {$\hat{q}$};

                \draw [thick] (2.23, 1.2) -| ++(19.7:-0.7) node [pos=0.3, above, inner sep = 0pt, outer sep = 1pt] {$s$} node[pos=0.75, left,inner sep = 0pt, outer sep = 1pt] {$q$};
				\draw [<-,line width = \fthick, shorten <=2pt, straightline!45] (2.23, 1.2) -- ++(19.7:-0.7) node[below, pos=0.12, inner sep = 0pt, outer sep = 3pt] {$\hat{q}$};
				
			\end{scope}

		\end{scope}

		\node [left, ndes] at (2,0,0) {$\mathrm{A}$};
		\node [right, ndes] at (1,2,0) {$\mathrm{B}$};

		\begin{scope}[every node/.style={thick, draw=black, circle, fill=white, inner sep = 2pt}]
			\node at (A1) {};\node at (B1) {};
		\end{scope}
		
		\coordinate (co) at (2.7,-0.3,0);
		\draw[->] (co)-- ++(-0.5,0,0) node[pos=1.2] {$y$};
		\draw[->] (co)-- ++(0,0.5,0) node[pos=1.2] {$x$};
		\draw[->] (co)-- ++(0,0,0.5) node[pos=1.2] {$z$};
		
	\end{tikzpicture}
	\caption{Notation for grid-shell topology and geometry optimization without self-weight. The blue lines show elements in the ground structure defined in the horizontal plane. The highlighted element is also shown (in brown) as it would be in the optimal solution, once the end-points are projected to their optimal elevations for the final structure. The force, $\hat{q}$, acting on the nodes is aligned to the final centre-line of the element, i.e. at an angle of $\theta$ to the horizontal.}
	\label{fig:straightVault}
	\end{figure}
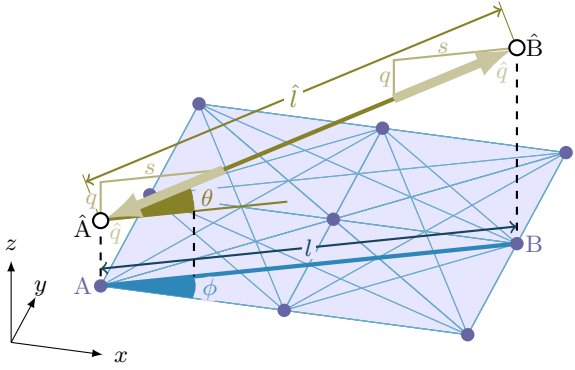
	
 \section{Previous Formulations}
	\label{sec:formulations}
	\subsection{Grid-shell topology and geometry optimization without self-weight}
	\label{sec:vault}
	Firstly, the weightless vault formulation of \cite{bolbotowski2022optimal} is recalled. For clarity, only the discretised formulation is discussed here. In this approach, the allowable design domain is defined in plan, and locations of applied loads and supports are also specified (Figure \ref{fig:summary}a). This 2D domain is discretised using a dense array of nodes (Figure \ref{fig:summary}b), and every pair of nodes is connected by a potential structural element (Figure \ref{fig:summary}c). Conceptually, nodes are allowed to move in the vertical direction to obtain the optimal shape (except where vertical support is provided at a node), and any applied forces move vertically with the associated node.
	
    When formulating the mathematical optimization problem, for each potential element two optimization variables are required, $s_i$ and $q_i$, representing the horizontal and vertical component of the axial force in element $i$, over the set of members $M$. The objective function is to minimize the total volume, calculated on a per-element basis and then summed over all elements in the ground-structure. For a single element, as shown in Figure \ref{fig:straightVault} with stress $\sigma_i$ the volume is given by $\hat{l}\frac{\hat{q}}{\sigma_i}$. Geometrical relationships from Figure \ref{fig:straightVault} can then be used to re-write this in terms of $s_i$ and $q_i$

    The required optimization problem is then given by,	
	\begin{subequations} \label{eqn:vault}
	\begin{align}
		\min \sum_{i\in M}& \frac{l_i}{\sigma_i} \left(s_i + \frac{q_i^2}{s_i}\right), \label{eqn:vaultObj}\\ 
		\text{s.t. } \mathbf{Bs} &= \mathbf{f}_{xy}, \label{eqn:vaultInplane} \\
		\mathbf{Dq} &= \mathbf{f}_z, \label{eqn:vaultOutplane}\\
		\mathbf{s} & \geq 0, \label{eqn:vaultComp}
	\end{align}
	\end{subequations}

    \noindent
	where $l_i$ is the horizontal length of potential structural element $i$, and the objective function to be minimized being the summation of all element volumes. The matrix $\mathbf{B}$ contains in-plane direction cosines ($\pm\cos\phi$ and $\pm\sin\phi$) such that (\ref{eqn:vaultInplane}) enforces in-plane equilibrium at each node subject to horizontal components of external forces $\mathbf{f}_{xy}$. Matrix $\mathbf{D}$ contains entries of 0, -1 and 1 such that (\ref{eqn:vaultOutplane}) enforces equilibrium in the vertical direction at each node with vertical components of external forces $\mathbf{f}_z$. Finally, the positivity of the $s$ variables ensures that all elements remain in compression. Conversely, the $q$ variables may be positive or negative, corresponding to elements with an upwards or downwards slope respectively. Note that, as only a single loading condition is permitted, for practical problems $\mathbf{f}_{xy}$ is typically zero.
		
	Problem (\ref{eqn:vault}) can be transformed into a convex conic problem by assigning an additional variable $r_i$ to each potential element, which is defined as
	\begin{equation}
		2r_i \geq \frac{q_i^2}{s_i}. \label{eqn:rVar}
	\end{equation}
	The variables $r_i$ are then used in the objective function. Note that this transformation is a slight relaxation of the problem (\ref{eqn:vault}), in that the original problem demands equality of (\ref{eqn:rVar}), however this would not be a convex constraint. However, as the introduction of any slack in (\ref{eqn:rVar}) would increase the objective without affecting any other constraints, it is easy to see that this relaxation will not be in effect at the optimum solution.
	
	Note that a feasible solution to the formulation (\ref{eqn:vault}) does not require that the resulting structure be single layered. However, it was proven in \cite{bolbotowski2022optimal} that an \textit{optimal} solution to the problem will permit a consistent, single-layered elevation function across the domain. A more intuitive demonstration for this was also given in \cite[][Fig 3]{he2025minimum}.  It is possible to obtain the required heights of each node of the gridshell in the final optimal solution by progressively moving along each non-zero element in the solution, calculating its slope by the ratio between $q_i$ and $s_i$ \cite{he2025minimum}. However, the elevations can also be obtained directly from the dual of problem (\ref{eqn:vault}). The dual problem can be obtained through conic duality principles \cite{boyd2004convex} or derived through kinematic principles \cite{bolbotowski2022optimal} and is written as, 

    \begin{subequations}
	\begin{align}
		\max_{\mathbf{u}, \mathbf{w}, \mathbf{t}_1, \mathbf{t}_2,\mathbf{t}_3} \mathbf{f}_z^T \mathbf{w} + \mathbf{f}_{xy}^T \mathbf{u} & \\
            \text{s.t.  }\mathbf{B}^T \mathbf{u} + \mathbf{t}_2 &= \frac{1}{\sigma}\mathbf{l}, \\
            \mathbf{D}^T \mathbf{w} + \mathbf{t}_3 &= \mathbf{0}, \\
            \mathbf{t}_1 &= \frac{2}{\sigma}\mathbf{l}, \\
            \big(2t_{1,i}t_{2,i} & \geq t_{3,i}^2 \big) \forall i \in M.		
	\end{align}
    \end{subequations}

	In this formulation, the variables $\mathbf{u}$ and $\mathbf{w}$ represent virtual displacements in the horizontal and vertical direction. The optimal elevation, $z$ at a point was then found by \citet{bolbotowski2022optimal} to be related to the virtual strain at the optimum by the relationship
    \begin{equation}
        z = \frac{1}{2} w.
        \label{eqn:lightVaultZ}
    \end{equation}

    However, a structure designed through this approach neglects the impact of the self-weight of the structure itself. This will lead to unsafe and under-designed structures, particularly as span lengths increase.

	\subsection{Truss topology optimization with continuous self-weight}
	\label{sec:cat}
		
	\begin{figure}
    \centering
		\begin{tikzpicture}[scale=1.4]
				\coordinate (A) at (0,{ln(cos(deg(0.7)))/0.14)});
				\coordinate (B) at (4, {ln(cos(deg(0.7-0.14*4)))/0.14)});	
                \path (A)--(B) coordinate [pos=0.45](T);
				\draw[ultra thick, smooth, samples=20, domain=-0:4, curveline]	plot({\x}, {(ln(cos(deg(0.7-0.14*\x)))/0.14} );
								
				\draw [thick, straightline, fill=straightline](T) ++ (0.4, 0)  arc(0:deg(0.42):0.4) node [right, pos = 0.6, yshift=1pt] {$\theta$};
				\draw [very thick, straightline, dotted, shorten <=2cm, shorten > = 2cm] (A) --(B);
				\draw [thick, straightline] (T)-- ++ (0.5,0);

				\draw [thick, curveline!50!black, fill =curveline!50!black ](A) --++ (0.5, 0)  arc(0:deg(0.7)-2:0.5) 
				node [above right, pos = 0, circle, inner sep = 0pt, fill opacity = 0.7, text opacity = 1, xshift = 2pt] {$\alpha_{A}$};
				\draw [thick, curveline!50!black, fill =curveline!50!black ](B) --++ (0.5, 0) arc(0:deg(0.7-0.14*4):0.5) node [below, pos = 0] {$\alpha_{B}$};
                \draw [thick] (B) --++ (0.6,0);
                \draw [thick, curveline] (B) --++ (8:0.6);
				
				\draw [straightline, very thick] (A) -- ++ ({deg(0.425)}: 1.4) node [pos=0.75, fill=white, inner sep = 1pt] {$\hat{q}$}; 
				
				\draw [straightline, very thick] (B) -- ++ ({deg(0.42)}: -1.4) node [pos=0.5, below, inner sep = 1pt] {$\hat{q}$}; 
				
				\draw [thick, curveline!70] (A) -| ++ (40: 1.66) node [pos=0.7, right, inner sep = 1.5pt, outer sep = 0pt] {$\hat{q}\cos\theta\tan\alpha_A$} node [pos= 0.25, below, inner sep = 1.5pt, outer sep = 0pt] {$\hat{q}\cos\theta$};
				\draw [thick, curveline!70, shorten <=10pt] (B) -| ++ (8: -1.28)  node [pos=0.7, left, inner sep = 1.5pt, outer sep = 0pt] {$\hat{q}\cos\theta\tan\alpha_B$} node [pos= 0.3, above, inner sep = 1.5pt, outer sep = 0pt] {$\hat{q}\cos\theta$};	
                \draw [thin, straightline!50] (B) ++ (8: -1.28) -- ++ (0,-0.4);
				
                \draw [latex-, line width = \fthick, curveline!70, shorten <=2pt] (A) -- ++ (40: 1.66);
				\draw [latex-, line width=\fthick, curveline!70, shorten <=2pt] (B) -- ++ (8: -1.28);
					
				
				\node [thick, draw=black, circle, fill=white, inner sep = 2pt] at (A) {};
				\node [thick, draw=black, circle, fill=white, inner sep = 2pt] at (B) {};
				\node [left] at (A) {A};
				\node [below] at (B) {B};

                \begin{scope}[>=latex]
                    \coordinate (co) at (4, -1.8);
                    \draw[->] (co)-- ++(0.6,0) node[pos=1.2] {$\dot{x}$};
                    \draw[->] (co)-- ++(0,0.6) node[pos=1.2] {$z$};
                \end{scope}
		\end{tikzpicture}
		\caption{Notation for the problem of truss topology optimization with continuous self-weight, a single element shown in elevation in the vertical plane containing the element. The element number subscript $i$ is omitted for clarity. The indicated forces acting on the nodes are aligned to the element centreline at the respective point.}
		\label{fig:catenary}
	\end{figure}
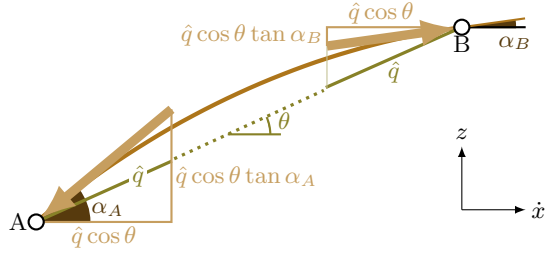

    As previously mentioned, it is common in truss optimization problems to model self-weight as forces applied directly onto the endpoints of an element, neglecting the effect within the element itself. Whilst this has been effective as an approximation for trusses of moderate span, it is not possible to use this approach within the framework outlined in Section \ref{sec:vault}. This is discussed in detail in \ref{sec:lumped}. Instead, a model must be applied which correctly accounts for the self-weight in a continuous manner, for this purpose the formulation presented in \citet{fairclough2018theoretically} is employed. 
    
	In this section, the catenary truss formulation of  \cite{fairclough2018theoretically} will be recalled in a format most conducive to adaptation to the vault problem. To simplify the upcoming adaption to compressive grid-shell problems, it will be assumed that only compressive members are to be permitted, and that the ground structure does not contain any elements which are exactly aligned in the vertical direction (as those require special treatment \cite{fairclough2018theoretically}). 
    
    It is here necessary to consider the 3D case, which was not outlined in \cite{fairclough2018theoretically}, but is easily obtained with just a few modifications. Specifically, each element can be thought of within its own local 2D coordinate system in the plane containing the element and the vertical axis, giving exactly the results in \cite{fairclough2018theoretically}, as shown in Figure \ref{fig:catenary}. The end-forces then have to be returned to the global coordinate system. Vertical forces are unchanged, but the horizontal force $\hat{q}\cos\theta$ must be resolved by using the in-plane angle $\phi$ as defined in Fig. \ref{fig:straightVault} to become $\hat{q}\cos\theta\sin\phi$ and $\hat{q}\cos\theta\cos\phi$ in the $x$ and $y$ directions respectively. 
    
    The resulting optimization problem thus becomes,
	\begin{subequations}
	\begin{align}
		\min_\mathbf{\hat{q}}  \sum_{i\in M}& \frac{\hat{q}_i \cos\theta_i}{\rho g} \left(\tan\alpha_{B,i} - \tan\alpha_{A,i}\right), \label{eqn:trussObj}\\
		\text{s.t. } \bar{\mathbf{B}}\mathbf{\hat{q}} &= \mathbf{f}_{xy}, \\
		\bar{\mathbf{D}}\mathbf{\hat{q}} &= \mathbf{f}_z, \\
		\hat{\mathbf{q}} &\geq \mathbf{0},
	\end{align}
    \label{eqn:catenaryFormulation}
	\end{subequations}
	where $\bar{\mathbf{B}}$ contains direction cosines in the $x$ and $y$ directions ($\pm\cos\theta_i\cos\phi_i$ and $\pm\cos\theta_i\sin\phi_i$ respectively). The matrix $\bar{\mathbf{D}}$ contains the vertical coefficients,\linebreak $\cos\theta_i\tan\alpha_{A,i}$ and $\cos\theta_i\tan\alpha_{B,i}$. The optimization variables, here denoted by $\mathbf{\hat{q}} = [\hat{q}_1, \hat{q}_2, ..., \hat{q}_m]^\mathrm{T}$, %
    are defined as the equivalent force directed straight between the points A and B. $\rho g$ is the unit weight of the material.

	To calculate the inclination angles $\alpha$, it is required to make reference to the definition of the catenary of equal stress from e.g. \cite[Eqn. 2.1]{fairclough2018theoretically} or \cite[Art. 453]{routh1896treatise}, the centre-line of the element follows a curve which can be expressed in a local 2D element coordinate system $(\dot{x}, z)$ as
    \begin{equation}
        \frac{\rho g}{\sigma} z = \log\left(\cos\left(-\frac{\rho g}{\sigma} \dot{x} + C_1\right)\right) + C_2
        \label{eqn:centerline}
    \end{equation}
    where $C_1$ and $C_2$ are constants which define the translation of the shape in the plane. The allowable stress of the material is represented by $\sigma$. Within this equation, the term $-\frac{\rho g}{\sigma} \dot{x} + C_1$ gives the inclination angle $\alpha$ for any point. By defining the end-points to be at $A=(0,z_A)$ and $B=(l,z_B)$, explicit relationships for $\tan\alpha_A$ and $\tan\alpha_B$ can be obtained.
	\begin{align}
		\tan\alpha_A &= -\frac{\cos(\frac{\rho g}{\sigma} l) - \exp\left({\frac{\rho g }{\sigma} (z_B-z_A)}\right)}{\sin(\frac{\rho g}{\sigma} l)},\label{eqn:alphasA} \\
	\tan\alpha_B &= \frac{\cos(\frac{\rho g}{\sigma} l) - \exp\left({-\frac{\rho g }{\sigma} (z_B-z_A)}\right)}{\sin(\frac{\rho g}{\sigma} l)}.\label{eqn:alphas}
	\end{align}
    
    It should be noted that the horizontal length of an element must be less than $\frac{\pi\sigma}{\rho g}$, otherwise the element is excluded from the ground structure, see \cite{fairclough2018theoretically}. Accordingly, $\sin(\frac{\rho g}{\sigma}l)$ is always a positive number ranging in the open interval $(0,1)$, while $\cos(\frac{\rho g}{\sigma} l)$ can be negative as it varies in $(-1,1)$.
    \color{black}

    \section{Grid-shell topology and geometry optimization with continuously distributed self-weight}
	\label{sec:vaultCat}

    In this section, the two methods above will be combined to give an approach suitable for obtaining the optimal grid-shell subjected to both external loading and its own self-weight. Firstly, in Section \ref{sec:buildingFormulation} the relevant optimization problem is constructed and cast as a conic problem, which can be easily solved to global optimality. As some relaxations are required to obtain this form, Section \ref{sec:num2phy} describes the procedure for reconstructing the physical structure from the solution of the optimization problem. Specifically, it is shown that the relaxations required do not affect the solution at the optimum points. 
    
	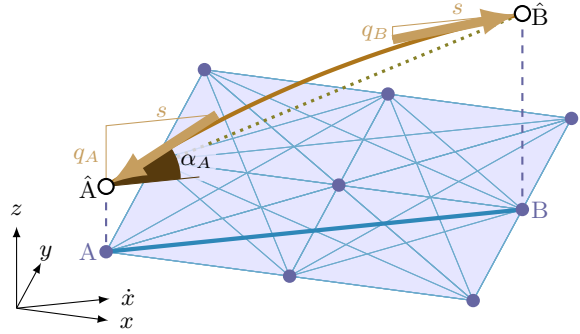
\begin{figure}
		\centering
		\tdplotsetmaincoords{60}{105}
		\begin{tikzpicture}[tdplot_main_coords,scale =2.5, >=latex]
			\coordinate (A1) at (2,0,0.4);
			\coordinate (B1) at (1,2,1.2);
			\draw[fill=blue!10, draw=none] 
			(0,0,0) -- (2,0,0) -- (2,2,0) -- (0,2,0) -- cycle;

			\foreach \xA in {0,1,2}{\foreach \yA in {0,1,2}{
				\foreach \xB in {0,1,2}{\foreach \yB in {0,1,2}{
					\draw[groundStruct!70] (\xA, \yA, 0) -- (\xB, \yB, 0);
				}}
			}}	
			\draw[ultra thick, groundStruct] (2,0,0) -- (1,2,0);

			\draw[thick, dashed, ndes] (A1) -- (2,0,0);	
			\draw[thick, dashed, ndes] (B1) -- (1,2,0);	
		
			\foreach \x in {0,1,2}{	\foreach \y in {0,1,2}{
					\node [circle, fill=ndes, draw =ndes, thick, inner sep=1.5pt] at (\x, \y, 0) {};	
			}}

			\begin{scope}[plane origin = {(2,0,0)}, plane y = {(2,0,1)}, plane x = {(2-1/2.23, 2/2.23, 0)}, canvas is plane]
				\draw[very thick, straightline, dotted] (A1) -- (B1);
				\draw[ultra thick, smooth, samples=20, domain=-0:2.23, curveline]	plot({\x}, {1.23+(ln(cos(deg(0.56-0.2*\x)))/0.2} );
				
				\node at (0.37, 0.51) [right, fill=blue!10, fill opacity = 0.8,text opacity = 1, inner sep = 1pt, circle] {$\alpha_A$};
				\draw[curveline!50!black] (A1) -- ++(0.5,0);
				\fill[curveline!50!black] (A1) -- ++(0.4,0) arc(0:deg(0.56)-2:0.4); 
					
				\draw[curveline!70] (A1) |- ++({deg(0.56)}: 0.7)
					node [pos=0.25, left, inner sep=2pt] {$q_A$} 
					node [pos=0.75, above, inner sep=1pt] {$s$};
				\draw[curveline!70, line width = \fthick, latex-, shorten <=2pt] (A1) -- ++({deg(0.56)}: 0.7);
				
				\draw[curveline!70] (B1) -| ++({deg(0.114)}: -0.7)
					node [pos=0.25, above, inner sep=2pt] {$s$} 
					node [pos=0.75, left, inner sep=1pt] {$q_B$};
				\draw[curveline!70, line width = \fthick, latex-, shorten <=2pt] (B1) -- ++({deg(0.114)}: -0.7);
			\end{scope}

			\begin{scope}[every node/.style={thick, draw=black, circle, fill=white, inner sep = 2pt}]
			\node at (A1) {};			
            \node at (B1) {};
			\end{scope}
			\node at (A1) [left] {$\hat{\mathrm{A}}$};
			\node at (B1) [right] {$\hat{\mathrm{B}}$};
            \node [left, ndes] at (2,0,0) {$\mathrm{A}$};
		  \node [right, ndes] at (1,2,0) {$\mathrm{B}$};

            \begin{scope}[>=latex]
            \coordinate (co) at (2.7,-0.3,0);
			\draw[->] (co)-- ++(-0.5,0,0) node[pos=1.2] {$y$};
    		\draw[->] (co)-- ++(0,0.5,0) node[pos=1.2] {$x$};
    		\draw[->] (co)-- ++(0,0,0.5) node[pos=1.2] {$z$};
            \draw[->] (co)-- ++(-0.2236,0.4472)  node[pos=1.2] {$\dot{x}$};
            \end{scope}
			
		\end{tikzpicture}
	\caption{Notation for catenary elements in the vault formulation. The forces acting on the nodes are as shown. $s$, $q_A$ and $q_B$ are the corresponding optimization variables.}
	\label{fig:singleElement}
	\end{figure}
	
	\subsection{Formulation of the optimization problem}
	\label{sec:buildingFormulation}

    To formulate the convex optimization problem, this subsection will first set out the required optimization variables (Section \ref{sec:vars}), and then derive the required objective function (Section \ref{sec:obj} and constraints (Sections \ref{sec:equilibrium} and \ref{sec:geometrical}). The required conic optimization problem can then be written (Section \ref{sec:complete}), and strategies to speed up solving discussed (Section \ref{sec:memAdding}). 
    
	\subsubsection{Optimization variables}
	\label{sec:vars}
	The key difference between elements in the catenary formulation in (Section \ref{sec:cat}) and the grid-shell optimization framework (Section \ref{sec:vault}) is that the vertical forces at each end of an element are no longer equal. Thus, the required optimization variables to combine these two approaches will be:
	\begin{itemize}
		\item $\mathbf{s} = [s_1, s_2, ..., s_m]^\T$, variables representing the horizontal component of force in each element. These variables are non-negative, with positive values representing forces acting in a compressive manner.
		\item $\mathbf{q}_A = [q_{A,1}, q_{A,2}, ..., q_{A, m}]^\T$, variables representing vertical component of the force at the start of each element. Positive values represent downwards force, i.e. Fig \ref{fig:singleElement} shows a positive value for $q_{A,i}$. 
		\item $\mathbf{q}_B = [q_{B,1}, q_{B,1}, ..., q_{B, m}]^\T$, variables representing vertical component of the force at the end of each element. Positive values represent downwards force, i.e. Fig \ref{fig:singleElement} shows a negative value for $q_{B,i}$. 
	\end{itemize}
	
	These variables are shown visually for a single element in Figure \ref{fig:singleElement}. Note that quantities defined in Figures \ref{fig:straightVault} and \ref{fig:catenary} are also required for this formulation. In particular, $l$ refers to the horizontal length of the element as in Figure \ref{fig:straightVault}, and $\alpha$, $\theta$ and $\hat{q}$ are as defined in Figure \ref{fig:catenary}.
	
    
    Conceptually, the objective function and constraints are unchanged from those of the previous formulations. Note that in problems (\ref{eqn:vault}) and (\ref{eqn:catenaryFormulation}) the objectives and constraints address the same physical concerns -- to minimize structural volume, whilst enforcing equilibrium horizontally and vertically. This is also the overall structure of the formulation derived herein. As in (\ref{eqn:vault}), the nodal elevations $z$ will not be explicitly present in the formulation as design variables. Instead, they will be cleverly recast after solving the optimization problem to be put forth, see Section \ref{sec:elevations} below for details.
    
    In the remainder of this section, the subscript indicating element number will be dropped for clarity.

	\subsubsection{Objective function}
        \label{sec:obj}
	In the catenary self-weight formulation in Section \ref{sec:cat}, the horizontal component of the element force was defined based on the inline force $\hat{q}$. Converting this to the variables required in the grid-shell formulation necessitates the relationship,
	\begin{equation}
	s = \hat{q} \cos\theta. \label{eqn:qtos}
	\end{equation}

    \noindent
	From Fig. \ref{fig:singleElement}, the following relationships can be observed,
	\begin{align}
		\frac{q_{A}}{s} = \tan\alpha_{A} \label{eqn:tanA},  \\
		-\frac{q_{B}}{s} = \tan\alpha_{B} \label{eqn:tanB}.
	\end{align}
 	Combining (\ref{eqn:qtos}), (\ref{eqn:tanA}) and (\ref{eqn:tanB}) with (\ref{eqn:trussObj}) gives the following expression for the volume of a single bar, $V_i$,
 	\begin{align}
 		V &=  \frac{s}{\rho g} \left(\frac{q_{B}}{s} + \frac{q_{A}}{s}\right)\nonumber, \\
 		&= \frac{1}{\rho g} (q_{B} + q_{A}).
 	\end{align}
 
 	This can also be intuitively understood as the total downwards force from the element (i.e. its weight) divided by the unit weight to give the volume. 
 
	\subsubsection{Equilibrium}
	\label{sec:equilibrium}
	Equilibrium is enforced in much the same manner as in the original vault formulation. In-plane and out-of-plane equilibrium are considered separately. In-plane equilibrium is entirely unchanged, and is enforced by the constraints
	\begin{equation}
		\mathbf{Bs} = \mathbf{f}_{xy},
	\end{equation}
	where $\mathbf{B}$ contains entries of $\pm\cos\phi$ and $\pm\sin\phi$.
	
	For out-of-plane equilibrium, slight changes are needed to allow for the presence of separate force variables in $\mathbf{q}_A$ and $\mathbf{q}_B$.
	\begin{equation}
		\mathbf{D}_A\mathbf{q}_A + \mathbf{D}_B\mathbf{q}_B = \mathbf{f}_{z},
        \label{eqn:eqeq_z}
	\end{equation}
	where $\mathbf{D}_A$ (resp. $\mathbf{D}_B$) contains all zeros except in position $ij$ where node $j$ is the start (resp. end) node of element $i$ where it contains 1. 
	
	\subsubsection{Geometrical coupling and its relaxation to a conic constraint}
	\label{sec:geometrical}
	
    The values of $s$, $q_A$ and $q_B$ should be such that the total forces at A and B are aligned to the centreline of a catenary of equal stress connecting the points A and B  whose elevations are equal to $z_A$ and $z_B$, respectively, cf. Figure \ref{fig:singleElement}.    This leads to a coupling condition for the vectors $\mathbf{s}, \mathbf{q}_A, \mathbf{q}_B$. 
        
        In the case of the grid-shells without the self-weight this coupling was simple: $q_A=-q_B= s\, \frac{z_B-z_A}{l}$, see \cite{bolbotowski2022optimal}.
        For an equally stressed catenary the relations are more involved, and they can be obtained by combining the two groups of equations for the tangents $\tan \alpha_A$, $\tan\alpha_B$: \eqref{eqn:alphasA},\,\eqref{eqn:alphas} and \eqref{eqn:tanA},\,\eqref{eqn:tanB}. These lead to,
        \begin{align}
		q_A &= -s \,\frac{\cos(\frac{\rho g}{\sigma} l) - \exp\left({\frac{\rho g }{\sigma} {(z_B-z_A)}}\right)}{\sin(\frac{\rho g}{\sigma} l)}, \label{eqn:couplingA} \\
	q_B & =- s\, \frac{\cos(\frac{\rho g}{\sigma} l) - \exp\left({-\frac{\rho g }{\sigma} (z_B-z_A)}\right)}{\sin(\frac{\rho g}{\sigma} l)}.\label{eqn:couplingB}	
	\end{align}
    
        It is crucial to observe that the elevation differences $z_B-z_A = z_{B,i} - z_{A,i}$ cannot  vary arbitrarily from element to element. In fact, if $\mathbf{z}$ is the vector of nodal elevations, then $z_{A,i}$, $z_{B,i}$ are the $i$-th entries of, respectively, $\mathbf{D}_A^\mathbf{T} \mathbf{z}$, $\mathbf{D}_B^\mathbf{T} \mathbf{z}$. Accordingly, \eqref{eqn:couplingA},\,\eqref{eqn:couplingB} not only entail local couplings between $s_i$, $q_{A,i}$, $q_{B,i}$ for each element $i$, but also a global compatibility condition for the vectors $\mathbf{s}, \mathbf{q}_A, \mathbf{q}_B$.

        A simple and natural fix would be to add the nodal elevations $\mathbf{z}$ as variables and explicitly enforce \eqref{eqn:couplingA},\,\eqref{eqn:couplingB}. This, however, would lead to a non-convex optimization problem and, as a result, to compromising the efficiency of the method. Instead, following the idea of \cite{bolbotowski2022optimal} for grid-shells without self-weight, conditions \eqref{eqn:couplingA},\,\eqref{eqn:couplingB} will undergo a suitable relaxation.

    The relaxation will be performed in two steps. The first consists in eliminating the slopes $z_B-z_A$. A slight rewriting of \eqref{eqn:couplingA},\,\eqref{eqn:couplingB} and putting $\bar{l} = \K l$ for brevity yields
         \begin{subequations}
         \begin{align}
            \label{eqn:comp1}
		\sin \bar{l}\, q_A + \cos \bar{l} \,s &= s \,\exp\left({\frac{\rho g }{\sigma} (z_B-z_A)}\right), \\
        \label{eqn:comp2}
        \sin \bar{l}\, q_B + \cos \bar{l} \,s &= s \,\exp\left({-\frac{\rho g }{\sigma} (z_B-z_A)}\right).
	\end{align}
            \label{eqn:comp}
        \end{subequations}
        Side-wise multiplication of the two equalities eliminates the $z$ values and leads to
        \begin{equation}
            \label{eqn:shape1}
            \big(\sin \bar{l}\, q_A + \cos \bar{l} \,s \big) \big(\sin \bar{l}\, q_B + \cos \bar{l} \,  s \big) = s^2.
        \end{equation}
        The above is a local coupling; when $s>0$, it gives the condition for the triple $s,q_A,q_B$, which corresponds to the shape of an equally stressed catenary \textit{for some} slope $z_B-z_A$, see Figure \ref{fig:singleElement}. The global compatibility with an elevation vector $\mathbf{z}$ is now forgotten (although it will be shown in Section \ref{sec:num2phy} that this global compatibility can be later recovered for optimal solutions).

        Adding the equation \eqref{eqn:shape1} would still deprive the formulation of convexity. It must be further relaxed to an inequality,
        \begin{equation}
            \big(\sin \bar{l}\, q_A + \cos \bar{l} \,s \big) \big(\sin \bar{l}\, q_B + \cos \bar{l} \,  s \big) \geq s^2. \label{eqn:conic_form}
        \end{equation}
        Indeed, the latter can be written as the standard rotated conic quadratic constraint: $2t_1 t_2 \geq t_3^2$, where $t_1,t_2$ are the bracketed terms on the left-hand side and $t_3 = \sqrt{2} s$. It is a convex constraint that can be efficiently tackled by  modern convex optimization software. It is worth emphasizing that imposing the conic constraint  $2t_1 t_2 \geq t_3^2$ canonically comes with the inequalities $t_1 \geq 0$, $t_2 \geq 0$. These are valid here: the two factors on the left-hand side of \eqref{eqn:conic_form} are always non-negative, which can be seen from \eqref{eqn:comp}.

	\subsubsection{Complete formulation}
	\label{sec:complete}
	Combining the equations found in Sections \ref{sec:obj}, \ref{sec:equilibrium} and \ref{sec:geometrical} gives the following conic optimization problem,	
	\begin{subequations}
		\label{eqn:primal}
	\begin{align}
		\min_{\mathbf{s}, \mathbf{q_A}, \mathbf{q_B}}& \frac{1}{\rho g} \mathbf{1}^{\mathbf{T}}(\mathbf{q}_A + \mathbf{q}_B), \\
		\text {s.t.: } & \mathbf{Bs} = \mathbf{f}_{xy}, \label{eqn:horizEq} \\
		&\mathbf{D}_A \mathbf{q}_A + \mathbf{D}_B \mathbf{q}_B = \mathbf{f}_z, \label{eqn:verticalEq} \\
        &\big(\sin \bar{l}\, q_A + \cos \bar{l} \,s \big) \big(\sin \bar{l}\, q_B + \cos \bar{l} \,  s \big) \geq s^2 ,  \label{eqn:primalCone}	\\
    &\mathbf{s}\geq 0,
	\end{align}
	\end{subequations}
	where the conic constraint is repeated for each potential element, while $\bar{l} = \K l_i$. Recall that the canonical form of the conic constraint implies the two extra inequalities: $\sin \bar{l}\, q_A + \cos \bar{l} \,s \geq 0$ and $\sin \bar{l}\, q_B + \cos \bar{l} \,s \geq 0$.
	
	    The dual of this problem can be derived using the classical systematic methods, see \cite{boyd2004convex},
    \begin{subequations}
		\label{eqn:dual}
		\begin{align}
			\max_{\mathbf{u}, \mathbf{w}, \mathbf{g}_1, \mathbf{g}_2, \mathbf{g}_3}  \mathbf{f}^\T_z \mathbf{w} + \mathbf{f}_{xy}^\T \mathbf{u},\qquad \qquad\quad \\
			\text{s.t.:} \quad  \mathbf{B}^\T \mathbf{u} + \mathbf{C} \mathbf{g}_1 + \mathbf{C} \mathbf{g}_2 +  \sqrt{2} \mathbf{g}_3 \leq  \mathbf{0},& \label{eqn:dualS}  \\
			\mathbf{D}_A^\T \mathbf{w} + \mathbf{S}\mathbf{g}_1  = \frac{1}{\rho g}\mathbf{1},& \label{eqn:dualqA} \\
			\mathbf{D}_B^\T \mathbf{w} + \mathbf{S}\mathbf{g}_2  = \frac{1}{\rho g} \mathbf{1},& \label{eqn:dualqB}\\
			2\,\mathbf{g}_1 \circ \mathbf{g}_2\geq \mathbf{g}_3^{\circ 2},& \label{eqn:dualCone}
		\end{align}\label{eqn:dualVault}
	\end{subequations}

    \noindent
	where $\mathbf{x}\circ \mathbf{y}$ represents element-wise product of vectors, i.e. its $i$-th  element is $x_i y_i$, and $\mathbf{x}^{\circ n}$ represents element-wise power, i.e. each element of vector $\mathbf{x}$ is raised to the power $n$. Meanwhile $\mathbf{S} = \mathrm{diag}$ $ ([\sin\bar{l}_1,$ $\sin\bar{l}_2, ..., \sin\bar{l}_m])$, and matrix $\mathbf{C} = \mathrm{diag}\allowbreak([\cos\bar{l}_1, \cos\bar{l}_2,$ $ ..., \cos\bar{l}_m])$. The optimization variables $\mathbf{w} = [w_1, w_2,$ $ ..., w_n]^\T$ and $\mathbf{u} = [u_1^x, u_1^y, u_2^x, ..., u_n^y]^\T$ contain virtual displacements in the vertical and horizontal directions respectively. The remaining optimization variables $\mathbf{g}_1$ to $\mathbf{g}_3 \in \mathbb{R}^{m}$ are auxiliary variables with no specific physical meaning, which are dual to each of the terms in (\ref{eqn:primalCone}).
	
	Note that, in practice, it is more convenient to set-up and solve the primal problem (\ref{eqn:primal}). However, consideration of various aspects of the dual will be necessary below. When using modern conic solvers, it is typically easy to obtain the solutions for both problems after solving either one.

	

    	\subsubsection{Member adding}
        \label{sec:memAdding}
	Member adding \cite{gilbert2003layout} is a specialization of column generation which can be used to significantly improve the speed and memory usage of a ground structure optimization problem, whilst guaranteeing that the same globally optimal volume will be obtained. 
    
    The procedure begins with an active ground structure consisting of a subset of all possible connections (adjacent connectivity is usually employed). Once the initial problem is solved, the constraints in the dual problem (\ref{eqn:dualVault}) can be checked for inactive elements, i.e. those which were not included in the initial ground structure.  Formally, for each element it is necessary to obtain a $g_1, g_2, g_3$ satisfying the relevant row of each constraint (\ref{eqn:dualVault}b-e). Candidate values for $g_1, g_2, g_3$ are obtained by solving the first three constraints, taking (\ref{eqn:dualS}) as an equality to give a system which can be solved by simple rearrangement. These candidate values are then checked within the conic constraint (\ref{eqn:dualCone}), and the element is thus categorized as violating or not violating.     
    
    If no elements were found to be violated, i.e. (\ref{eqn:dualCone}) was satisfied for each set of $g$, then the obtained values of $g$, combined with the incumbent solution provide a feasible solution to (\ref{eqn:dualVault}) for the full ground structure. A feasible primal solution is easily obtained by inserting zero values for $q_A, q_B$ and $s$ for inactive elements. These now form a pair of primal-feasible and dual-feasible solutions, with the same objective. Thus the strong duality theorem \cite{boyd2004convex} proves that the obtained solution is optimal for the fully connected ground structure. 
    
    This situation rarely occurs after the solution of the first sub-problem. Several iterations are usually required, adding some violated elements to the ground structure at each iteration. When one or more sets of candidate values for $g$ do not is not satisfy (\ref{eqn:dualCone}), then the corresponding element is considered for addition to the ground structure in the next iteration. 

\subsection{From numerical solution to physical structure}
\label{sec:num2phy}

This section discusses how a solution to the pair of problems (\ref{eqn:primal}) and (\ref{eqn:dual}) is used to construct the corresponding structural form. Reconstruction of the optimal grid-shell is possible under two conditions stated and discussed in  Section \ref{sec:assumptions}. Under those conditions, primal and dual optimal solutions meet the optimality criteria listed  
in Section \ref{sec:properties}. These, in turn, will pave a way to a rigorously justified algorithm for recasting the nodal elevations $\mathbf{z}$. The global compatibility conditions \eqref{eqn:couplingA}, \eqref{eqn:couplingB} will be evidenced, rendering the pair  (\ref{eqn:primal}), (\ref{eqn:dual}) a convex reformulation of the original, non-convex, optimal form-finding problem\footnote{The original, non-convex, form-finding problem is the primal problem (\ref{eqn:primal}), but where the relaxed geometry constraint (\ref{eqn:primalCone}) is replaced by equations (\ref{eqn:comp}) for each element, and $z$ elevation for each node is explicitly included as an optimization variable.} rather than merely its relaxation.

\subsubsection{Criteria for optimal fully-stressed grid-shell structures}
\label{sec:assumptions}

In order that the grid-shell form is possible to construct based on the solutions of the problems \eqref{eqn:primal}, \eqref{eqn:dual}, two criteria must be met:
\begin{itemize}
    \item the primal problem \eqref{eqn:primal} is feasible,
    \item the solution $(\mathbf{u},\mathbf{w})$ of the dual problem \eqref{eqn:dual} satisfies 
     \begin{equation}
            \mathbf{w} <  \frac{1}{\rho g}
            \label{eqn:w_bound}
        \end{equation}
    (note that, \textit{a priori},  $\mathbf{w} \leq  \frac{1}{\rho g}$ by the dual conic constraints).
\end{itemize}
These criteria are of \textit{a posteriori} nature since, in practice, they can be verified only after solving the pair of problems \eqref{eqn:primal}, \eqref{eqn:dual}.  Whilst the first condition is trivially necessary for the correctness of the method, the second one is rather abstract. Nevertheless, both conditions can be translated to natural structural requirements, as explained below.

Firstly, it is worthwhile to mention that the analysis of the two criteria simplifies greatly when there are no upward loads, i.e.
\begin{equation}
        \mathbf{f}_z \leq 0.
\end{equation}
Note that this loading scenario is the most frequent in practice.
It turns out that, in this case, the second criteria, \eqref{eqn:w_bound}, is automatically satisfied whenever the problem \eqref{eqn:primal} is feasible. This can be made rigorous with a rather long and technical  proof, available in the supplementary materials. But intuitively, the term $\mathbf{f}^\T_z \mathbf{w}$ in the dual objective favours negative virtual deflections $\mathbf{w}$ for downward loads, and hence for these cases it is found that $\mathbf{w} \leq 0 < \frac{1}{\rho g}$. 

Regardless of the loading applied, if the first criteria does not hold, i.e. the primal problem (\ref{eqn:primal}) is infeasible, then there is no possible structure capable of satisfying the specified scenario. This is because the relaxations introduced in Section \ref{sec:geometrical} only increase the possible feasible solutions, they do not exclude any feasible structures.  

To assure the feasibility of \eqref{eqn:primal}, it is naturally necessary to guarantee a proper connectivity of the ground structure with respect to the loads and supports, minding that only compression is allowed. This requirement is no different than the one for the weightless variant of the problem in \eqref{eqn:vault}. However, when the self-weight is present, infeasibility may also be caused by excessive weight-to-stress ratio $\frac{\rho g}{\sigma}$ for the spans involved. Indeed, note that elements $i$ whose horizontal length $l_i$ is  $\frac{\pi \sigma}{\rho g}$ or more are automatically excluded from the ground structure, cf. \cite{fairclough2018theoretically}. Nonetheless, guaranteeing that $l_i < \frac{\pi \sigma}{\rho g}$ for the whole ground structure is not a sufficient condition for feasibility, the specific geometry and connectivity of the problem must also be considered.

It turns out that the condition \eqref{eqn:w_bound} is also be related to keeping the ratio $\frac{\rho g}{\sigma}$ below a certain, problem-specific, limit. Intuitively, observe that as $\rho g \to 0$,  $\frac{1}{\rho g}\to \infty$, and so (\ref{eqn:w_bound}) does not constrain the value of $\mathbf{w}$.

Thus, both criteria correspond to requiring feasibility on the equivalent weightless problem, plus a requirement that the material weight-to-stress ratio is `sufficiently small' for the spans and geometry concerned. Finding the critical value for what is `sufficiently small' is difficult as this depends on all of the problem data including supports, domain geometry and loading (including whether these are upwards or downwards). However as a first approximation and intuition, this often relates to when some distance in the domain (e.g. between adjacent supports) reaches the maximum span of the catenary $\frac{\pi \sigma}{\rho g}$.

\smallskip

If upward loads are present, then one may obtain numerical solutions for which the condition \eqref{eqn:w_bound} fails, without causing infeasibility. Such a numerical solution does not allow for constructing a corresponding structure that is a `pure' grid-shell, as the original non-convex form finding problem is ill-posed. However, a physically meaningful structure can still be proposed if lumped masses are introduced at nodes for which $w = \frac{1}{\rho g}$. Although this structure is less practical to construct than the pure grid-shell structures, it provides a valid bound on the minimum material usage. 
Section \ref{sec:lumped_mass_relaxation} below outlines this case.

\subsubsection{Reconstruction of the optimal grid shell from an optimal primal and dual solution}
\label{sec:properties}

To show that the relaxations introduced in Section \ref{sec:geometrical} are never active in optimal solutions of \eqref{eqn:primal}, \eqref{eqn:dual}, one must appeal to additional properties that are exhibited at optimality. Under the \textit{a posteriori} criteria in Section \ref{sec:assumptions}, the optimality conditions will pave the way to a globally-optimal, fully-stressed grid-shell. Below, the essential features of the optimal primal and dual variables $\mathbf{s}, \mathbf{q}_A, \mathbf{q}_B$ and $\mathbf{u}, \mathbf{w}$ are listed; rigorous derivation of these from duality principles can be found in \ref{sec:analysis}.
    \begin{enumerate}
        \item[(I)] The primal conic constraint \eqref{eqn:primalCone} is always satisfied as an equality,
        \begin{equation}
            \big(\sin \bar{l}\, q_A + \cos \bar{l} \,s \big) \big(\sin \bar{l}\, q_B + \cos \bar{l} \,  s \big) = s^2.
            \label{eqn:no_lumped_mass}
        \end{equation}
        \item[(II)] Whenever $s \neq 0$, the complementary slackness conditions give:
        \begin{align}
            &\big( \tfrac{1}{\rho g} - w_A \big) \big( \sin \bar{l}\, q_A + \cos \bar{l} \,s \big) \nonumber \\
            &\quad= \big( \tfrac{1}{\rho g} - w_B \big) \big( \sin \bar{l}\, q_B + \cos \bar{l} \,s \big) \label{eqn:slack}.
        \end{align}
        \item[(III)] For every element with $s=0$ there is also $q_A = q_B =0$.
    \end{enumerate}

     A few comments about the foregoing properties are in order.
    The property (I) shows that for each non-zero element ($s>0$) the forces $s,q_A,q_B$ are aligned with the shape of an equally stressed catenary as in Figure \ref{fig:singleElement}. 
    However, in the case when $s=0$, the equality \eqref{eqn:no_lumped_mass} only implies that $q_A q_B=0$, which allows that, e.g., $q_A=0$, $q_B >0$. This does not adhere to the conditions \eqref{eqn:couplingA}, \eqref{eqn:couplingB}. Physically, it corresponds to the scenario where $q_B$ does not come from a bar but represents a lumped mass at the end $B$, cf. Section \ref{sec:lumped_mass_relaxation} below. The property (III) (proved separately from (I)) rules out such a possibility.\color{black}

    Finally, the property (II), combined with (I), will be used below to reconstruct the nodal elevations $\mathbf{z}$ that validate the global compatibility conditions \eqref{eqn:couplingA},\,\eqref{eqn:couplingB}.

    \label{sec:elevations}
    
    In the original vault formulation \cite{bolbotowski2022optimal} without self-weight, the required elevation at each node may be obtained from the virtual deflections in the vertical direction, as found in the optimal dual variable $\mathbf{w}$, through a simple formula: $z = -\frac{w}{2}$.  For the case with self-weight the required relationship turns out to be:	
	\begin{align}
		z &= \frac{\sigma \log(1- \rho g  w)}{2 \rho g}. 
        \label{eqn:zNodes}
	\end{align}
    This formula provides an extra intuition behind the second criteria in Section \ref{sec:assumptions}, i.e. the inequality \eqref{eqn:w_bound}: it is necessary to guarantee that the argument of the logarithm above is positive. The very same formula was used for optimal archgrids with self-weight by \citet[Eqn. 16]{rozvany1980optimal}. To justify it in the broader framework developed in this paper, for each non-zero element ($s>0$) the compatibility relations \eqref{eqn:couplingA},\,\eqref{eqn:couplingB} must be evidenced. To that aim, the properties (I),\,(II) of the optimal solutions will be used.
    
    Using the formula \eqref{eqn:zNodes} leads to:
    \begin{align*}
        z_B-z_A &= \frac{\sigma}{2\rho g} \Big( \log(1- \rho g \,  w_B)- \log(1- \rho g  \,w_A) \Big) \\
        & = \frac{\sigma}{2\rho g} \log \left( \frac{1 - \rho g\, w_B}{1 - \rho g\, w_A
        } \right)\\
        & = \frac{\sigma}{2\rho g} \log \left( \frac{\sin \bar{l}\, q_A + \cos\bar{l} \,s }{\sin \bar{l}\, q_B + \cos\bar{l} \,s } \right),
    \end{align*}
    where to pass to the last line the complementary slackness condition \eqref{eqn:slack} was employed. Next, the equality \eqref{eqn:no_lumped_mass} furnishes $\sin \bar{l}\, q_B + \cos\bar{l} \,s  = s^2/\big(\sin \bar{l}\, q_A + \cos\bar{l} \,s \big)$, thus allowing to continue,
     \begin{align*}
        \frac{\rho g}{\sigma}(z_B-z_A) &= \frac{1}{2} \log \left( \Big(\frac{\sin \bar{l}\, q_A + \cos\bar{l} \,s }{s } \Big)^2 \right) \\
        & =  \log \left( \frac{\sin \bar{l}\, q_A + \cos\bar{l} \,s }{s }   \right).
    \end{align*}
    Inverting the logarithm leads to the condition \eqref{eqn:comp1}. The condition \eqref{eqn:comp2} can be checked similarly, and, together, they are equivalent to \eqref{eqn:couplingA},\,\eqref{eqn:couplingB}. Readily, the solutions, and the method itself by extension, are rigorously validated for data that satisfy the criteria in Section \ref{sec:assumptions}.


    Once the optimal elevations of the nodes are known from (\ref{eqn:zNodes}), centerlines for each element can be obtained using the centerline equation \eqref{eqn:centerline}, see also \cite{fairclough2018theoretically}. This allows for graphical representation of the structure. Further equations from \cite{fairclough2018theoretically} also allow the cross-section area at any point along an element to be obtained based on the value of the primal variable $s=\hat{q}\cos\theta$.

    \color{black}

    \color{black}

    \subsubsection{A lumped mass interpretation for cases where the optimal structure is not fully stressed}
    \label{sec:lumped_mass_relaxation}


    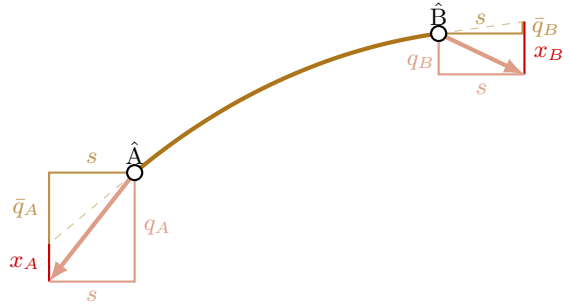
\begin{figure}
		\centering
		\begin{tikzpicture}[scale=1]
			\coordinate (A) at (0,{ln(cos(deg(0.7)))/0.14)});
			\coordinate (Atan) at ([shift=(40:-1.47)] A);
			\coordinate (Atot) at ([yshift=-0.5cm] Atan);
			\coordinate (B) at (4, {ln(cos(deg(0.7-0.14*4)))/0.14)});
			\coordinate (Btan) at ([shift=(8:1.14)] B);
			\coordinate (Btot) at ([yshift=-0.7cm] Btan);	
			
			\draw[ultra thick, smooth, samples=20, domain=0:4, curveline]	plot({\x}, {(ln(cos(deg(0.7-0.14*\x)))/0.14} );
			
			\draw [dashed, curveline!50] (A) -- (Atan);
			\draw [dashed, curveline!50] (B) -- (Btan);

			\draw [thick, curveline!80] (A) -| (Atan) node [pos=0.25, above] {$s$} node [pos=0.75, left] {$\bar{q}_A$};
			\draw [curveline!50!lumped!50, ultra thick, latex-] (Atot) -- (A);
			\draw [thick, lumped] (Atot) -- (Atan) node [pos=0.5, left] {$x_A$};
			\draw [thick, curveline!50!lumped!50] (Atot) -| (A) node [pos=0.75, right] {$q_A$} node [pos=0.25, below] {$s$};
			
			\draw [thick, curveline!80] (B)-|([xshift=-0.7pt]Btan) node [pos=0.25, above] {$s$} node [pos=0.75, right] {$\bar{q}_B$};
			\draw [curveline!50!lumped!50, ultra thick, latex-] (Btot) -- (B);
			\draw [thick, lumped] (Btot) -- (Btan) node [pos=0.4, right] {$x_B$};
			\draw [curveline!50!lumped!50, thick] (B) |- (Btot) node [pos=0.35, left, inner sep = 1pt] {$q_B$} node [pos=0.75, below] {$s$};
			\node [thick, draw=black, circle, fill=white, inner sep = 2pt] at (A) {};
			\node [thick, draw=black, circle, fill=white, inner sep = 2pt] at (B) {};
			\node [above] at (A) {$\hat{\mathrm{A}}$};
			\node [above] at (B) {$\hat{\mathrm{B}}$};
		\end{tikzpicture}
	\caption[A single catenary element AB in the catenary vault formulation, including representation of the required `lumped mass' relaxation.]{A single catenary element AB in the catenary vault formulation, including representation of the required `lumped mass' relaxation. Note that, as in Figure \ref{fig:singleElement}, $\bar{q}_A$ is shown with a positive value, whilst $\bar{q}_B$ is shown with a negative value, meanwhile $x$ and $s$ are always restricted to non-negativity.}
	\label{fig:lumps}
    \end{figure}

    This section will discuss the case where an optimal solution is obtained to (\ref{eqn:primal}), (\ref{eqn:dual}) which does not comply with the a posteriori criteria (\ref{eqn:w_bound}), i.e. $w=\frac{1}{\rho g}$ for at least some nodes. Note that, as discussed in Section \ref{sec:assumptions}, this case occurs only when there are upwards loads present and the self weight is relatively large (or, equivalently, spans are long). 

    As $w \to \frac{1}{\rho g}$, equation (\ref{eqn:zNodes}) implies that $z\to -\infty$. So the original, non-convex problem attains the optimum only at the limit. Nonetheless, the convex problem (\ref{eqn:primal}) attains the same optimum value with no issue as $z$ is not explicitly present, however the reconstruction process in Section \ref{sec:properties} no longer applies. 
    
    The original non-convex form-finding problem inherently assumes that all material must be fully stressed. The relaxation of moving from the equality \eqref{eqn:shape1} to the inequality \eqref{eqn:conic_form} can be understood as removing this assumption, and in so doing unlocks a more interpretable physical structure for these cases. 


    Assume that, for some element, \eqref{eqn:conic_form} is a strict inequality. This means that the vertical forces $q_A$ and $q_B$ are not aligned to the end-points of a catenary of equal stress with horizontal length $l$ and thrust $s$, see Figure \ref{fig:singleElement}. From the direction of the inequality, the values of $q_A$ and/or $q_B$ can only be larger (in the downwards direction) than expected from the definition of the equally stressed catenary. Physically, such surplus can be understood as putting lumped masses $x_A$, $x_B \geq 0$ at the respective endpoints (which always act downwards, i.e. are positive). This is illustrated graphically in Figure \ref{fig:lumps}, from which the following relations may be observed,
     \begin{align}
         q_A = \bar{q}_A + x_A, \qquad q_B = \bar{q}_B + x_B,
     \end{align}
     where $\bar{q}_A$, $\bar{q}_B$ are the aligned forces, i.e. they satisfy $\big(\sin \bar{l}\, \bar{q}_A + \cos \bar{l} \,s \big) \big(\sin \bar{l}\, \bar{q}_B + \cos \bar{l} \,  s \big) = s^2$. 
    
     Note that the objective function is calculated based on the total vertical forces $q$, not the aligned force $\bar{q}$. This means that any lumped masses which occur in a solution are correctly counted as part of the structural volume. 
    
    There are now two possible ways in which upwards loads may be resisted by the structure. These correspond to (\ref{eqn:w_bound}) being true or false respectively:
    \begin{itemize}
        \item  \textit{Grid-shell regions}, as described above, where fully-stressed frameworks transmit the load back to some supports, and $w<\frac{1}{\rho g}$.
        \item  \textit{Counterweight regions} where the weight of material acting at each node is exactly equal to the upward force applied at that node, permitting vertical equilibrium to be attained without recourse to supports, and  $w = \frac{1}{\rho g}$.
    \end{itemize}    
    Note that these definitions operate node-wise, i.e. it is possible for the minimum material solution to contain some grid-shell regions and some counterweight regions. 
    
    Supplementary material section S2 proves these further features for these optimal solutions:
    \begin{itemize}  
        \item [(SI)] No real element (i.e. with $s>0$) can connect between a node in a counterweight region and a node in a grid-shell region. 
        \item [(SII)] Within counterweight regions, in the usual case where $\mathbf{f}_{xy} = \mathbf{0}$, there is an optimal solution consisting only of lumped masses, which allows elevations of each node to be chosen arbitrarily. Otherwise, the procedure to obtain valid elevations is shown in the supplementary material.         
    \end{itemize}    
    So for cases containing both region types, the optimal structure would be disconnected between the grid-shell and counterweights. Disconnected `floating' regions of material are likely to be impractical, and are clearly in unstable equilibrium. Nonetheless, this solution can still be used as a lower bound on the material required to support the given scenario. Extra elements added to ensure connectivity would require only nominal cross-sections. Furthermore, the solution in the grid-shell regions acts independently and can still be utilized even if the counterweight areas are deemed impractical.

    

    \section{Examples}
	\label{sec:example}


    A series of numerical examples are now presented to demonstrate the efficacy of the method and to demonstrate characteristics of optimal vaults under significant self-weight. 

    To allow illustration and comparisons between structures with negligible impact from self-weight and those where self-weight is dominating, each example is considered across a varying range of values for unit weight ($\rho g$) of the material used. Note that, throughout the derivation section, it is the strength-to-weight ratio of the material which is used in calculations. Thus, a proportional increase in both unit weight and allowable stress implies only a uniform scaling of the thickness of each element. In other words, increasing the unit weight could be equivalently thought of as increasing the span (for fixed material properties) or reducing the allowable stress (for a constant span and material weight). Presentation in terms of unit weight is chosen here for ease of rendering and visually comparing the resulting forms -- in this way, changes in structure elevations and element thicknesses in the images are solely due to the impact of the self-weight.
    
    To highlight this, results are presented in normalised form. For context, values given at $\rho g = 1 \frac{\sigma}{L}$ correspond to $L = 4.4 m$ for the material values observed in the Armadillo vault \citep[][Stress = 0.1 MPa, density 2320 kg/m$^3$]{mele2016form}. The impact of self weight can be significantly reduced by increasing the allowable design stress, however this is likely to be less desirable in practical terms as it may increase the impact of second-order effects such as buckling. 

    Problems have been solved using MOSEK \cite{mosek}, interfaced via the CVXPY programming interface \cite{cvxpy}. The CVXPY interface has a substantial processing overhead when adding many rotated quadratic cones. Therefore, the rotated quadratic cone (\ref{eqn:primalCone}) is re-written as the equivalent standard quadratic form
    \begin{equation}
        (q_A + q_B)\sin\bar{l} + 2 s\cos\bar{l}     \geq \sqrt{
        \left(\sin\bar{l} (q_A - q_B)\right)^2 +  \left(2s\right)^2}.
    \end{equation}
    Example python scripts and Rhino/Grasshopper files are provided \cite{codes}, and the use of CVXPY means that alternative open-source solvers can easily be used if MOSEK license is not available. 
    
	\subsection{One-way spanning barrel vault}

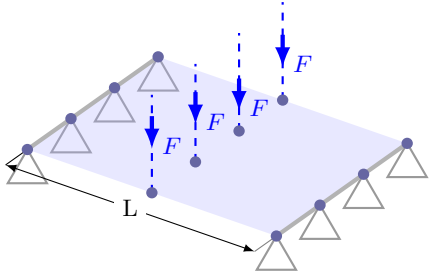
\begin{figure}
		\centering 
		
		\tdplotsetmaincoords{60}{125}
		
		\begin{tikzpicture}[tdplot_main_coords,scale =1, >=latex]
			
			\foreach \x in {0,1, 2, 3}{
				\foreach \y in {0,4}{
					\node [below, outer sep = 0, regular polygon, regular polygon sides=3, fill=white, draw =black!40, thick, inner sep = 3pt] at (\x, \y, 0) {};
				};		
			};
			\fill[fill=blue!30, fill opacity = 0.3] 
			(0,0,0) -- (3,0,0) -- (3,4,0) -- (0,4,0) -- cycle;	
			\draw[ultra thick, black!30] (0,0,0) -- (3,0,0);
			\draw[ultra thick, black!30] (0,4,0) -- (3,4,0);
			
			\begin{scope}[black]
				
				\draw [<->] (3.5, 0, 0) -- (3.5, 4, 0)	node [pos=0.5, fill=white] {L};
				\draw [thin] (3,0,0)--(3.5,0,0); \draw [thin, black!50] (3,4,0) -- (3.5,4,0);
			\end{scope}

			\foreach \x in {0,1, 2, 3}{			
				\draw[blue, thick, dashed] (\x, 2,0 ) -- (\x, 2, 1.5);
				\draw[blue, ultra thick, latex-] (\x, 2, {0.5-0.2*sin(100*\x)}) -- (\x, 2, {1-0.2*sin(100*\x)}) node [pos=0.1, right] {$F$};
				
				\foreach \y in {0,2,4}{
					\node [circle, fill=ndes,  thick, inner sep=1.5pt] at (\x, \y, 0) {};
				};
			};		
		\end{tikzpicture}

		\caption{Barrel vault: Problem setup, consisting of nodal supports at two edges of a rectangular domain with span $L$, with point loads $F$ applied to the centre/apex line.}
		\label{fig:validationExample}
	\end{figure}

	To validate the formulation proposed here, a very simple example will be considered. The example is as shown in Fig. \ref{fig:validationExample}, and consists of a ground structure of 12 nodes, with point loads applied at the mid-span and pinned supports along 2 edges.

	\begin{figure}[t]
		\begin{tikzpicture}
			\begin{axis}[xlabel={\hspace{10pt}Height, $h$},  minor x tick num = 1, ylabel= {Volume,$V$}, ymax = 0.18, xmax = 1.27, ymin = 0, xmin = 0, width=7.2cm, clip mode = individual, name = plt, 
				ytick = {0, 0.05, 0.1, 0.15}, yticklabels={0, 5$\frac{FL}{\sigma}$, 10$\frac{FL}{\sigma}$, 15$\frac{FL}{\sigma}$}, ytick pos = left, ylabel near ticks,
				xtick = {0, 0.2, 0.4, 0.6, 0.8, 1.0}, xticklabels={0, 0.2$L$, 0.4$L$, 0.6$L$, 0.8$L$, 1.0$L$}, minor y tick num = 4]
				
				\addplot [only marks, mark = o, draw =blue, fill=white] coordinates {
					(0.5, 0.04002) (0.26, 0.048891)	(0.9, 0.047139) (0.4, 0.132337) (0.56, 0.114204) (0.9, 0.149761)	};

				\foreach \rval in {1, 500, 1000, 1500, 2000}{
					\renewcommand{\r}{k \rval}
					\addplot[no marks, thick, black, forget plot] table[x=h,y=\rval,col sep=comma] {BarrelVault/ResultsExhaustive.csv};
					
				}
				\foreach \r in {100, 200, 300, 400, 600, 700, 800, 900, 1100, 1200, 1300, 1400, 1600, 1700, 1800, 1900}{
					\addplot[no marks, gray, forget plot] table[x=h,y=\r,col sep=comma] {BarrelVault/ResultsExhaustive.csv};
				}
				
				\addplot [only marks, mark = x, thick] table[x=h, y=vol, col sep = comma]{BarrelVault/ResultsVault.csv};

				\coordinate (lightOptimal) at (axis cs:0.5, 0.04002);
				\coordinate (lightLow) at (axis cs:0.26, 0.048891);
				\coordinate (lightTall) at (axis cs:0.9, 0.047139);
				
				\coordinate (heavyLow) at (axis cs:0.4, 0.132337) ;
				\coordinate (heavyOptimal) at (axis cs:0.56, 0.114204); 
				\coordinate (heavyTall) at (axis cs:0.9, 0.149761);
				
				\node [right, fill = white, inner sep = 1pt] at (axis cs: 1, 0.050031) {$\rho g \approx 0$};
				\node [right, fill = white, inner sep = 1pt] at (axis cs: 1, 0.070076) {$\rho g = 0.5 \frac{\sigma}{L}$};
				\node [right, fill=white, inner sep = 1pt] at (axis cs: 1, 0.104453) {$\rho g = 1 \frac{\sigma}{L}$};	
				\node [right, fill = white, inner sep = 1pt] at (axis cs: 1, 0.169967) {$\rho g = 1.5 \frac{\sigma}{L}$};
				
			\end{axis}
			
			\begin{scope}{every path/.append style={thick, blue}}
				\node (lLw) [below right, inner sep = 0.5pt, draw =blue] at (plt.outer south west) {\includegraphics[trim = 800 90 1500 250, clip, width=2.3cm]{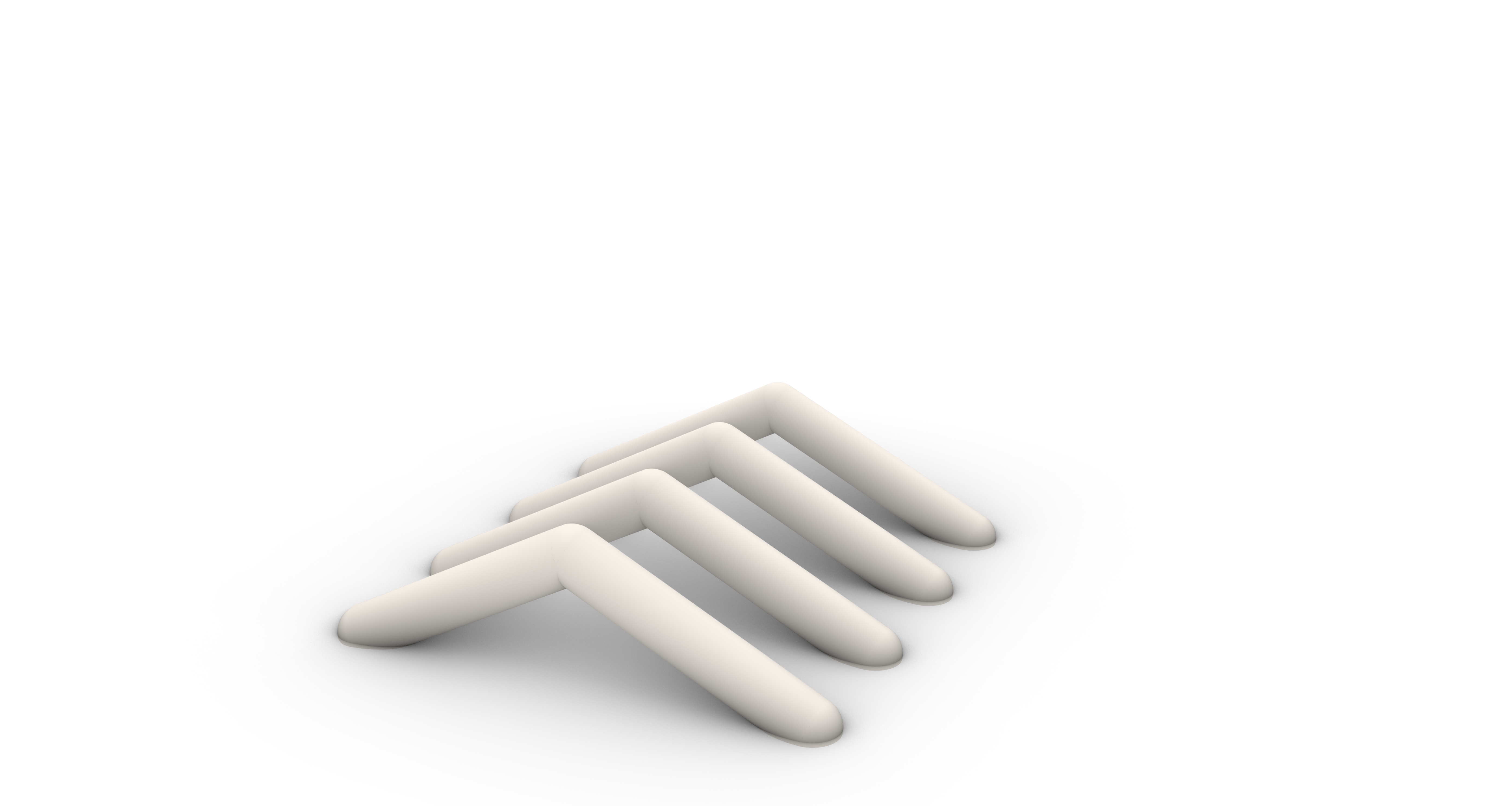}};
				\draw [, draw =blue] (lLw) -- (lightLow);
				
				\node (lOpt) [below, inner sep = 0.5pt, draw =blue] at (plt.outer south) {\includegraphics[trim = 800 90 1500 250, clip, width=2.3cm]{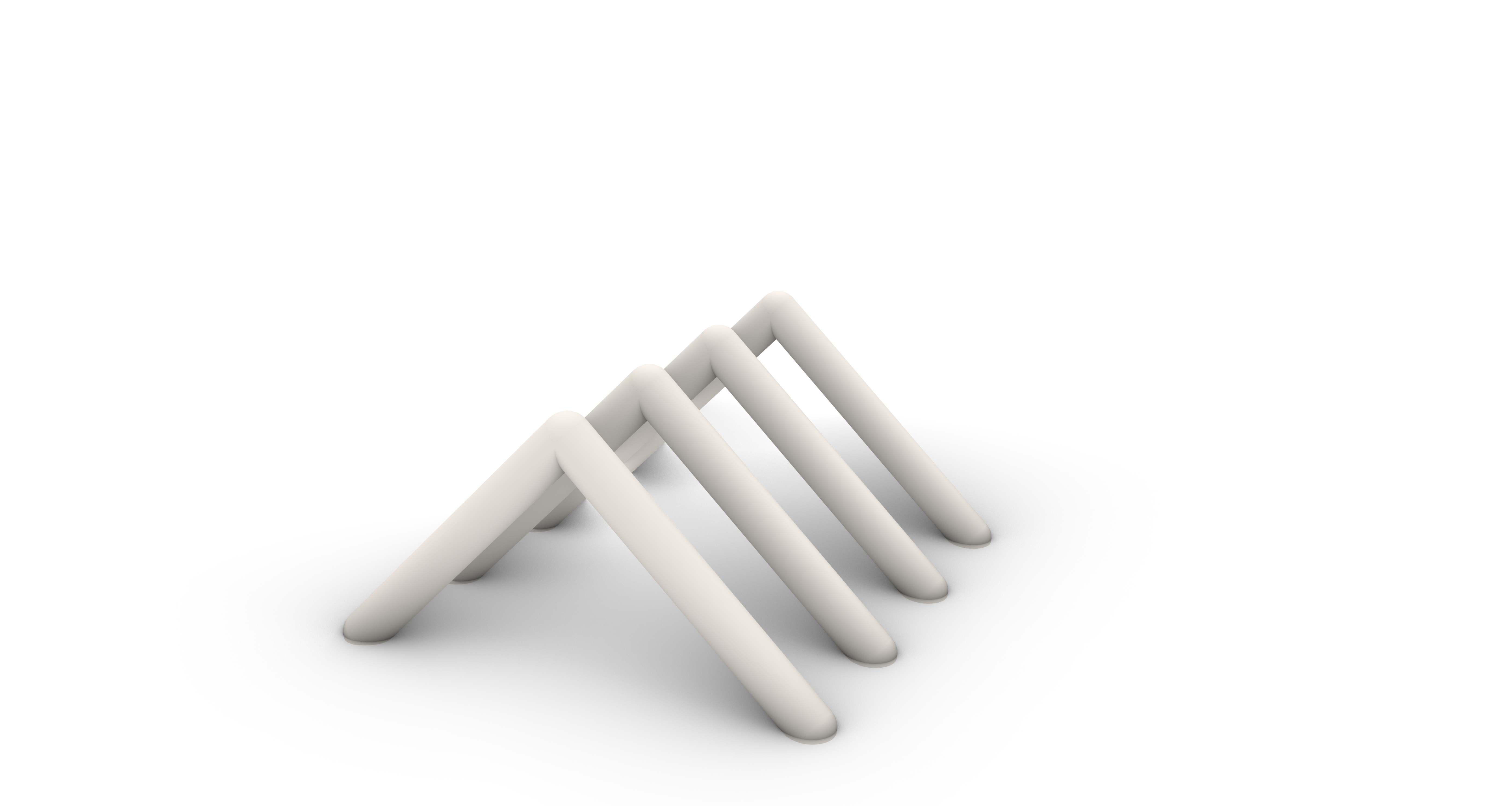}};
				\draw [, draw =blue] (lightOptimal) -- (lOpt);
				
				\node (ltl) [below left, inner sep = 0.5pt, draw =blue] at (plt.outer south east) {\includegraphics[trim = 800 90 1500 250, clip, width=2.3cm]{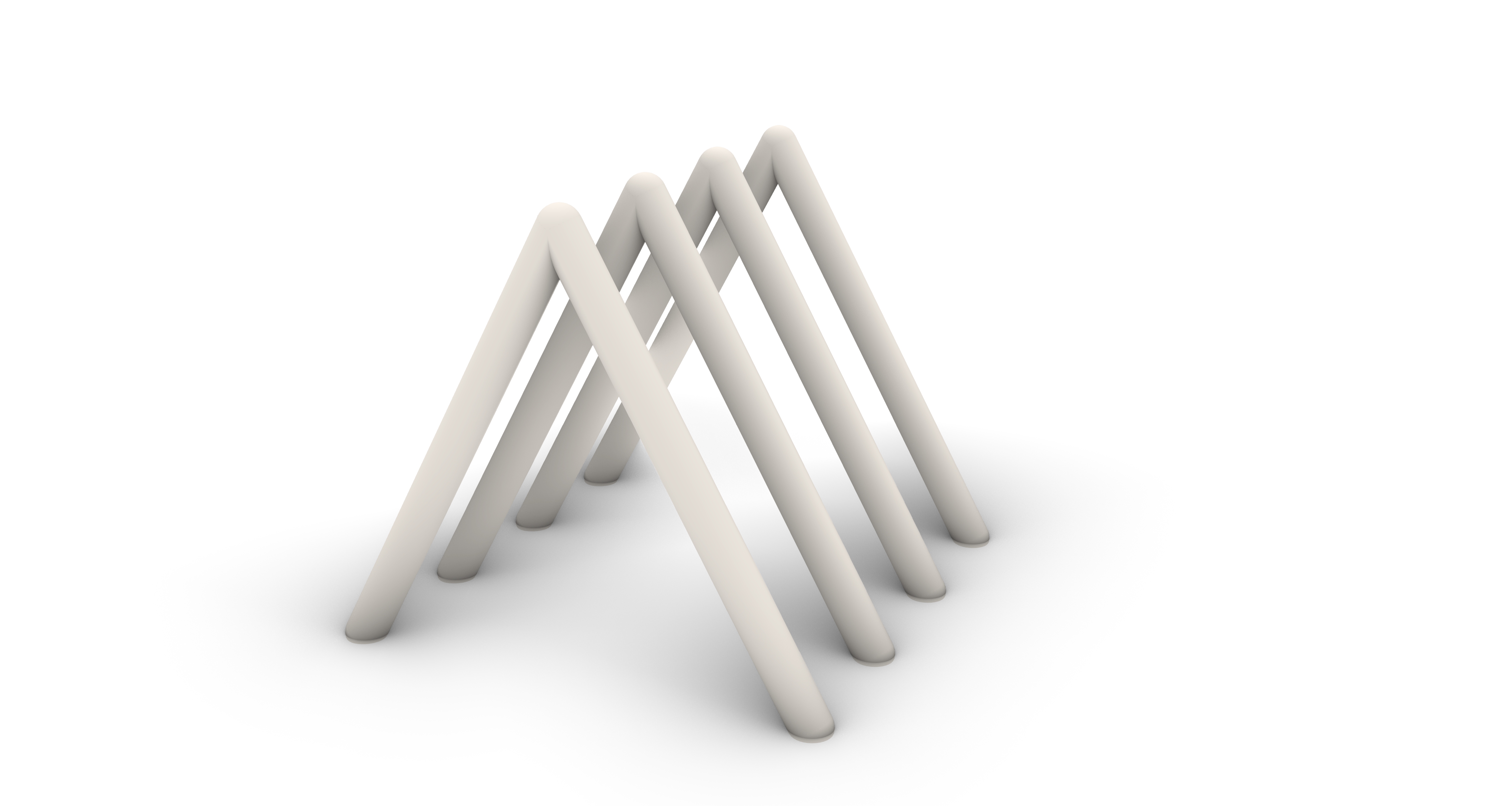}};
				\draw [, draw =blue] (ltl) -- (lightTall);
				
				\node (hLw) [above right, inner sep = 0.5pt, draw =blue] at ([yshift=3pt] plt.outer north west) {\includegraphics[trim = 800 0 1500 250, clip, width=2.3cm]{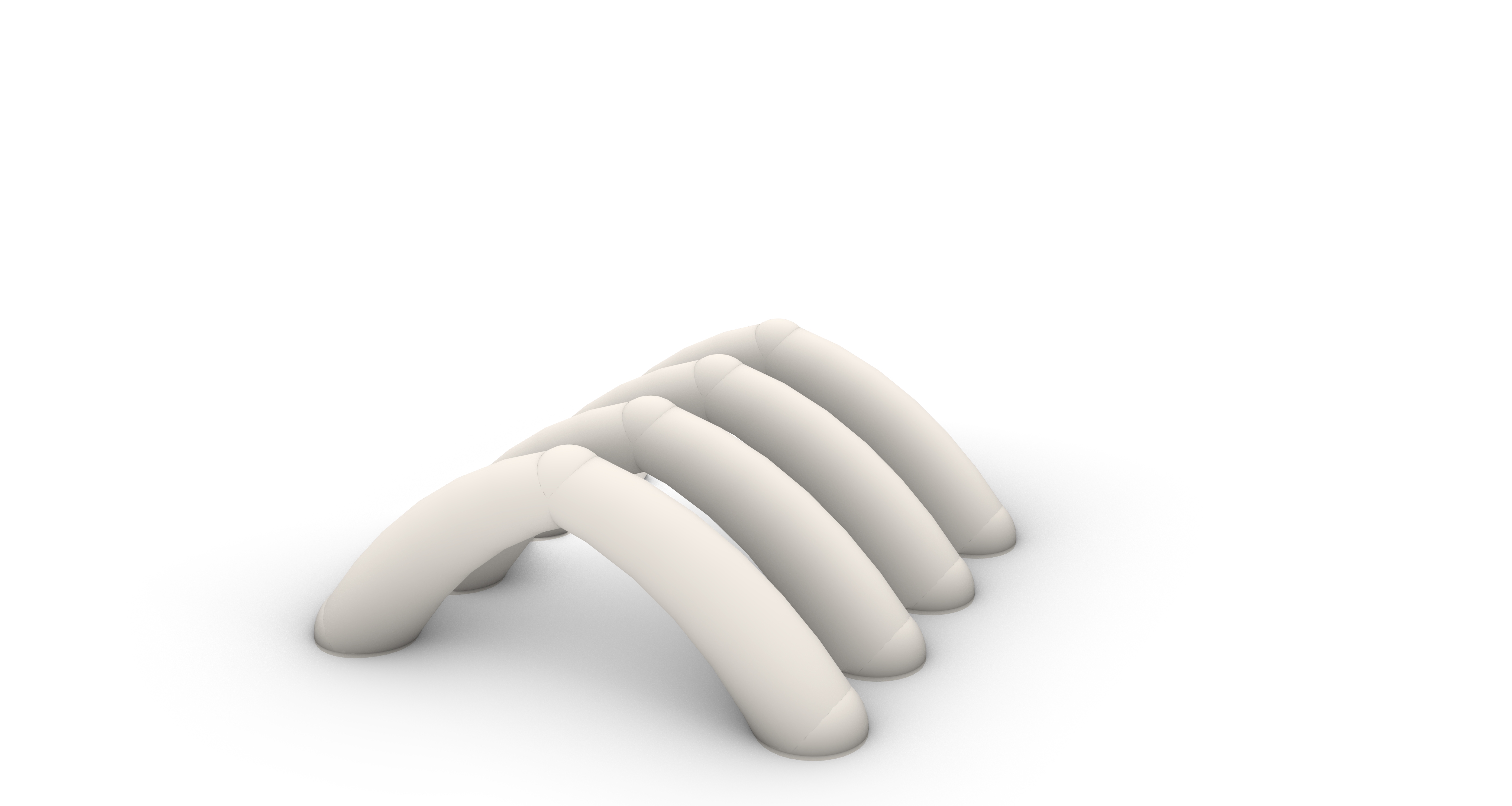}};
				\draw [, draw =blue](hLw) -- (heavyLow);
				
				\node (hOpt) [above, inner sep = 0.5pt, draw =blue] at ([yshift=3pt] plt.outer north) {\includegraphics[trim = 800 0 1500 250, clip, width=2.3cm]{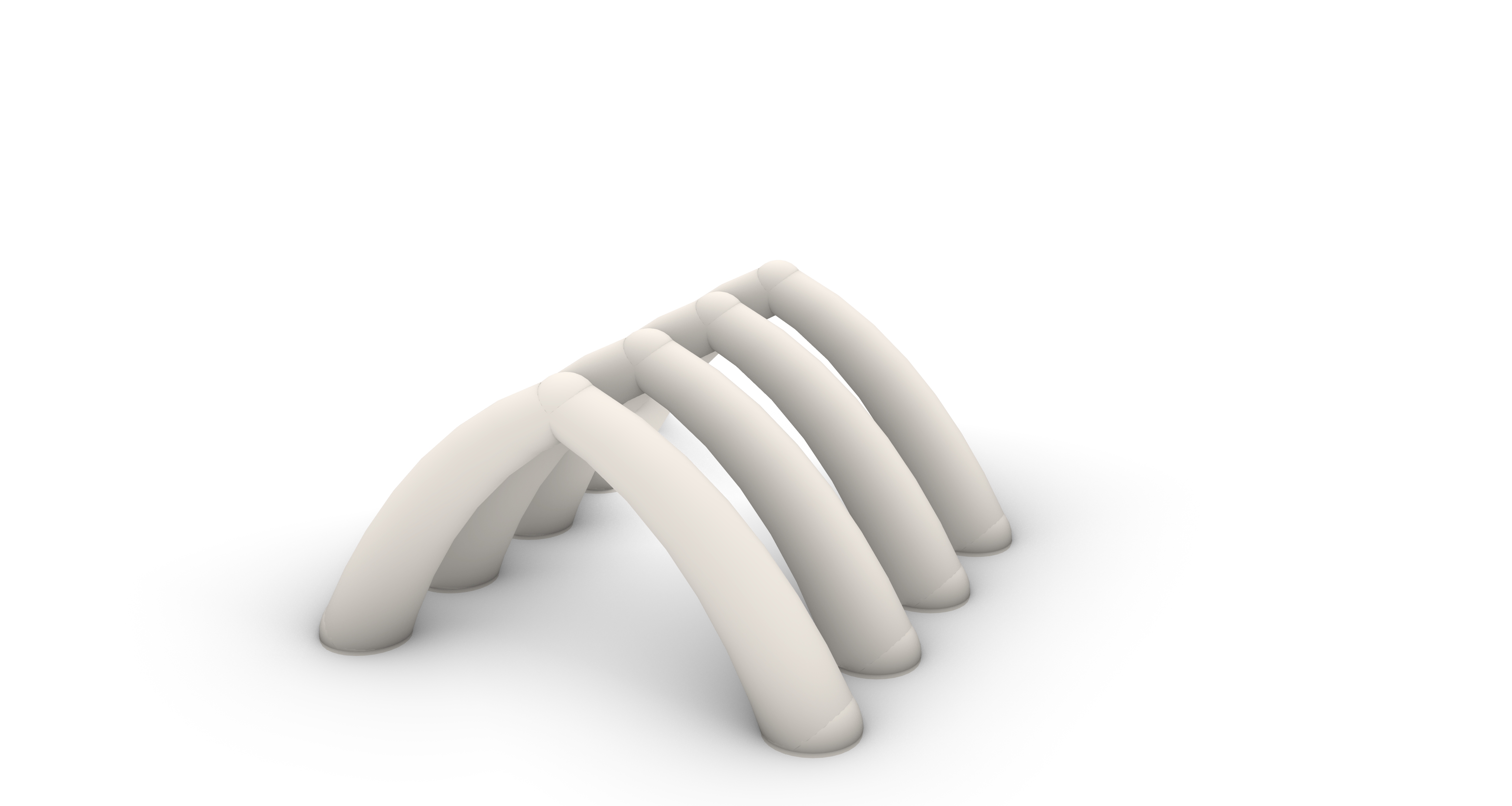}};
				\draw [, draw =blue] (heavyOptimal) -- (hOpt);
				
				\node (htl) [above left, inner sep = 0.5pt, draw =blue] at ([yshift=3pt] plt.outer north east) {\includegraphics[trim = 800 0 1500 250, clip, width=2.3cm]{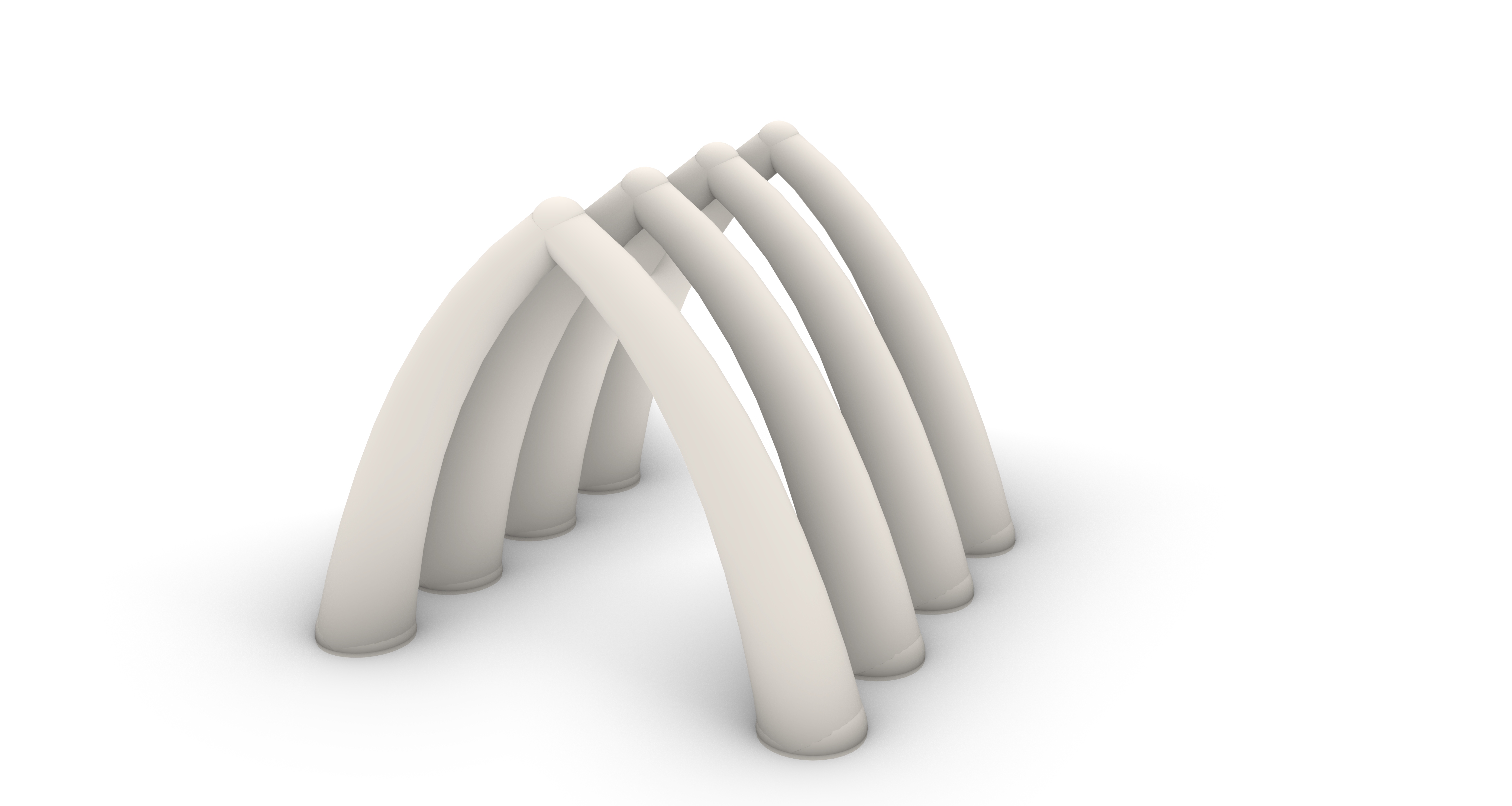}};
				\draw [draw =blue] (htl) -- (heavyTall);
				
			\end{scope}
		\end{tikzpicture}
		\caption[Barrel vault: Enumerated solutions and selected rendered forms]{Barrel vault: Enumerated solutions for problem heights at 0.02$L$ spacing, and material unit weights in 0.1$\frac{\sigma}{L}$ increments. Selected rendered forms shown for $\rho g = 0.001\frac{\sigma}{L} \approx 0$ and $\rho g = 1.5\frac{\sigma}{L}$. Black crosses mark the obtained solutions using formulation (\ref{eqn:primalCone}) for the same values of unit weight.}
    \label{fig:BarrelExhaustive}
	\end{figure}
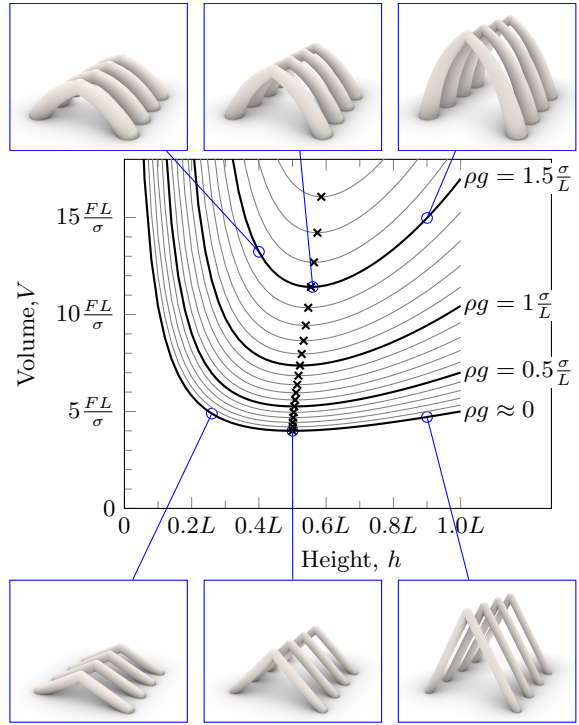

    	As this is essentially a 1-dimensional problem, the expected solution will be a series of four identical parallel pointed arches. Because of this, it is possible to enumerate all possible solutions to this problem through the changing of a single height variable. This will give a suitable comparison by which to validate the numerical solutions. Here, the enumerated solutions have been obtained through the 3D ground structure method using Peregrine \cite{peregrine}, which implements the catenary self-weight model from \cite{fairclough2018theoretically}. This has been solved for pre-determined loading points at various heights as shown in Figure \ref{fig:BarrelExhaustive}. 
     
     This procedure has been repeated for materials ranging from very lightweight to very heavy. The black crosses in Figure \ref{fig:BarrelExhaustive} show the optimal height and volume obtained by the vault formulation proposed here. This allows more precise identification of the optimal height compared to the finite step size used in the exhaustive approach. 
     
     From Figure \ref{fig:BarrelExhaustive} it can be seen that the problem is relatively in-sensitive to changes in elevation when self-weight is negligible, particularly with respect to structures that are taller than the optimal height. However, when self-weight is more significant, the importance of selecting the optimal height is increased, with larger volume increases observed for structures which are either too tall or too short. Furthermore, there are significant differences in the results between the case with negligible self-weight and when self-weight is significant. Adding self-weight requires significantly more material to be used and the curvature of the elements to be increased, demonstrating the importance of considering self-weight in the form-finding stage.

    \begin{figure}
        \includegraphics[trim = 400 00 300 200, clip, width=\linewidth]{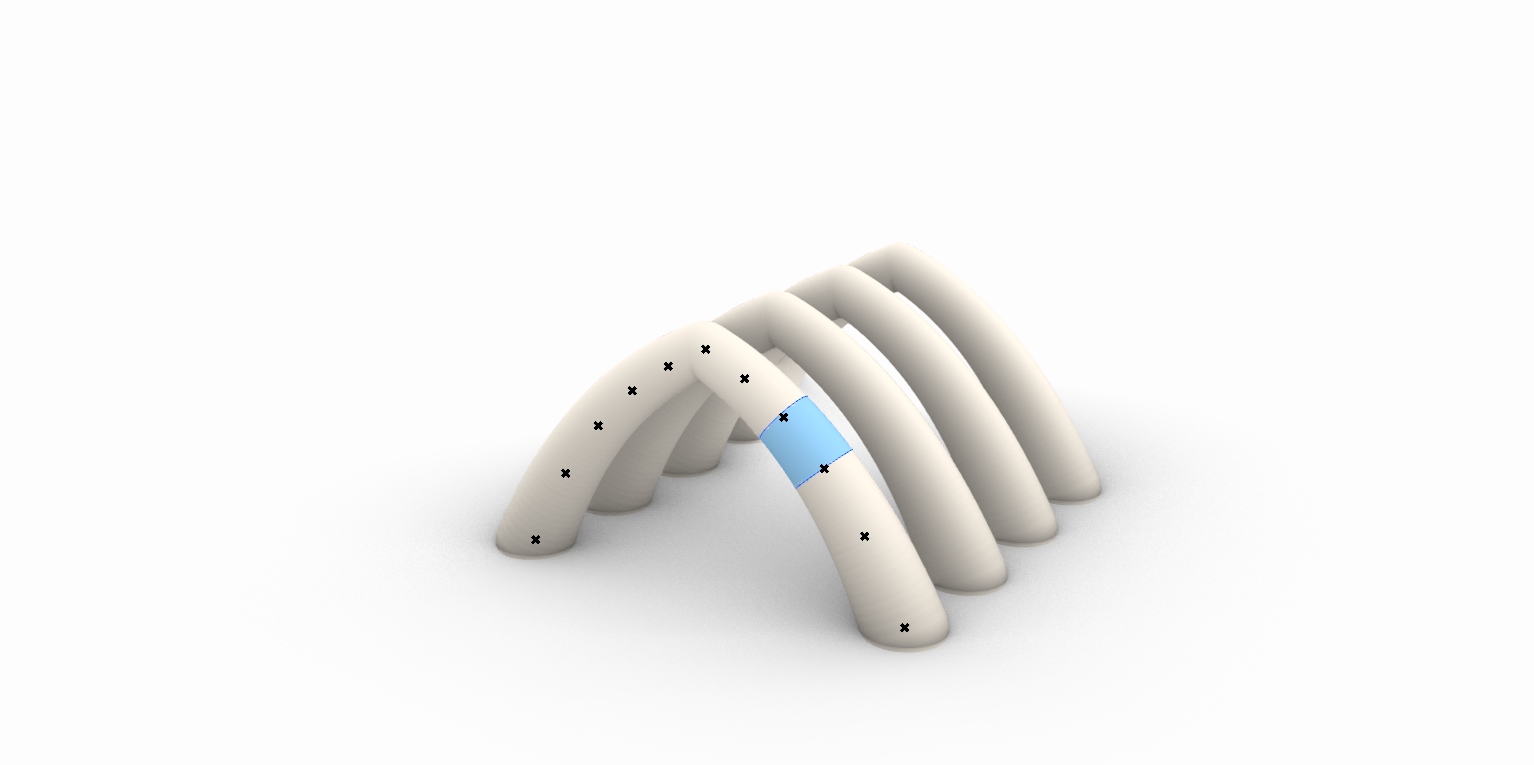}

        \caption{Barrel Vault: Output from catenary vault formulation at higher resolution showing how an individual element (blue) forms part of the arch and the correctly determined heights of intermediate nodes (black).}
        \label{fig:BarrelHighRes}
    \end{figure}

     Figure \ref{fig:BarrelHighRes} shows the same example but with additional nodes in the ground structure. From this it can be seen that the shape of the catenary of equal stress is successfully obtained as a series of curved segments, attaining the same material usage, as expected. This is in contrast to the lumped mass formulation (\ref{sec:lumped}) where the optimal volume changes with nodal resolution.

    \renewcommand{\topfraction}{0.8}
	\begin{figure}[t]
    \begin{tikzpicture}
        \begin{axis}[xlabel={Unit weight, $\rho g$},ylabel={Volume, $V$}, xmin = 0, xmax = 2, ymin = 0, ymax=160, scale only axis,          height = 5cm, legend pos = north west,
            ytick = {0, 40, 80, 120, 160}, yticklabels = {0, 0.5$\frac{FL}{\sigma}$, 1$\frac{FL}{\sigma}$, 1.5$\frac{FL}{\sigma}$, 2$\frac{FL}{\sigma}$}, ylabel = {Volume, $V$}, ylabel shift = -10pt,
            xtick = {0, 0.5, 1, 1.5, 2}, xticklabels={0, 0.5$\frac{\sigma}{L}$, 1$\frac{\sigma}{L}$,1.5$\frac{\sigma}{L}$, 2$\frac{\sigma}{L}$}, xlabel = {Unit weight, $\rho g$},
            width = 0.7\linewidth]


            \addplot[forget plot, no marks, black!30, domain = 0.06:2, samples = 50] plot (\x, 10/\x) node[pos= 0.55, sloped, fill=white, inner sep = 1pt, xshift = 5pt, yshift = 2pt] {\footnotesize $W=0.125F$};
            \addplot[forget plot, no marks, black!30, domain = 0.13:2, samples = 50] plot (\x, 20/\x) node[pos= 0.55, sloped, fill=white, inner sep = 1pt, xshift = 5pt, yshift = 1pt] {\footnotesize $W=0.25F$};
            \addplot[forget plot, no marks, black!30, domain = 0.26:2, samples = 50] plot (\x, 40/\x) node[pos= 0.55, sloped, fill=white, inner sep = 1pt] {\footnotesize $W=0.5F$};
            \addplot[forget plot, no marks, thick, black!30, domain = 0.53:2] plot (\x, 80/\x) node[pos= 0.55, sloped, fill=white, inner sep = 1pt] {\footnotesize $W=F$};            
            \addplot[forget plot, no marks, black!30, domain = 1:2] plot (\x, 160/\x) node[pos= 0.55, sloped, fill=white, inner sep = 1pt] {\footnotesize $W=2F$};

            \addplot[no marks, thick, black, ] table[x=normalised,y=unrestricted,col sep=comma] {Square_experiments/TopologyUnnorm.csv};
            \addlegendentry{Downward load}

            \addplot[no marks, blue, double] table[x=normalised,y=upwards,col sep=comma] {Square_experiments/TopologyUnnorm.csv};
            \addlegendentry{Upward load}

            \addplot[forget plot, no marks, very thick, white] table[x=normalised,y=Archgrid,col sep=comma] {Square_experiments/TopologyUnnorm.csv};
            \addplot[no marks, thick, black, dashed] table[x=normalised,y=Archgrid,col sep=comma] {Square_experiments/TopologyUnnorm.csv};
            \addlegendentry{Archgrid}
            
        \end{axis}
        \end{tikzpicture}
        \centering (a)

        \begin{tabular}{@{}c@{}c@{}}
            \includegraphics[width=0.5\linewidth, trim = 30cm 3cm 25cm 0, clip]{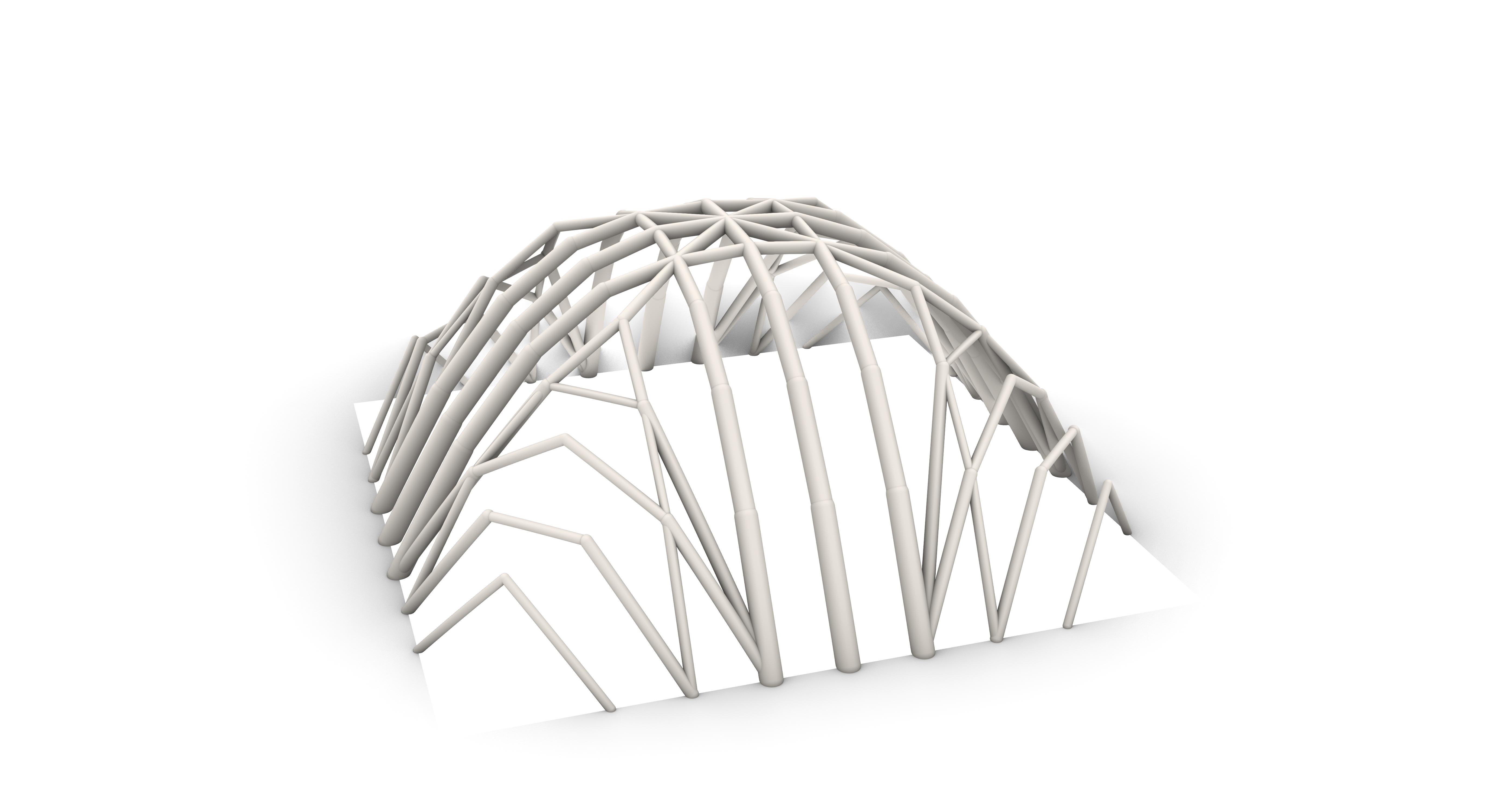} 
            &
            \includegraphics[width=0.5\linewidth, trim = 30cm 3cm 25cm 0, clip]{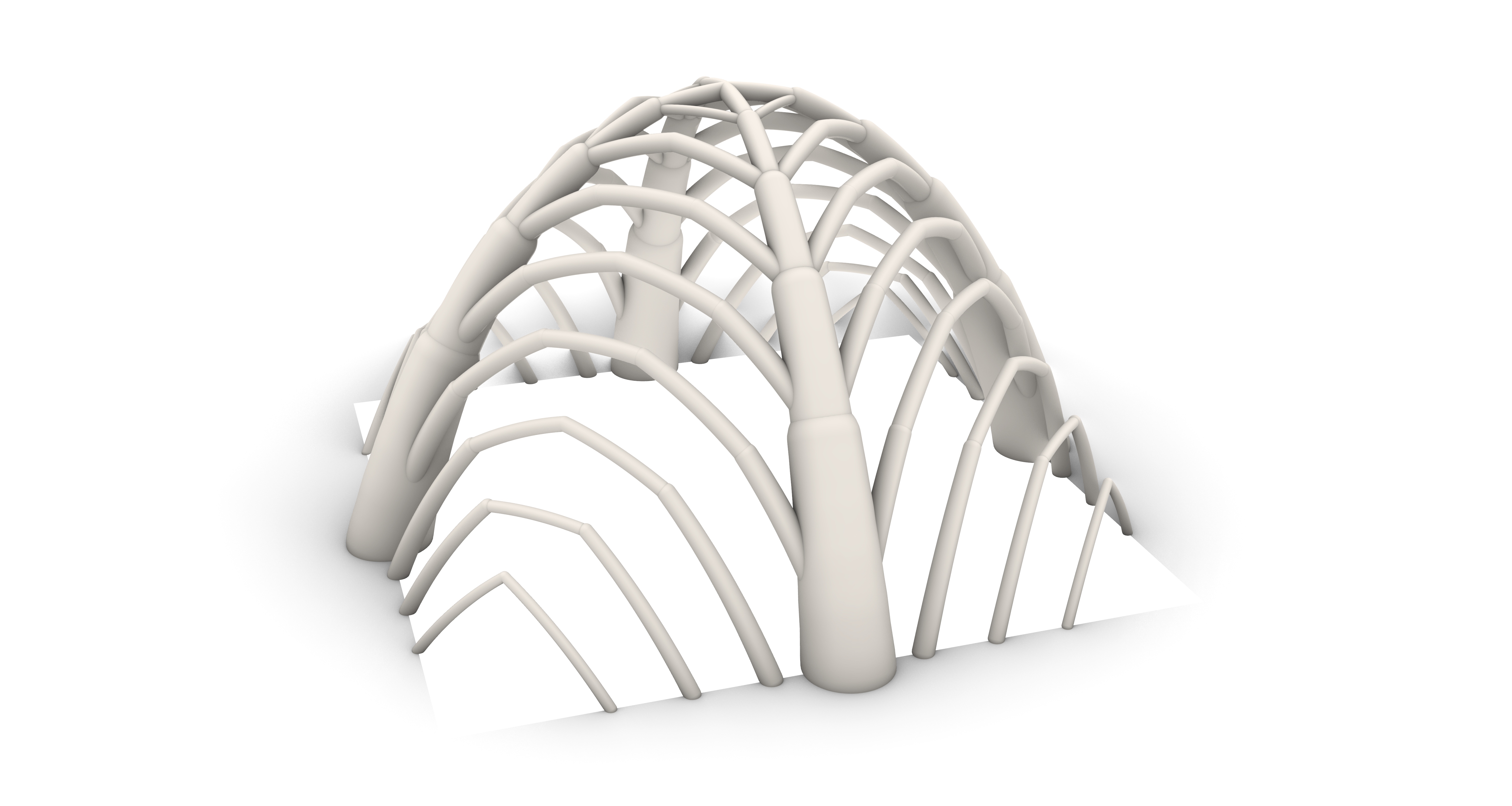}
            \\[-2pt]
            (b) & (c) \\
            \includegraphics[width=0.5\linewidth, trim = 30cm 17cm 25cm 0, clip]{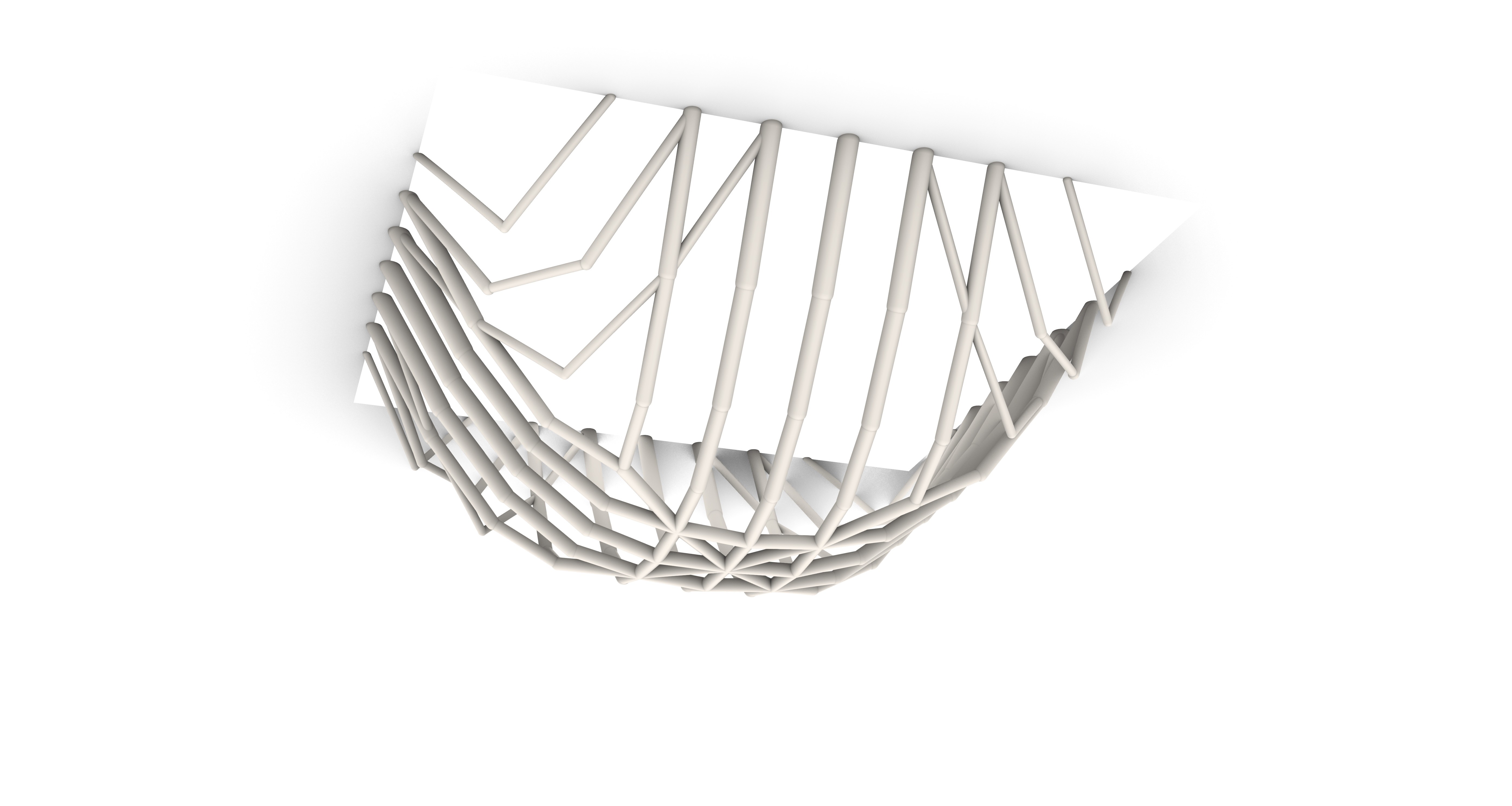}
            &
            \includegraphics[width=0.5\linewidth, trim = 30cm 17cm 25cm 0, clip]{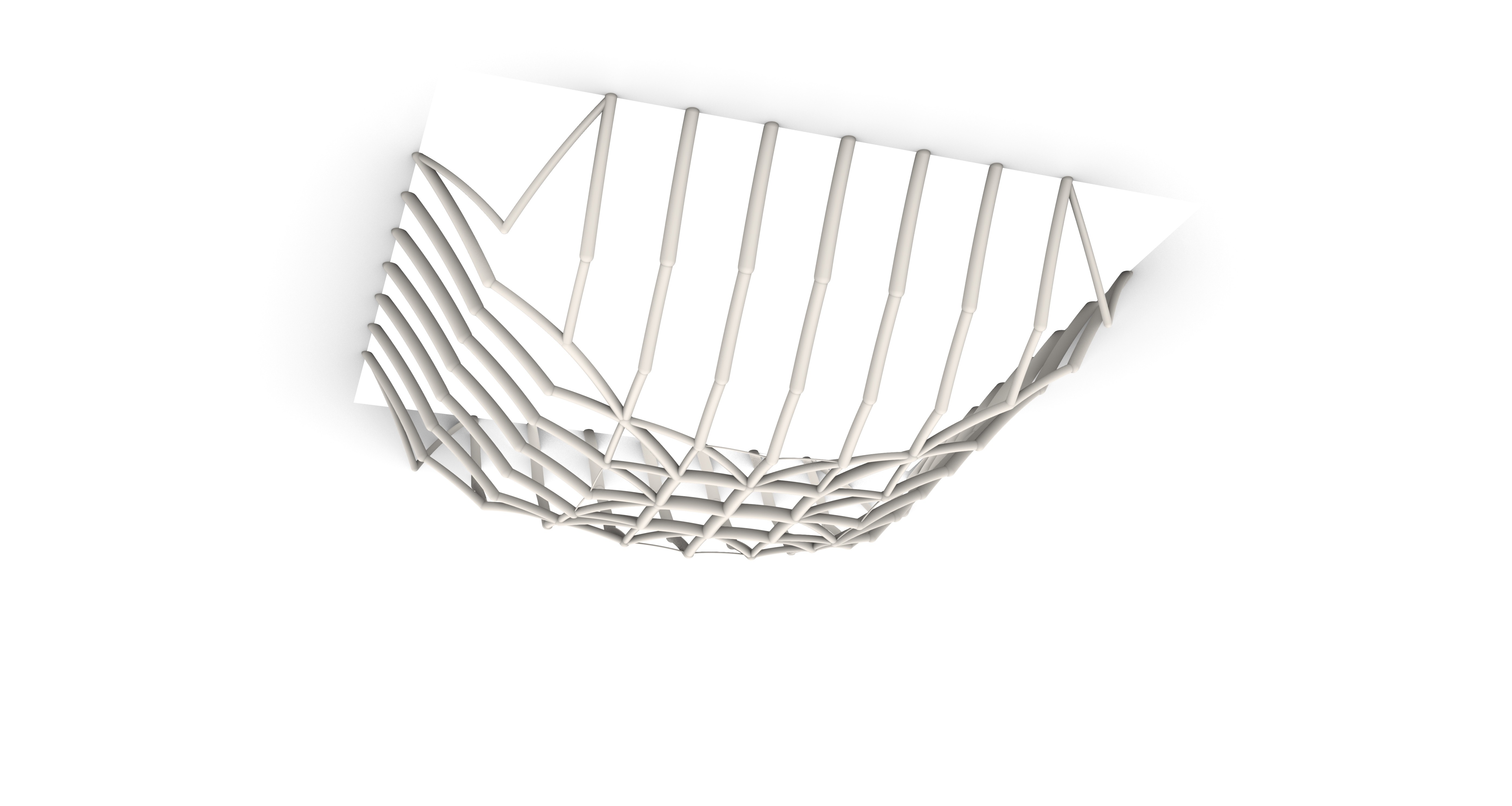}\\[-5pt]
            (d) & (e)
        \end{tabular}
        \caption{Square domain example with distributed loads and pinned edges: (a) Optimal volumes for materials of various unit weights. For context, relationships between the structure weight $W = \rho g V$ and the total external loading $F$ are given. (b)-(e) optimal solutions; (b), (c) for downward load and (d), (e) for upward load; (b), (d) for $\rho g = 0.2\frac{\sigma}{L}$ and (c), (e) for $\rho g = 2\frac{\sigma}{L}$.}
        \label{fig:squareVol}    
	\end{figure}
	
	\begin{figure*}
        \centering
		\begin{tikzpicture}
			\begin{axis}[xlabel={Unit weight, $\rho g$}, ylabel=Volume penalty (vs. full connectivity), xmin = 0, xmax = 2, ymin = 0.99, ytick={1, 1.1, 1.2, 1.3}, yticklabels={+0\%, +10\%, +20\%, +30\%}, xtick = {0, 0.5, 1, 1.5, 2}, xticklabels = {0, $0.5\frac{\sigma}{L}$,$1\frac{\sigma}{L}$,$1.5\frac{\sigma}{L}$,$2\frac{\sigma}{L}$}, scale only axis, width=6.5cm, name = to]
				\addplot[no marks, thick, black, dashed] table[x=normpg,y=Archgrid,col sep=comma] {Square_experiments/fixedtopology.csv};
				\label{plt:archgrid}
				
				\addplot[no marks, double, blue!50!black] table[x=normpg,y=Diagonals,col sep=comma] {Square_experiments/fixedtopology.csv};
				\label{plt:Xtopology}
				
				\addplot[no marks, thick, cyan!50] table[x=normpg,y=Heavy,col sep=comma] {Square_experiments/fixedtopology.csv};
				\label{plt:heavy}
				
				\addplot[no marks, thick, blue] table[x=normpg,y=Light,col sep=comma] {Square_experiments/fixedtopology.csv};
				\label{plt:light}
			\end{axis}

            \node [below, draw, outer sep = 40pt, align = left] at (to.south){
                    \\
			\begin{tabular}{cccc}
                Name and&& \multicolumn{2}{c}{Optimal results} \\
                legend & Topology & $\rho g = 0.2 \frac{\sigma}{L}$ & $\rho g = 2 \frac{\sigma}{L}$ \\[10pt]
				\raisebox{0.06\linewidth}{\shortstack{Opt-light \\ \ref{plt:light}}} 
                &\includegraphics[height = 0.125\linewidth]{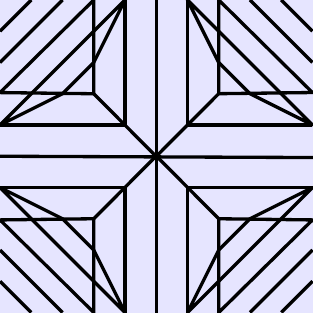}
                & \includegraphics[height=0.125\linewidth,trim = 15cm 0 12.5cm 0, clip]{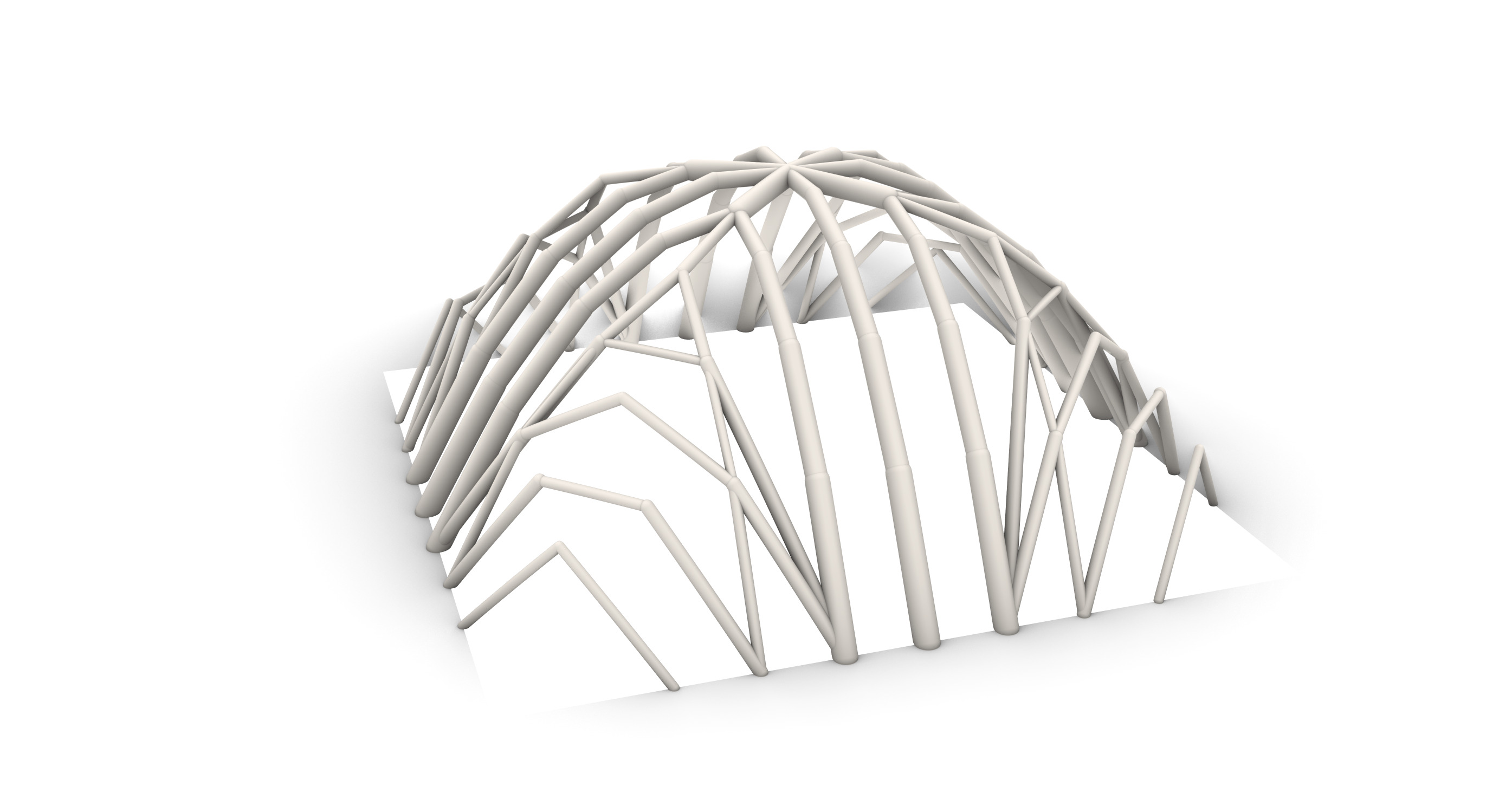}
                & \includegraphics[height=0.125\linewidth, trim = 15cm 0 12.5cm 0, clip]{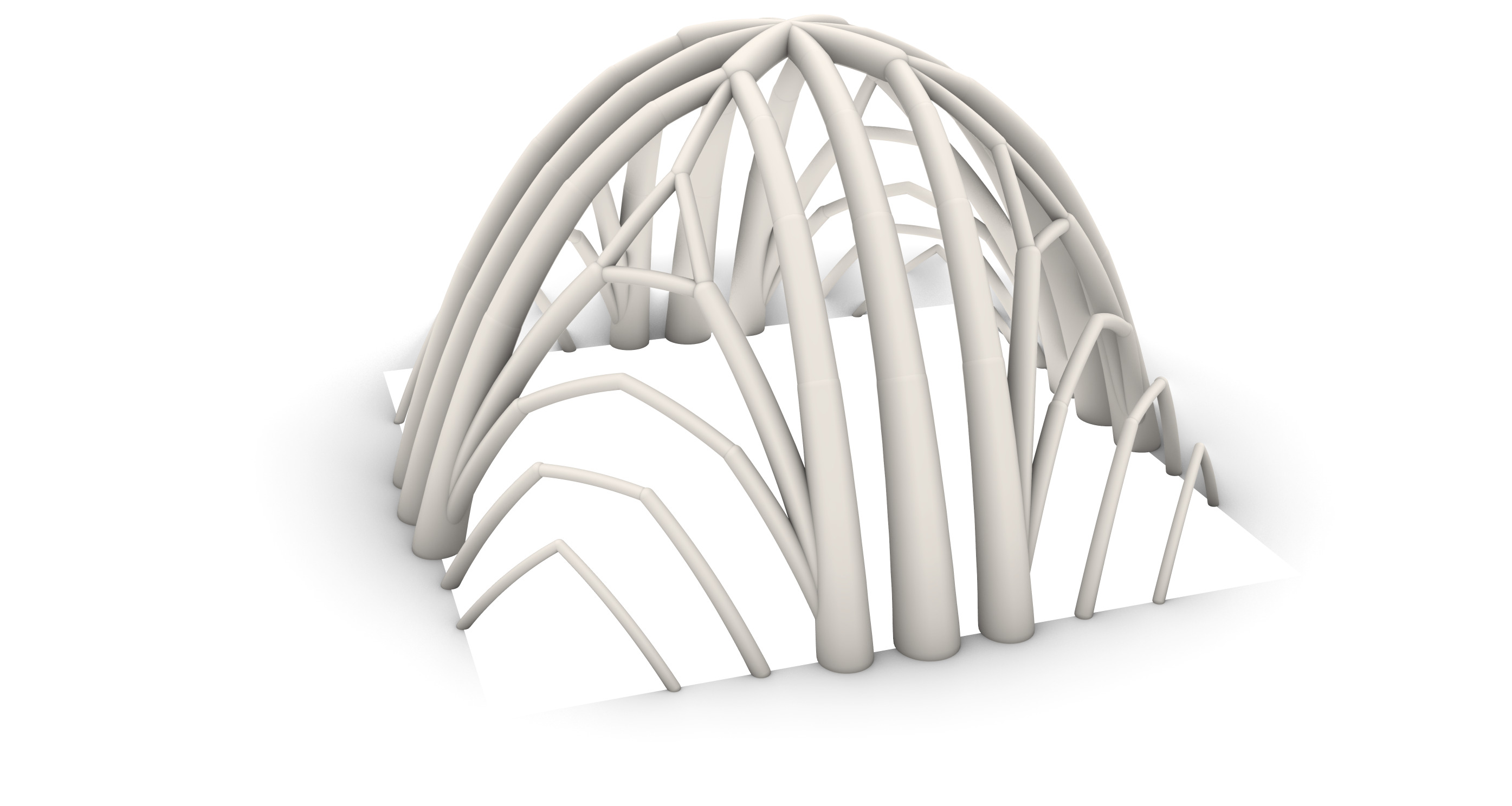}\\
				\raisebox{0.06\linewidth}{\shortstack{Opt-heavy \\ \ref{plt:heavy}}} 
                &\includegraphics[height = 0.125\linewidth]{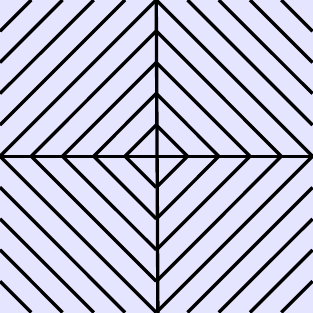}
                & \includegraphics[height=0.125\linewidth,trim = 15cm 0 12.5cm 0, clip]{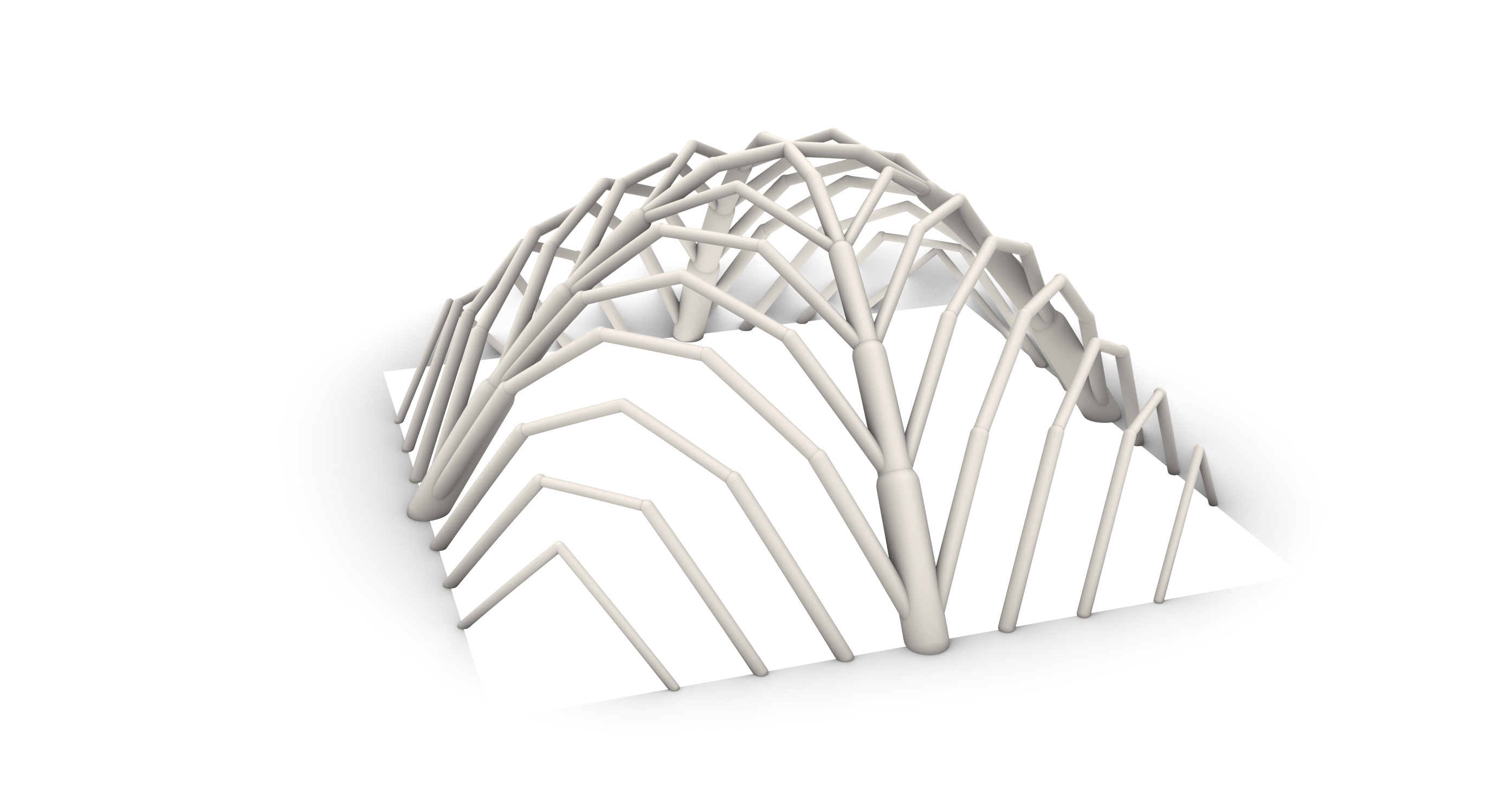}
                & \includegraphics[height=0.125\linewidth, trim = 15cm 0 12.5cm 0, clip]{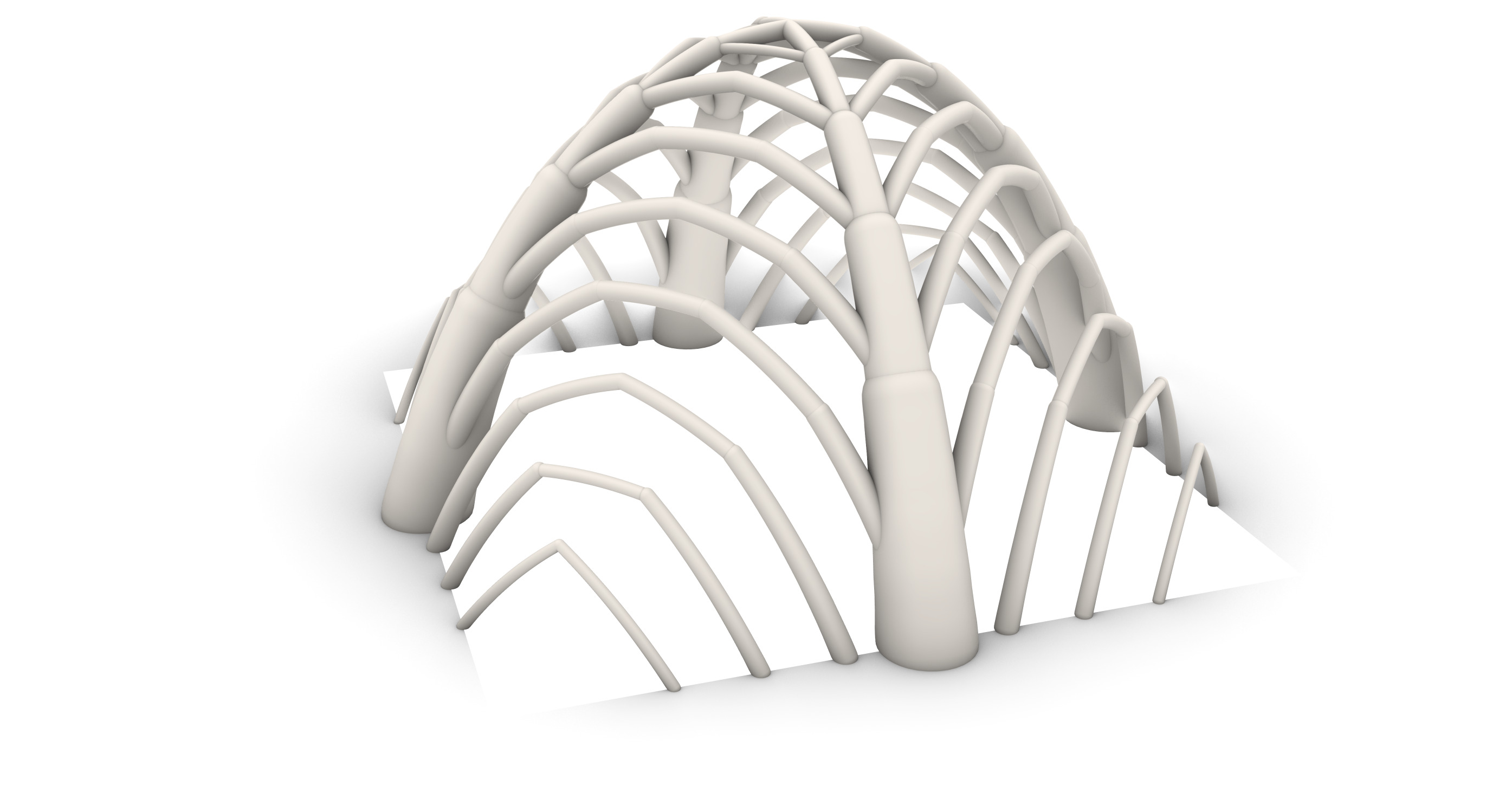} \\ %
                \raisebox{0.06\linewidth}{\shortstack{Archgrid \\ \ref{plt:archgrid}}}
                &\includegraphics[height = 0.125\linewidth]{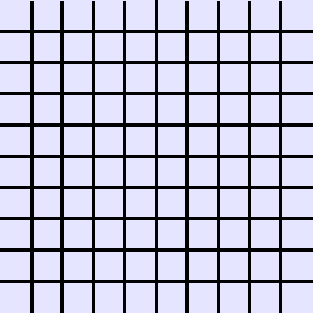}
                & \includegraphics[height=0.125\linewidth,trim = 15cm 0 12.5cm 0, clip]{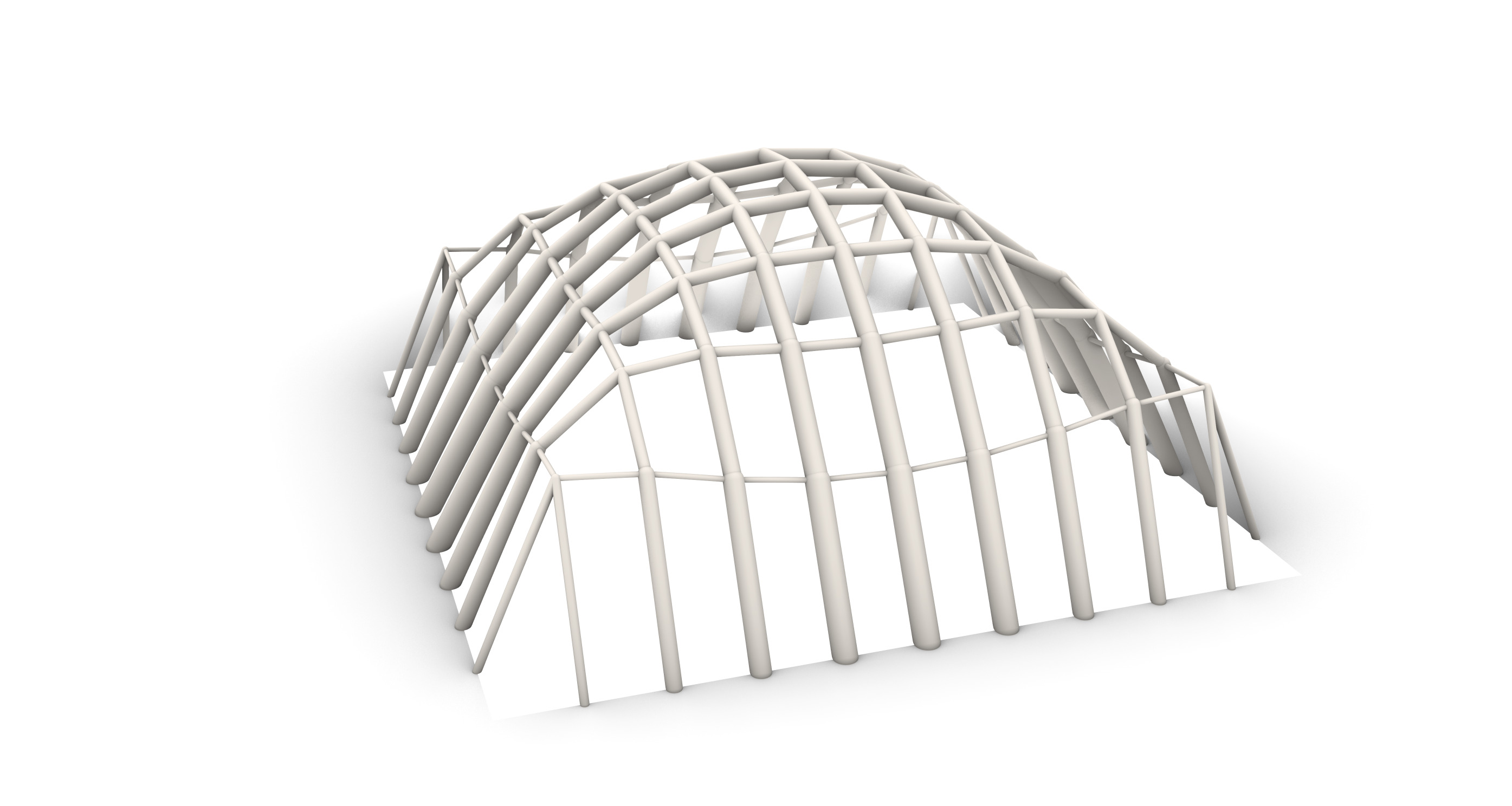}
                & \includegraphics[height=0.125\linewidth, trim = 15cm 0 12.5cm 0, clip]{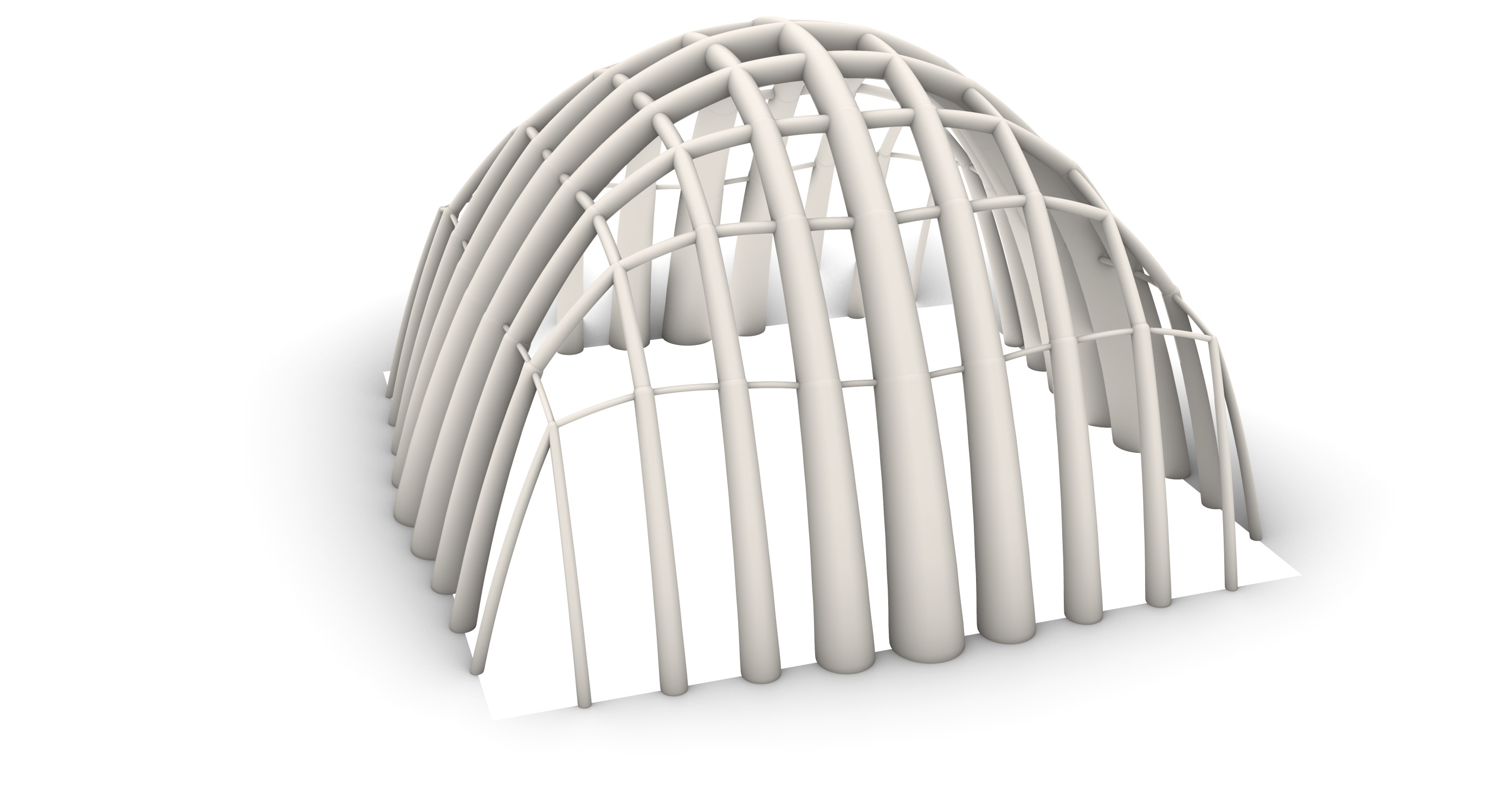} \\
				\raisebox{0.06\linewidth}{\shortstack{Diagonals \\ \ref{plt:Xtopology}}}
                &\includegraphics[height = 0.125\linewidth]{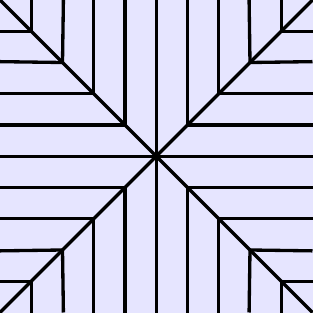}
                & \includegraphics[height=0.125\linewidth,trim = 15cm 0 12.5cm 0, clip]{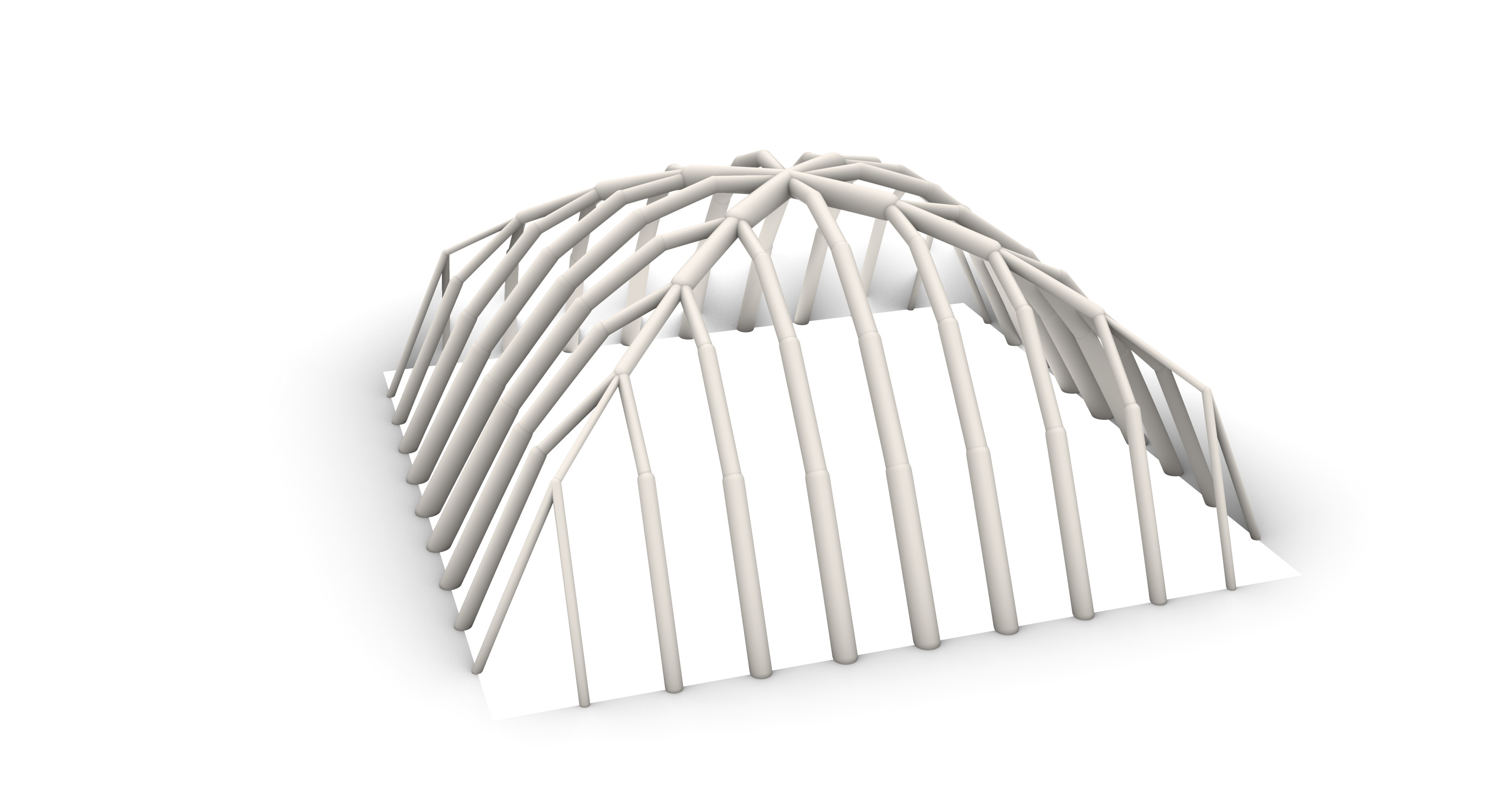}
                & \includegraphics[height=0.125\linewidth, trim = 15cm 0 12.5cm 0, clip]{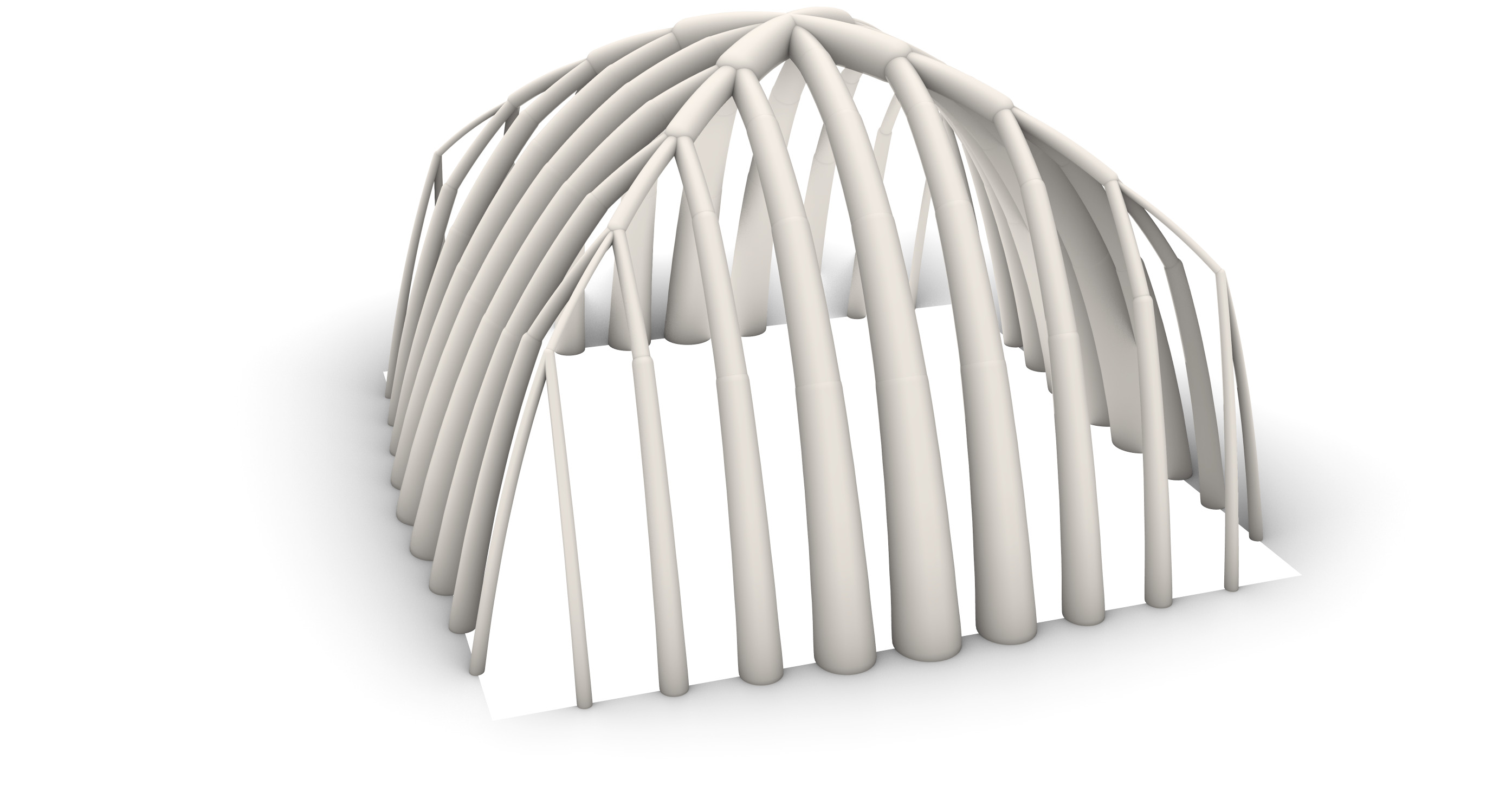}\\
			\end{tabular}
            };
   \end{tikzpicture}
	\caption[Square based example: Results with restricted topologies]{Square based example: Results with restricted topologies. 'Topology' images show the available elements in the restricted ground structures. Opt-light is based on the optimized structure at low self-weight values (e.g. Figure \ref{fig:squareVol}b), manually simplified to remove equally-optimal elements around the center of the domain. Opt-heavy is the topology identified as optimal at high self-weight values (e.g. Figure \ref{fig:squareVol}c). Archgrid allows only elements aligned to the $x$ or $y$ axes. Diagonals allows the two main diagonals, plus elements perpendicular to the closest edge of the domain, based on extending the pattern observed at the center of the Opt-light topology. Note that some ground-structure elements are not used in the optimized results, e.g. the elements connecting to the corner points in the Diagonals result. }
    \label{fig:topology}
	\end{figure*}

    \subsection{Square domain with distributed load}
    \label{sec:distributed}
	The next problem to be considered is a square domain with pin supports at all edge nodes, with the domain discretized by an 11 $\times$ 11 grid of nodes (121 nodes total). Point loads of equal magnitude are added at all unsupported internal nodes. The problem has again been solved with materials ranging from very light to very heavy and the total volumes are shown in Figure \ref{fig:squareVol}. The problem has been solved with the load applied in both an upward or downward direction. It can be seen that when self-weight is less significant, the upward and downward loadings produce structures with similar volumes, and where the forms are close to  mirror images of one another (Figure \ref{fig:squareVol}b and d). However, as self-weight becomes more significant, the forms diverge (Figure \ref{fig:squareVol}c and e). In both cases, the arching/sagging of the elements becomes more pronounced, with the downward load this causes an increase in height and with upward load the height reduces. For this example, the structure for upward loading undergoes only minor changes until $\rho g \approx 15 \frac{\sigma}{L}$, when it abruptly switches to a counterweight solution across the entire domain near-simultaneously, which continues to be the optimal solution for $\rho g \to \infty$. For an example with upward loading which demonstrates the separation of the solution into regions, see Figure S1 in the supplementary materials. 

    For downward loads, the volume increases as self-weight becomes more significant. The volume tends to infinity, with an asymptote at $\rho g \approx 4\frac{\sigma}{L}$, beyond which the problem is infeasible. \citet{rozvany1980optimal} considered a problem similar to this, but with a fixed topology of elements parallel to the domain edges -- an archgrid. This can be easily modeled using the proposed approach by simply restricting the elements available in the ground structure, and the results of this are also shown in Figure \ref{fig:squareVol}, where it can be seen that optimizing the topology permits a significant material saving of almost 25\% in cases with significant self-weight ($\rho g = 2\frac{\sigma}{L}$).

    To further explore the influence of different topologies, Figure \ref{fig:topology} explores results with the shown selected restricted topologies. It can be seen that the decision of which topology is preferable varies significantly for different values of self-weight, indicating that some forms are better suited to carrying self-weight loads and others to carrying uniform external loading. For example, the Opt-heavy form provides the minimum material result for larger values of self-weight, but when used to design a structure with a lightweight material it is close to the worst performing topology tested. Meanwhile the converse is true for the Opt-light form. Even between the less efficient forms of the Archgrid and Diagonals topologies, there is a switch in preferable topology as self-weight increases. 
    
    Generally, the impact of selecting an incorrect topology appears to increase when self-weight is more significant, with only 3.5\% between the options tested when self-weight is negligible, and 30\% variations when $\rho g = 2\frac{\sigma}{L}$. Furthermore, Figure \ref{fig:topology} shows the importance of optimizing topology and elevations simultaneously, since the optimal elevation profiles vary between the different topologies tested.

    \begin{figure}
        \centering
        \pgfplotsset{major grid style={thick}} 
        
        {\raggedleft
        \begin{tikzpicture}
            \begin{semilogyaxis}[
            xlabel = Number of vertical nodal divisions, xmin=0, xmax = 42,
            ymin = 0.0005, ymax = 0.1, ytick = {0.0001, 0.001, 0.01, 0.1}, yticklabels = {+0.01\%, +0.1\%,+1\%, +10\%},
            ylabel = Volume error {\small (vs. result in Fig. \ref{fig:squareVol}c)},
            width=0.85\linewidth]
                \addplot[only marks, mark = x, thick] table[x=vertical_num,y=increase,col sep=comma] {Square_experiments/transmissibleResults.csv};

                \draw [black!30] (8,0.0005) -- (8,0.1) node [ outer sep = 6pt, inner sep = 1pt, fill=white, sloped, pos=0, right] {\footnotesize $d_{z} = d_{xy}$};
                \draw [black!30] (16,0.0005) -- (16,0.1) node [outer sep = 6pt, inner sep = 1pt, fill=white, sloped, pos=0, right] {\footnotesize$d_{z} = \frac{1}{2}d_{xy}$};
                \draw [black!30] (32,0.0005) -- (32,0.1) node [outer sep = 6pt, inner sep = 1pt, fill=white, sloped,pos=0, right] {\footnotesize$d_{z} = \frac{1}{4}d_{xy}$};
            \end{semilogyaxis}
        \end{tikzpicture}}

        (a) \vspace{10pt}
        
        \begin{tikzpicture}
            \begin{semilogyaxis}[xlabel = Number of vertical nodal divisions, ylabel = Time increase {\small (vs. result in Fig. \ref{fig:squareVol}c)},  xmin = 0, xmax = 42, legend pos = north west, ymin = 1,  ytick = {1, 10, 100, 1000, 10000, 100000}, yticklabels ={$\times 1$, $\times 10$, $\times 100$, {$\times 1,000$}, {$\times 10,000$}, {$\times 100,000$}}, ylabel shift = -13pt, every axis y label/.style={at={(-0.25,-0.08)}, rotate=90, anchor = west}, ymax = 100000, width = 0.85\linewidth]

                \addplot[only marks, mark = x, thick] table[x=vertical_num,y=time_norm,col sep=comma] {Square_experiments/transmissibleResults.csv}; 

                \addplot[black!30, no marks, forget plot] coordinates{(0,1/0.5965) (1000, 1/0.5965)} node [xshift = 2pt, pos = 0, above right] {1 second};
                \addplot[black!30, no marks, forget plot] coordinates{(0,60/0.5965) (1000, 60/0.5965)} node [xshift = 2pt,pos = 0, above right] {1 min};
                \addplot[black!30, no marks, forget plot] coordinates{(0,3600/0.5965) (1000, 3600/0.5965)} node [xshift = 2pt,pos = 0, below right] {1 hour};
                \addplot[black!30, no marks, forget plot] coordinates{(0,21600/0.5965) (1000, 21600/0.5965)} node [xshift = 2pt,pos = 0, below right] {6 hours};
            \end{semilogyaxis}
        \end{tikzpicture}

        (b)

        \noindent\begin{tabular}{@{}c@{}c@{}}
            \includegraphics[width=0.5\linewidth, trim=20cm 0 25cm 0, clip]{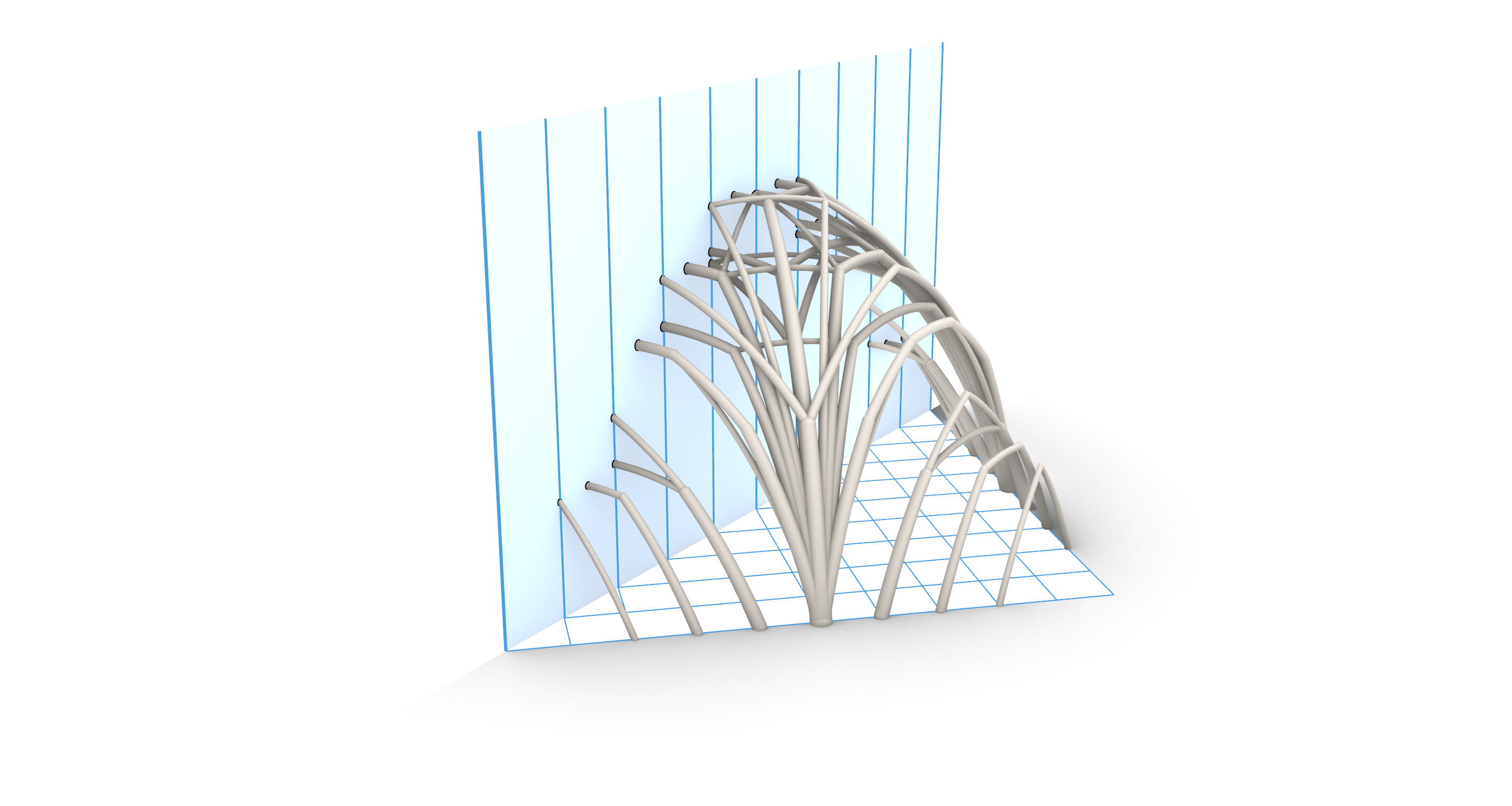}
            &
            \includegraphics[width=0.5\linewidth, trim=20cm 0 25cm 0, clip]{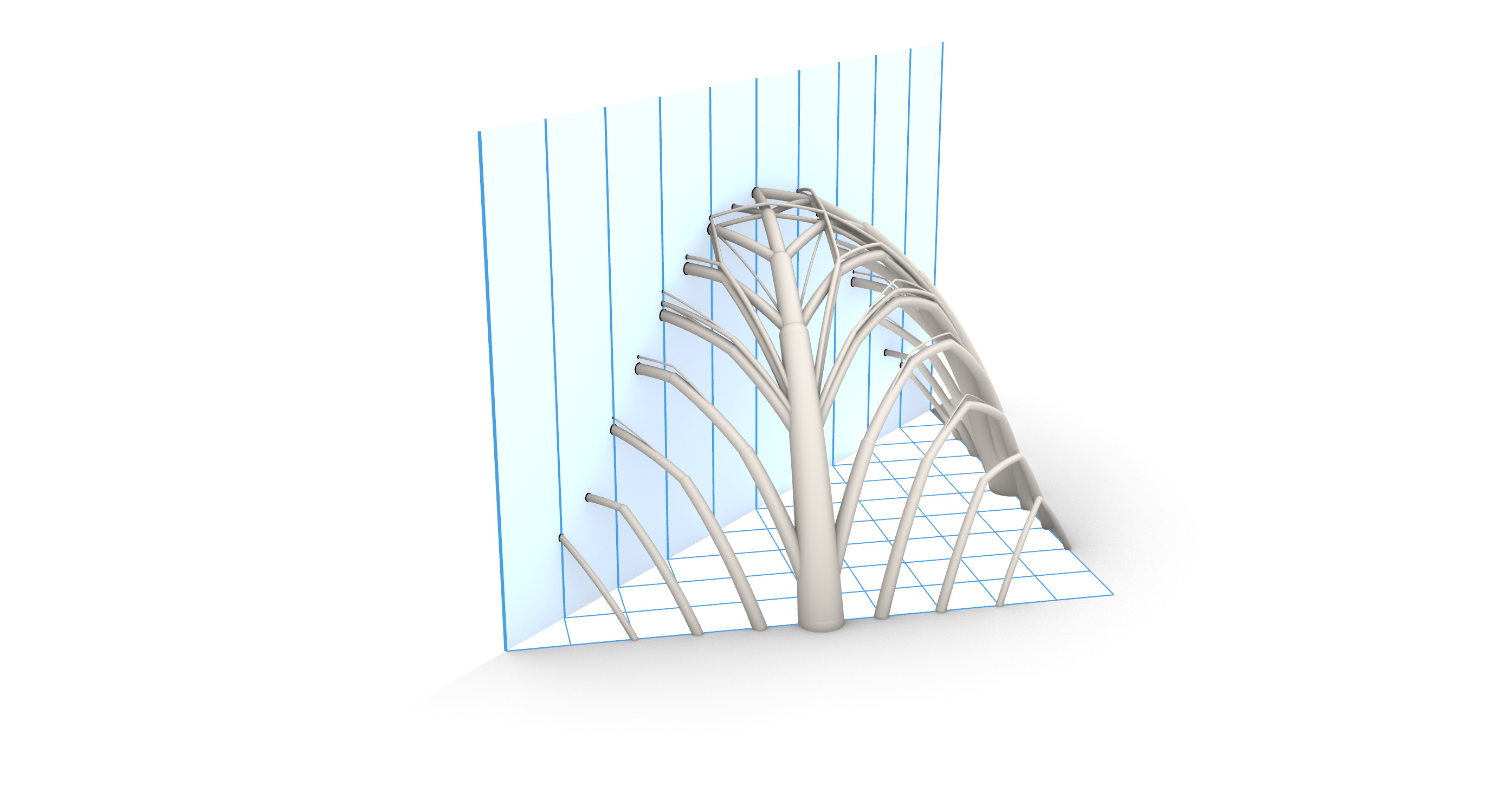} \\
            (c) & (d)
        \end{tabular}
        
        \caption{Square based example: Comparison of vault formulation with transmissible load approach for $\rho g = 2\frac{\sigma}{L}$, and maximum permitted height $0.8L$. (a) Volume, with indication of nodal spacings in the horizontal directions, $d_{xy}$ compared to the vertical nodal spacing $d_z$ (b) Time required and (c)-(d) forms obtained using the transmissible load approach. The blue plane cuts the domain corner-to-corner and the lines of the nodal grid are shown to highlight the non-single-surface nature of the results. (c) $d_z = d_{xy}$ i.e. 8 nodal divisions, (d) $d_z = \frac{1}{5}d_{xy}$ i.e. 40 nodal divisions.}
        \label{fig:transmissible}
    \end{figure}

    To show the benefit of the proposed approach against existing methodologies, results will be compared for accuracy and speed against the 3D truss layout optimization method with transmissible loads \cite{lu2021transmissible}. The comparison is carried out with a horizontal discretization of $0.1L$ as previously, and with an unrestricted topology (fully connected ground structure). To use the 3D truss layout optimization approach, the domain must also be defined and discretized in the vertical direction, and the choice of this discretization can greatly influence the speed and accuracy of the solution. Here a design domain with height $0.8L$ has been used (note that the corresponding result in Figure \ref{fig:squareVol}c has maximum height $0.604L$), and resulting volumes and solution time with a range of vertical spacings are shown in Figure \ref{fig:transmissible}. The results show good convergence of the volume towards the vault formulation result, providing further validation of the approach proposed here. However, the transmissible load approach generally produces a non-single-surface structure, leading to additional difficulty in interpreting the results, as shown in Figure \ref{fig:transmissible}c and d, where multiple elements are seen on each vertical grid-line. Furthermore, the time required to solve the problem using the vault formulation presented here was less than 0.6 seconds, whereas the transmissible load approach required several orders of magnitude more computation time (up to 6 hours at the discretizations tested here) to produce a less accurate result \footnote{Note that member adding has been used for the transmissible loading approach as well as the vault formulation, otherwise the required times would likely be 1-2 orders of magnitude larger, and the memory requirements would preclude solving on a standard laptop.}. 

\subsection{Square domain with point load}

\begin{figure}
        \begin{tikzpicture}
            \begin{axis}[xmin = 0, xmax = 2.01, ymin = 0, ymax = 4840, legend pos = north west, 
            ytick = {0,968, 1936, 2904,3872, 4840}, yticklabels = {0, 10$\frac{FL}{\sigma}$, 20$\frac{FL}{\sigma}$, 30$\frac{FL}{\sigma}$, 40$\frac{FL}{\sigma}$, 50$\frac{FL}{\sigma}$}, ylabel = {Volume, $V$},
            xtick = {0, 0.5, 1, 1.5, 2}, xticklabels={0, 0.5$\frac{\sigma}{L}$, 1$\frac{\sigma}{L}$,1.5$\frac{\sigma}{L}$, 2$\frac{\sigma}{L}$}, xlabel = {Unit weight, $\rho g$}, width = 0.65\linewidth, scale only axis]
            
                \addplot[forget plot, no marks, black!30, domain = 0.024:2, samples = 50] plot (\x, 96.8/\x) node[pos= 0.65, sloped, fill=white, inner sep = 1pt] {\footnotesize $W=F$};
                \addplot[forget plot, no marks, black!30, domain = 0.12:2, samples = 50] plot (\x, 484/\x) node[pos= 0.5, sloped, fill=white, inner sep = 1pt] {\footnotesize $W=5F$};
                \addplot[forget plot, no marks, black!30, domain = 0.24:2, samples = 50] plot (\x, 968/\x) node[pos= 0.5, sloped, fill=white, inner sep = 1pt] {\footnotesize $W=10F$};
                \addplot[forget plot, no marks, black!30, domain = 0.48:2, samples = 50] plot (\x, 1936/\x) node[pos= 0.5, sloped, fill=white, inner sep = 1pt] {\footnotesize $W=20F$};
                \addplot[forget plot, no marks, black!30, domain = 0.8:2, samples = 50] plot (\x, 3872/\x) node[pos= 0.7, sloped, fill=white, inner sep = 1pt] {\footnotesize $W=40F$};

                \addplot[no marks, thick] table[x=pg,y=vol_11x11,col sep=comma] {Square_experiments/SingleLoadResults.csv}; 
                \addlegendentry{$V_\text{11$\times$11}$}

                \addplot[no marks, dashed] table[x=pg,y=vol_simple,col sep=comma] {Square_experiments/SingleLoadResults.csv}; 
                \addlegendentry{$V_\text{5-nodes}$}

                \node (hv) [draw=blue, circle, inner sep = 1.5pt, ] at (1.85,2520.037266) {};
                \node (lt) [draw=blue, circle, inner sep = 1.5pt, ] at (1.5, 884.63) {};
                
            \end{axis}

            \draw [blue] (lt) -- (0.3\linewidth, 5.3cm);
            \node (pic1) at (0, 5.15cm) [above right, draw=blue, thick, inner sep = 0, outer sep = 0] {\includegraphics[width=0.3\linewidth, trim = 18cm 7cm 11cm 3cm, clip]{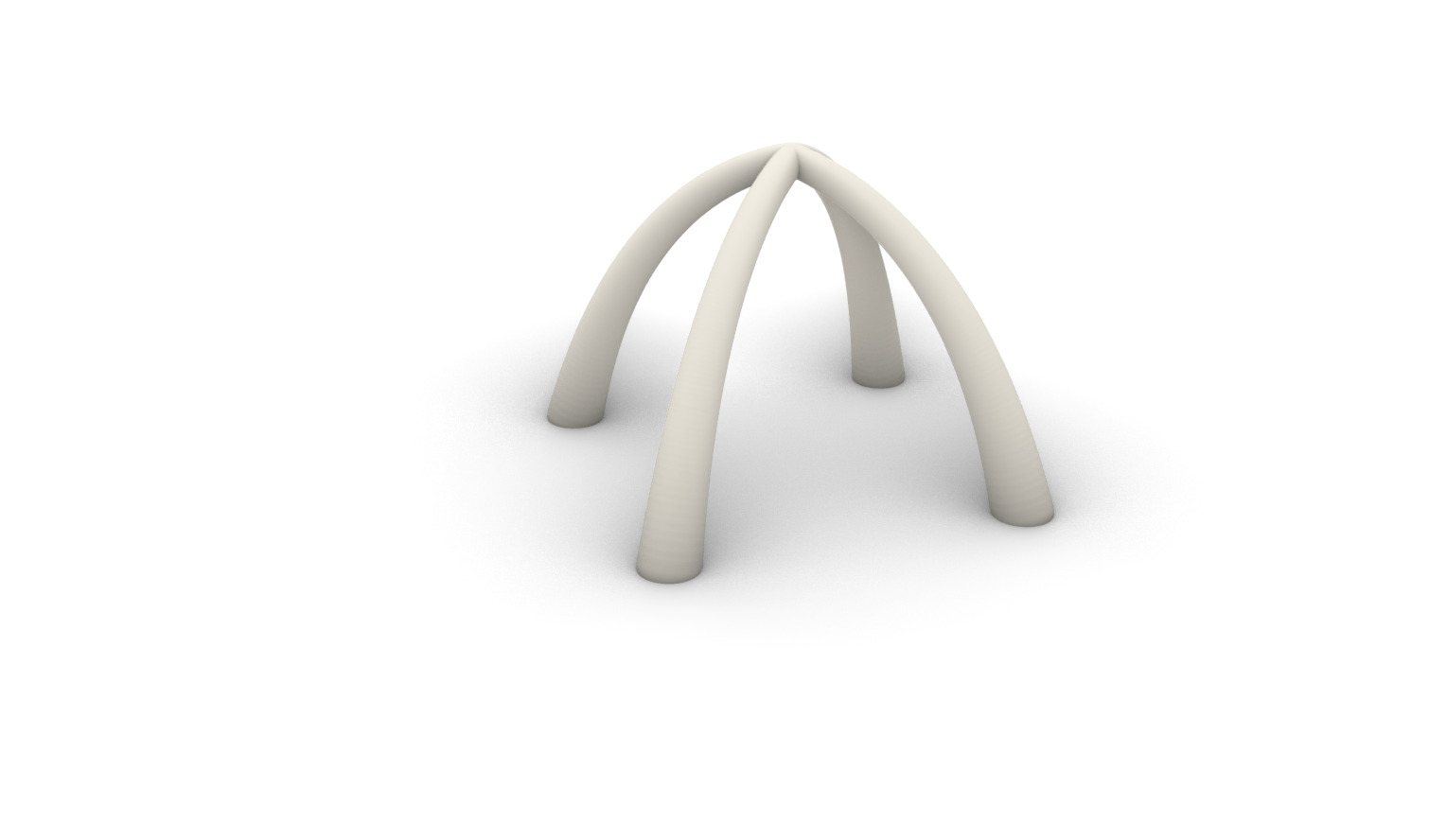}};

            \node (pic2) at (0.65\linewidth, 5.15cm) [above left, draw=blue, thick, inner sep = 0, outer sep = 0] {\includegraphics[width=0.3\linewidth, trim = 18cm 7cm 11cm 3cm, clip]{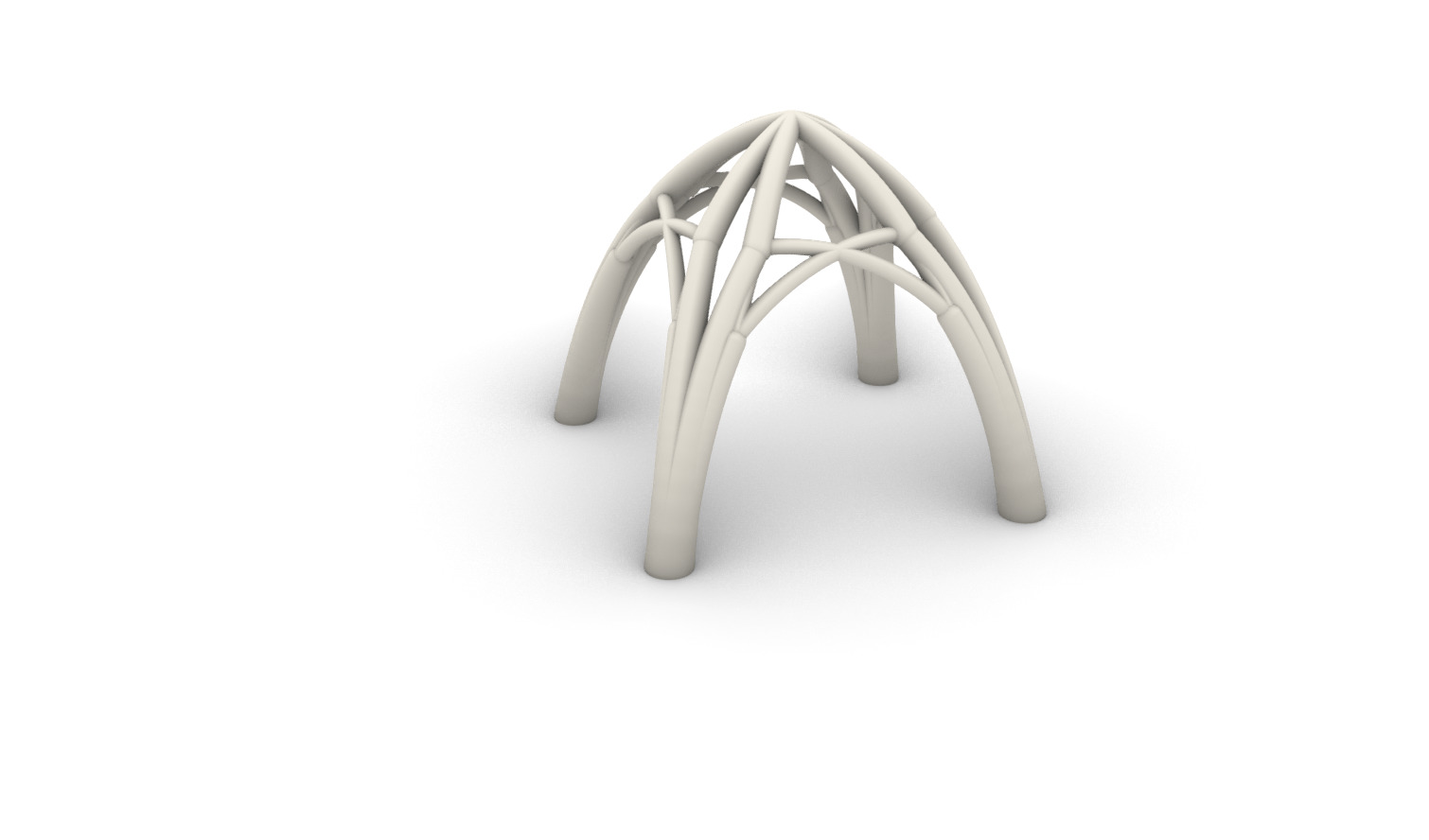}};
            \draw [blue] (hv) -- (pic2);

        \end{tikzpicture}
    \caption{Square example with point load: Optimal volumes for fully connected 11$\times$11 grid, $V_{11\times11}$ and for minimal 5 node (load and supports only) problem, $V_\text{5-nodes}$. Example results shown for $\rho g = 0.75\frac{\sigma}{L}$ (identical for 5-nodes and 11$\times$11 result) and  $\rho g=1.85\frac{\sigma}{L}$ (for 11$\times$11 grid). For context, relationships between the structure weight $W = \rho g V$ and the total external loading $F$ are given.}
    \label{fig:pointLoadPlotsA}
\end{figure}
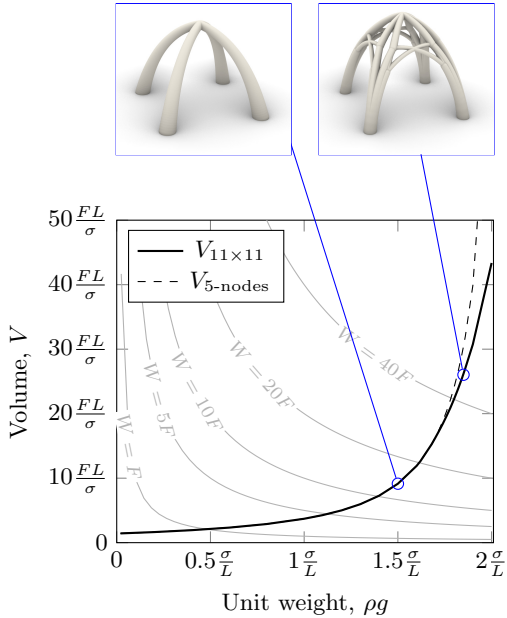

    The next problem again concerns a square domain with sides of length $L$. However, for this example, the supports are available only at the four corners, and the load is applied as a point load at the center of the domain acting downward. The resulting volumes based on an 11$\times$11 grid of nodes with a fully connected ground-structure are plotted in Figure \ref{fig:pointLoadPlotsA} as $V_\text{11$\times$11}$. It can be seen that the self-weight dominates at shorter spans than in Section \ref{sec:distributed}, with a much more rapid increase in volume observed; this is to be expected as there are fewer supports and there is no longer a proportion of the loading applied very close to the supports. When self-weight is significant, these solutions make use of nodes other than the loads or supports, see example in Figure \ref{fig:pointLoadPlotsA}. 

    When self-weight is not considered, it has been supposed that an optimal topology must exist that involves only joints at points that are loaded or supported (Conjecture 9.1 in \cite{bolbotowski2022optimal}) although there may be equally optimal solutions with extra joints. In this case, this would imply an equally optimal solution containing 5 nodes -- the loaded point plus the four supports -- and a simple $\times$-shaped structure. To verify that the more complex solutions are beneficial, the problem was re-analyzed using a restricted ground structure containing just those 5 nodes, and these results are also shown in Figure \ref{fig:pointLoadPlotsA} and Table \ref{tab:pointLoad} as $V_\text{5-nodes}$. It can be seen that the simpler topology uses significantly more material than the optimized topology, with the optimal topology requiring around half (53.7\%) of the material needed for the simpler design when $\rho g = 2 \frac{\sigma}{L}$. However, for lower values of self-weight ($\rho g < 1.65 \frac{\sigma}{L}$) both problems give the same solution, again implying a change in optimal topology when self-weight is more significant.

\begin{figure}[t!]
\centering

        \begin{tikzpicture}
            \draw [white] (-3,0)--(3,0);
        
            \node at (-2.5cm, 0) [inner sep = 0, outer sep = 0, above left]{\includegraphics[height=2.5cm, trim = 8.5cm 0cm 1.45cm 0cm, clip]{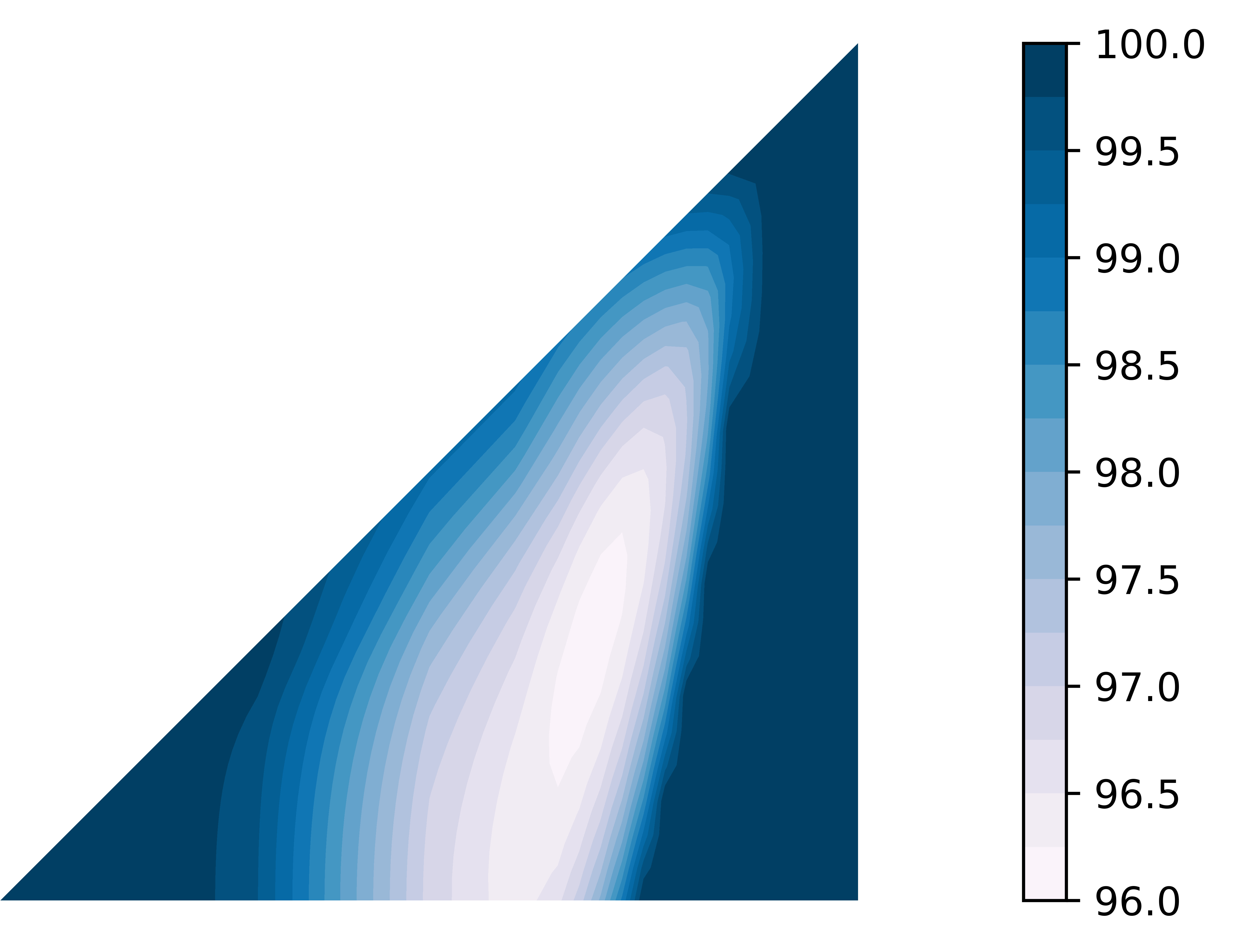}};
            \node [draw, circle, inner sep = 1pt, fill = black!20] at (6.9/4, 3.036/4) {};

            \node at (-2.5cm, 0) [right, yshift =3pt] {$V_\text{13-nodes}$};
            \node at (-2.5cm, 2.5cm) [right, yshift = -3pt] {$V_\text{5-nodes}$};
            
            \node [inner sep = 0, outer sep = 0, above right]{\includegraphics[width=2.5cm, trim = 0 0 3.3cm 0, clip]{Square_experiments/PointLoad/exhaustive}};
            \node [draw, circle, inner sep = 1pt, fill = black!20] at (6.9/4, 3.036/4) {};
        \end{tikzpicture}

    (a)

        \begin{tikzpicture}
            \begin{axis}[ticks = none, xmin = -10, xmax = 10, ymin = -10, ymax = 10, width = 5cm, height = 5 cm, scale only axis,axis x line = none, axis y line = none]
                
                \fill [blue!5] (-10,-10)--(10,-10)--(10,10)--(-10,10)--cycle;
                \fill [blue!20] (0,0)--(10,0)--(10,10)--cycle;
                \addplot[no marks] table[x=x_best,y=y_best,col sep=comma] {Square_experiments/SingleLoadResults.csv}; 

                \node [draw, circle, inner sep = 1.5pt, fill = white] at (6.74,6.2) {};
                \node [draw, circle, inner sep = 1.5pt, fill = black!20] at (6.9, 3.036) {};
                \node [draw, circle, inner sep = 1.5pt, fill = black!50] at (6.97,0) {};
                \node [draw, circle, inner sep = 1.5pt, fill = black] at (8.06,0) {};

                \addplot [only marks, mark = x, very thick] coordinates {(0,0)};
                \node [below] at (0,0) {Loaded point};
            \end{axis}

            \begin{scope}[black!30]
                \draw [ultra thick] (0,-0.5) -- (0,0)--(-0.5,0);
                \draw [ultra thick] (5,-0.5) -- (5,0)--(5.5,0);
                \draw [ultra thick] (5,5.5) -- (5,5) -- (5.5,5);
                \draw [ultra thick] (0,5.5) -- (0,5) -- (-0.5,5);
                \foreach \k in {0.05, 0.15, 0.25, 0.35, 0.45}{
                    \draw (0,-\k) -- ++ (-0.1,-0.1);
                    \draw (-\k, 0) -- ++ (-0.1,-0.1);
    
                    \draw (5,-\k) -- ++ (0.1,-0.1);
                    \draw (5+\k, 0) -- ++ (0.1,-0.1);
    
                    \draw (5,5+\k) -- ++ (0.1,0.1);
                    \draw (5+\k, 5) -- ++ (0.1,0.1);
    
                    \draw (0,5+\k) -- ++ (-0.1,0.1);
                    \draw (-\k, 5) -- ++ (-0.1,0.1);
                }
            \end{scope}
        \end{tikzpicture}

            (b)
        \caption[]{Square example with point load: Optimization of 13-node structures. 
        (a) Contour plot of volumes for all possible locations of the additional node for $\rho g = 1.76 \frac{\sigma}{L}$, with optimal location indicated.
        (b)  Optimal location of additional node for $\rho g = 1.68\frac{\sigma}{L}$ (white), $\rho g = 1.76\frac{\sigma}{L}$, $\rho g = 1.85\frac{\sigma}{L}$ and $\rho g = 2\frac{\sigma}{L}$ (black). }
        \label{fig:pointLoadPlotsB}
    \end{figure}

    The results given for $V_\text{11$\times$11}$ are the globally optimal solutions for that discretization of the domain. However, moving the (non-loaded) nodes may further improve the solution, as in geometry optimization which is commonly employed for truss problems. Here a simple, semi-manual approach is used for demonstration purposes. From observation of the optimal 11$\times$11 forms, it is assumed that the optimal topology involves the 5 loaded/supported points plus an additional point in each of the 8 symmetrical segments of the domain defined by the horizontal, vertical and diagonal symmetry of the problem; one of these segments is highlighted in Figure \ref{fig:pointLoadPlotsB} (duplicate nodes were removed for degenerate cases where the extra node is located on a symmetry plane). The position of all the extra nodes is thus controlled by just two variables, the horizontal and vertical position. 

    \begin{table}
        \caption{Square example with point load: Optimal volumes for selected values of self-weight, $\rho g$ as highlighted in Figures \ref{fig:pointLoadPlotsA}, \ref{fig:pointLoadPlotsB} and/or \ref{fig:pointLoadForms}. Results given for $V_{\text{5-nodes}}$ allowing nodes at supports and loaded points only; $V_{\text{11$\times$11}}$ using a fully connected $11\times11$ grid at fixed positions; and $V_{\text{13-nodes}}$ with nodes at supports, loaded point and one optimally selected additional node per segment as shown in \ref{fig:pointLoadPlotsB}.}
        \label{tab:pointLoad}
        \begin{tabular}{@{}cccc@{}}
        \hline \noalign{\smallskip}
        $\rho g \left(\frac{\sigma}{L}\right)$ & $V_{\text{5-nodes}} \left(\frac{FL}{\sigma}\right)$ & $V_{\text{11$\times$11}}\left(\frac{FL}{\sigma}\right)$ & $V_{\text{13-nodes}} \left(\frac{FL}{\sigma}\right)$ \\
        \noalign{\smallskip}\hline \noalign{\smallskip}
        $1.65$ & 13.8394&	13.8394&	13.8394 \\
        $1.68$ & 15.2528&	15.2516&	15.2415 \\
        $1.72$ & 17.5301&	17.3435&	17.3056\\
        $1.76$ & 20.4014&	19.6510&	19.6100\\
        $1.80$ & 24.0981&	22.2817&	22.2180\\
        $1.85$ & 30.4425&	26.1884&	26.0334 \\
        $2.00$ & 80.7391&	43.3682&	43.3575\\
        \noalign{\smallskip} \hline
        \end{tabular}
    \end{table}

\begin{figure}
        \begin{tabular}{@{}c@{}c@{}c@{}}
        \raisebox{0.5cm}{\rotatebox{90}{$\rho g = 1.65\frac{\sigma}{L}$}} &
        \includegraphics[width = 0.47\linewidth, trim = 45cm 15cm 33cm 15cm, clip]{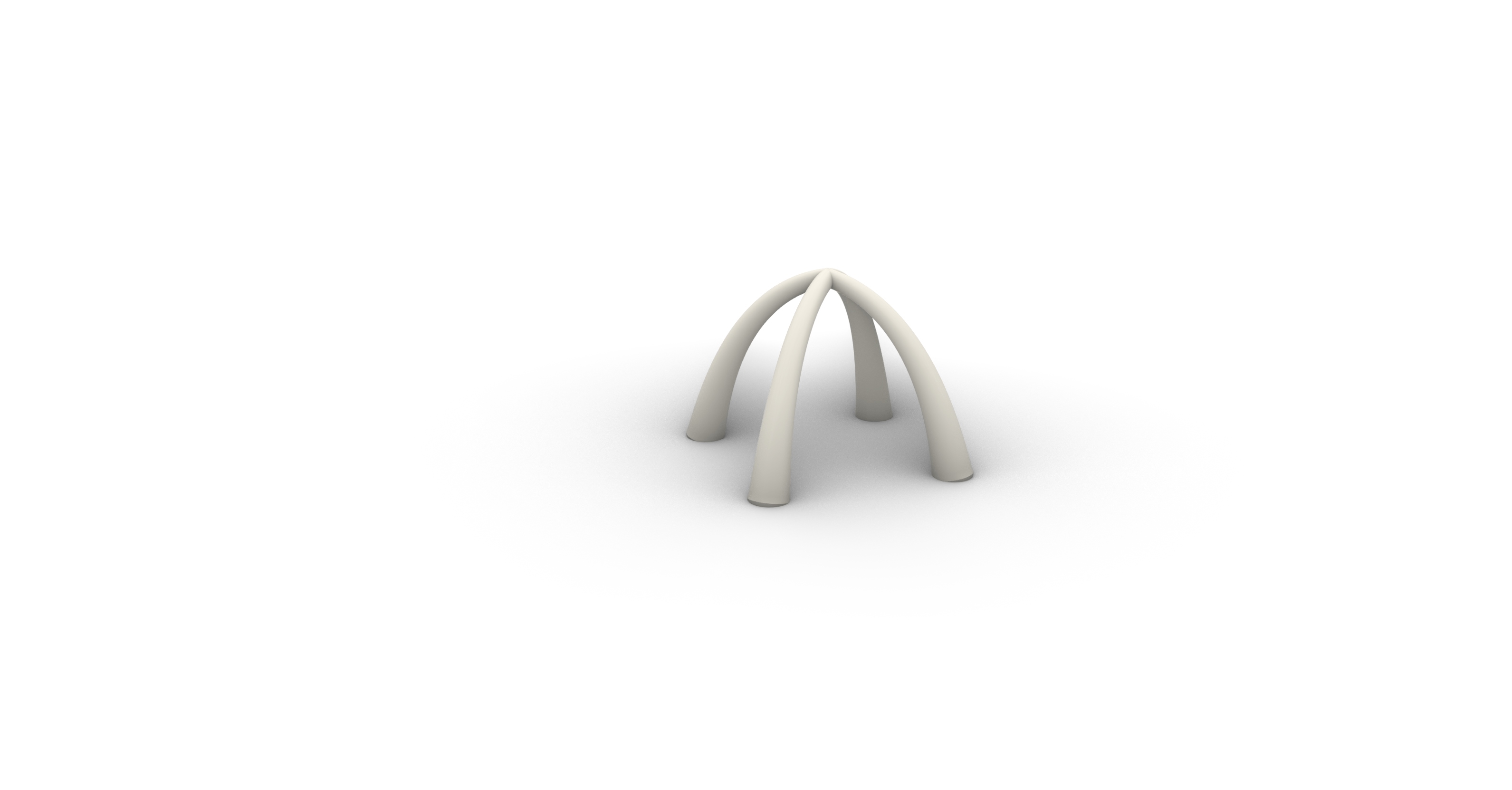} &
        \includegraphics[width = 0.46\linewidth, trim = 20cm 0 20cm 0, clip]{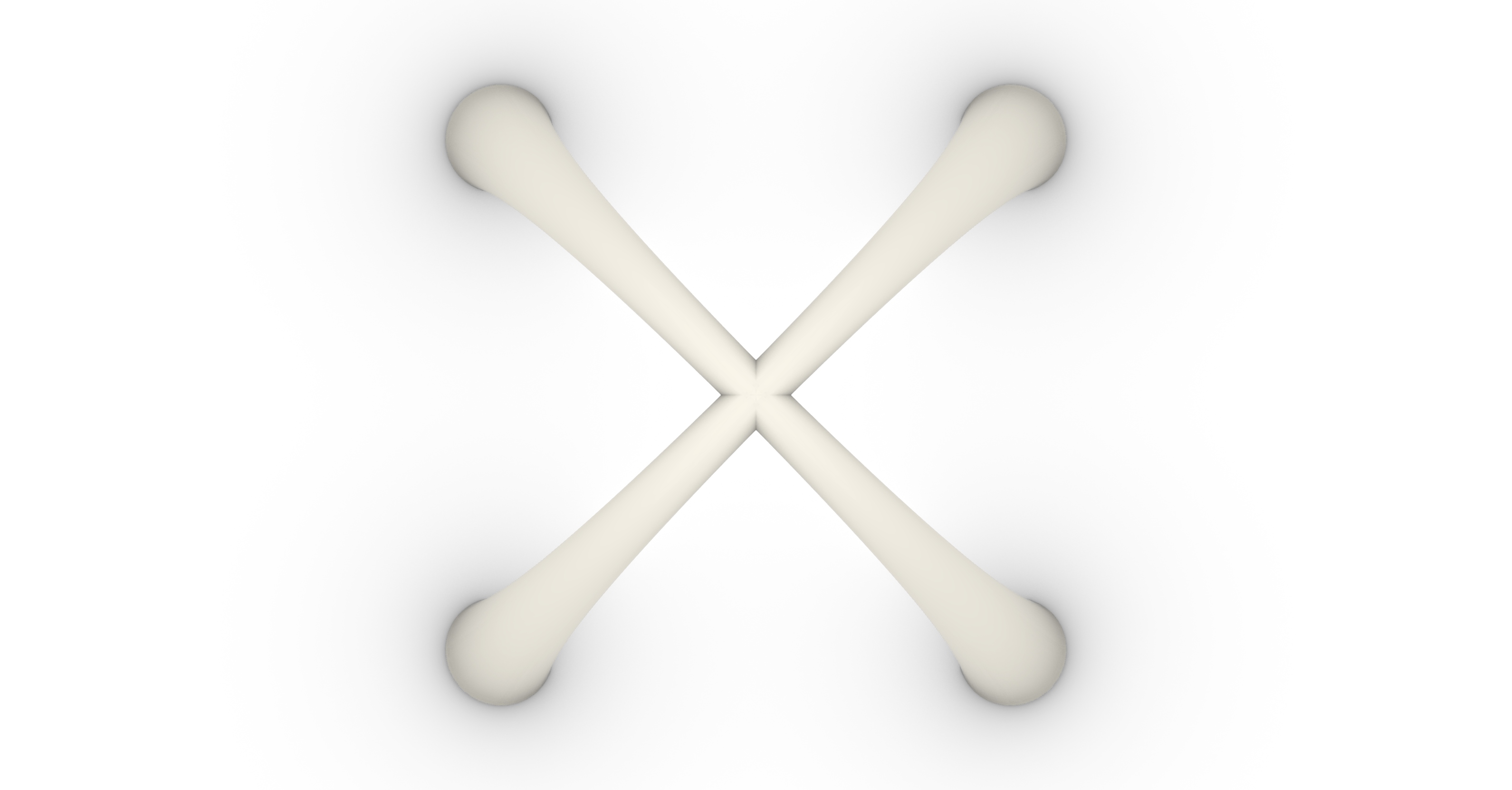} \\
        \raisebox{0.5cm}{\rotatebox{90}{$\rho g = 1.72\frac{\sigma}{L}$}} &
        \includegraphics[width = 0.47\linewidth, trim = 45cm 15cm 33cm 15cm, clip]{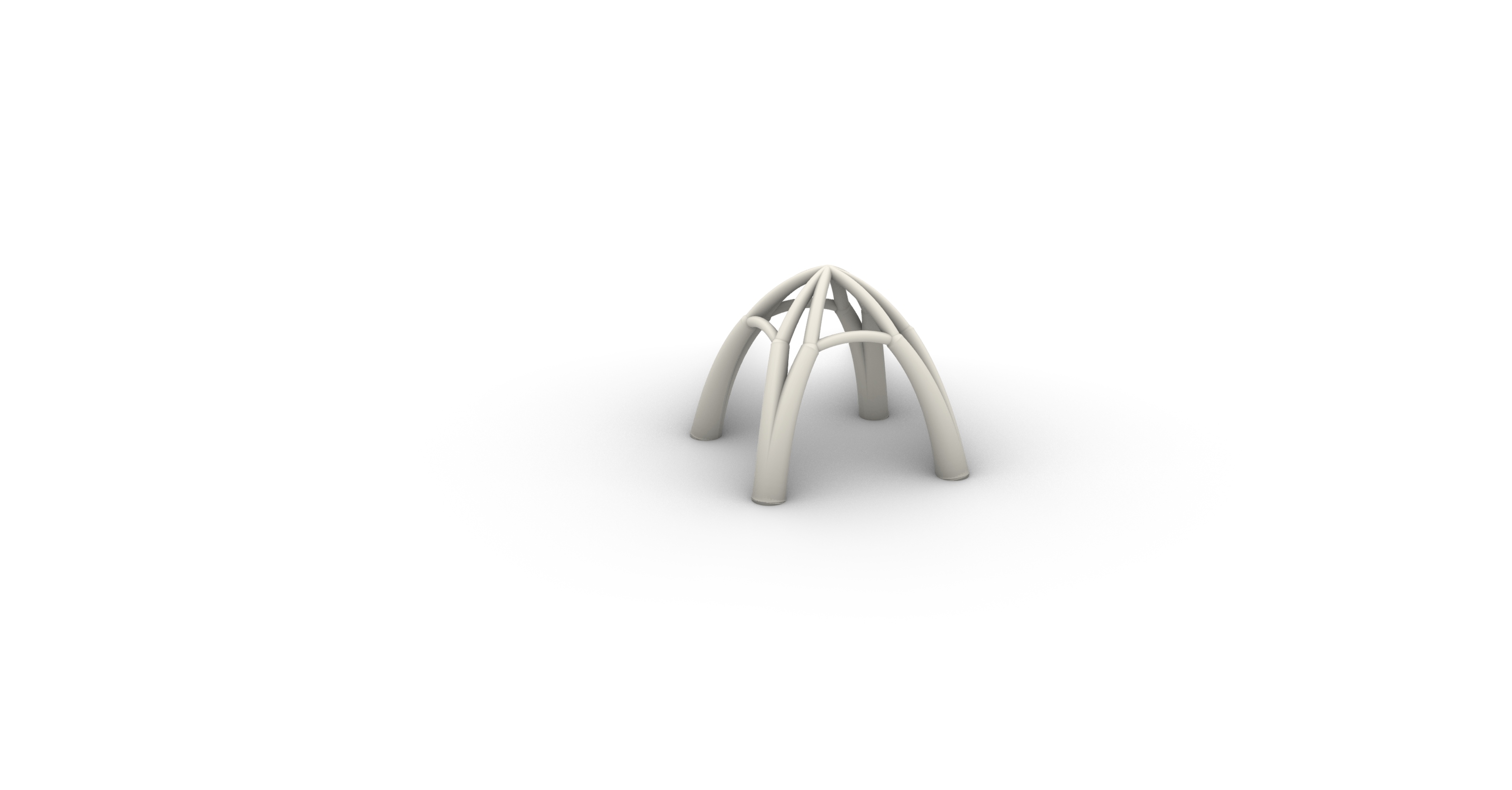} &
        \includegraphics[width = 0.46\linewidth, trim = 20cm 0 20cm 0, clip]{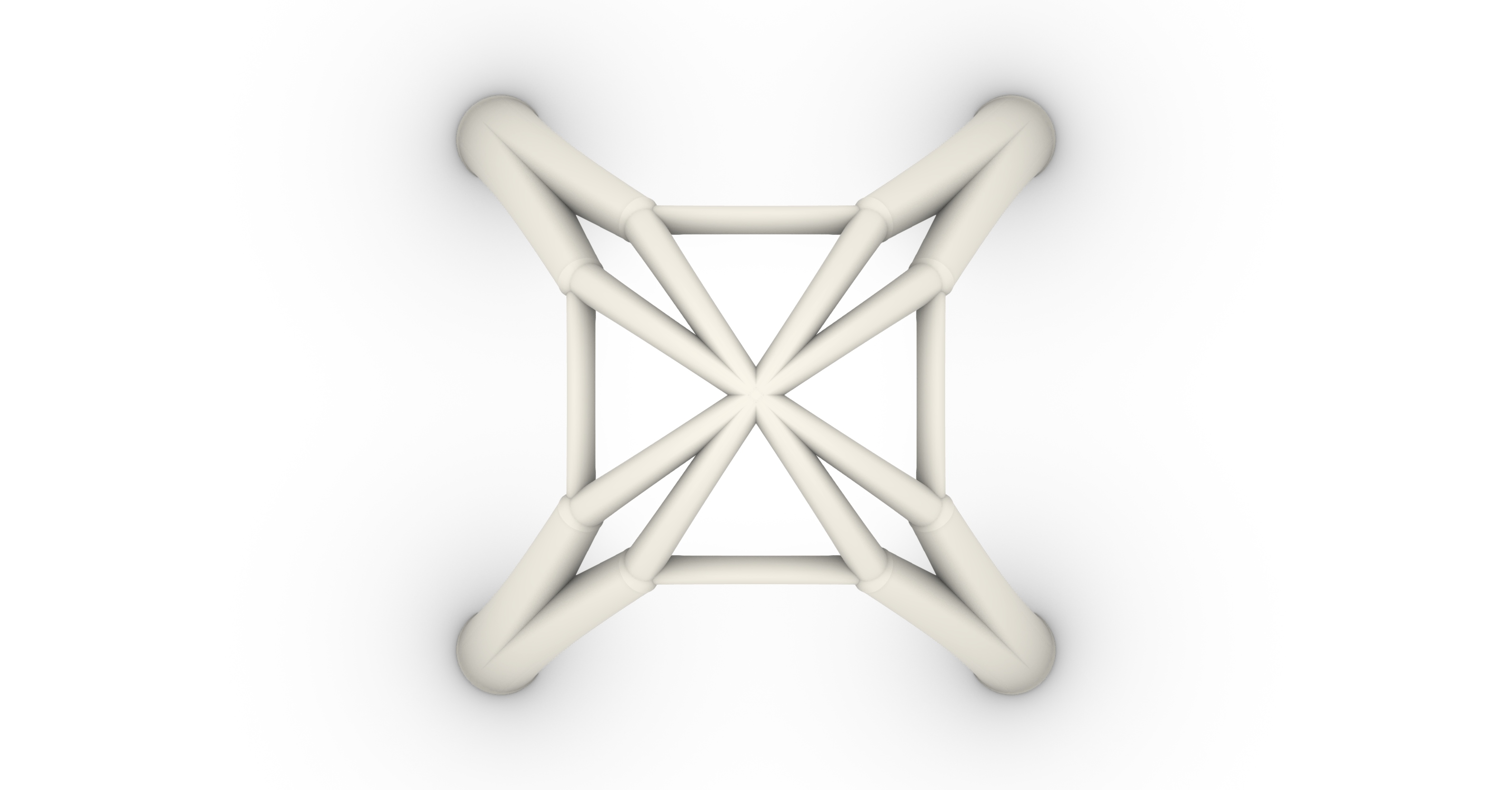} \\
        \raisebox{0.5cm}{\rotatebox{90}{$\rho g = 1.76\frac{\sigma}{L}$}} &
        \includegraphics[width = 0.47\linewidth, trim = 45cm 15cm 33cm 15cm, clip]{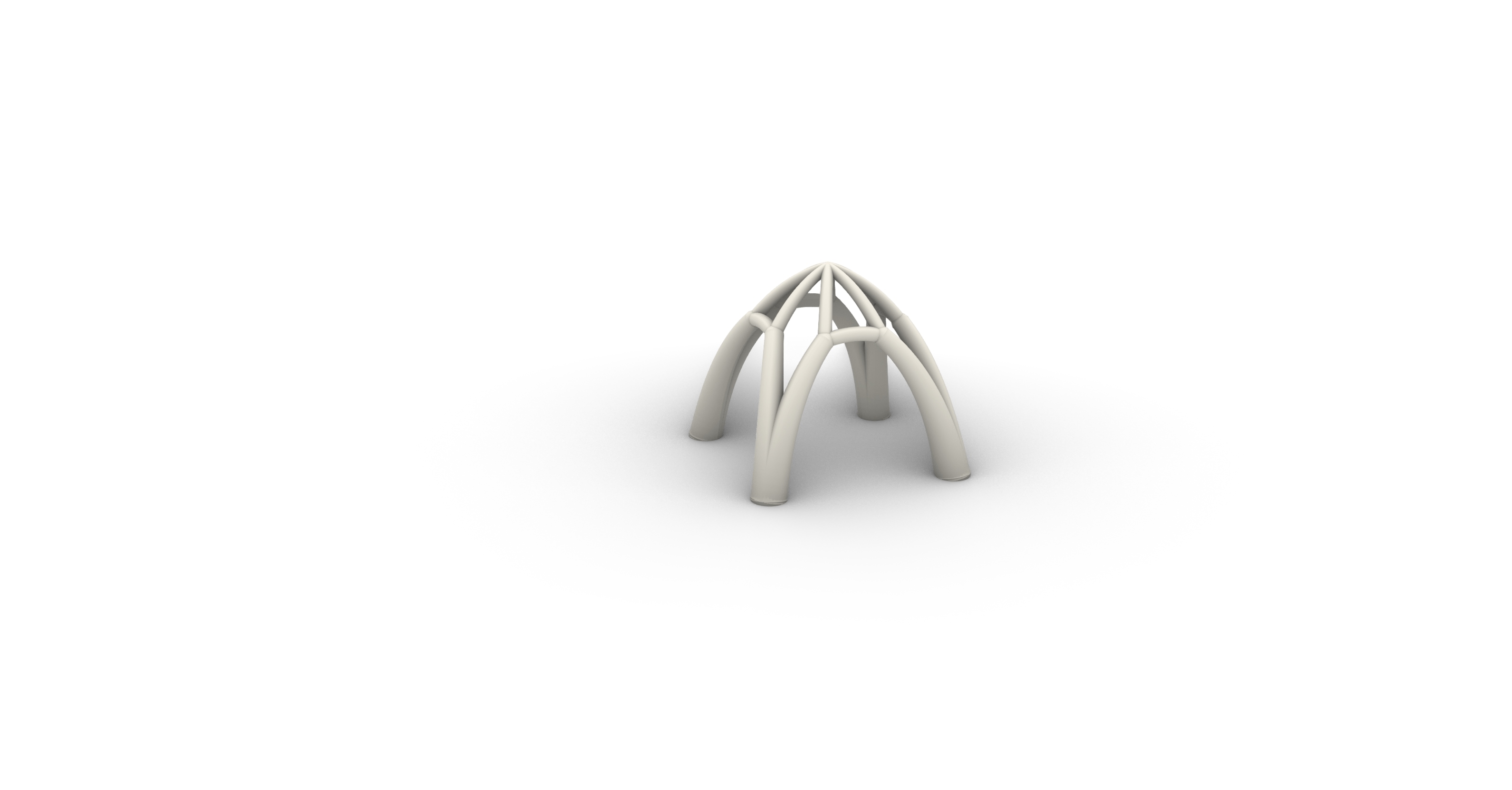} &
        \includegraphics[width = 0.46\linewidth, trim = 20cm 0 20cm 0, clip]{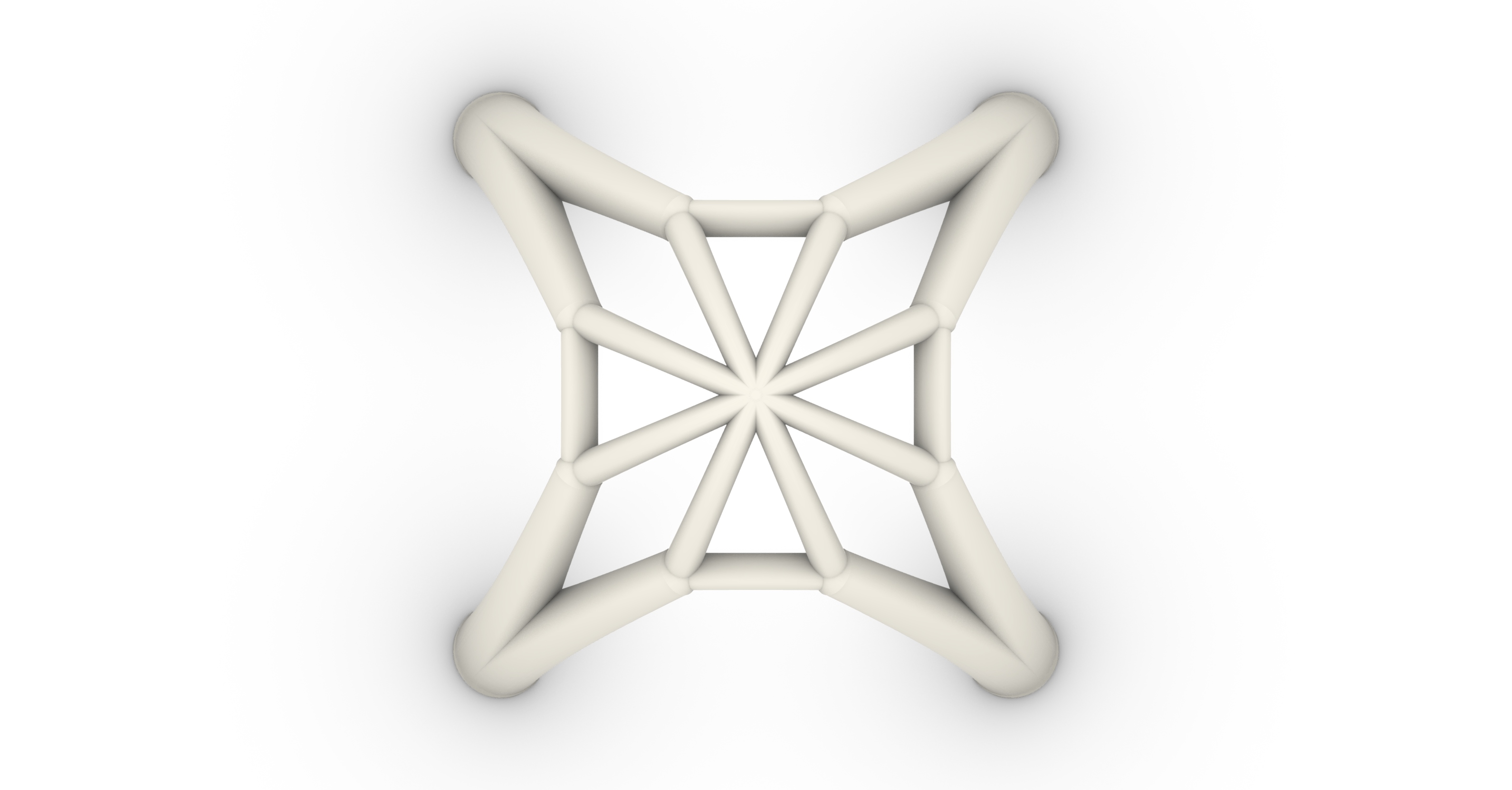} \\
        \raisebox{0.5cm}{\rotatebox{90}{$\rho g = 1.80\frac{\sigma}{L}$}} &
        \includegraphics[width = 0.47\linewidth, trim = 45cm 15cm 33cm 15cm, clip]{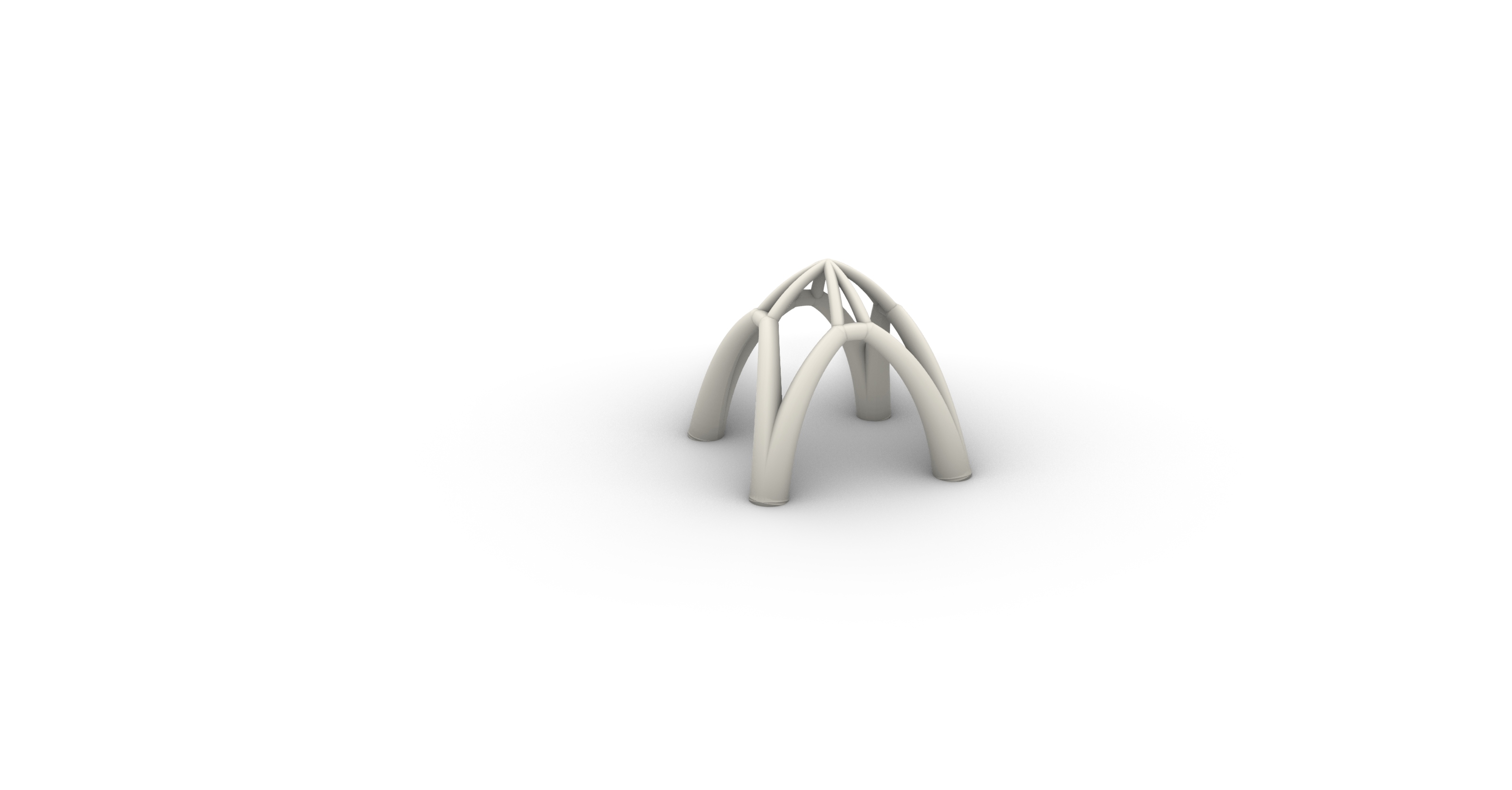} &
        \includegraphics[width = 0.46\linewidth, trim = 20cm 0 20cm 0, clip]{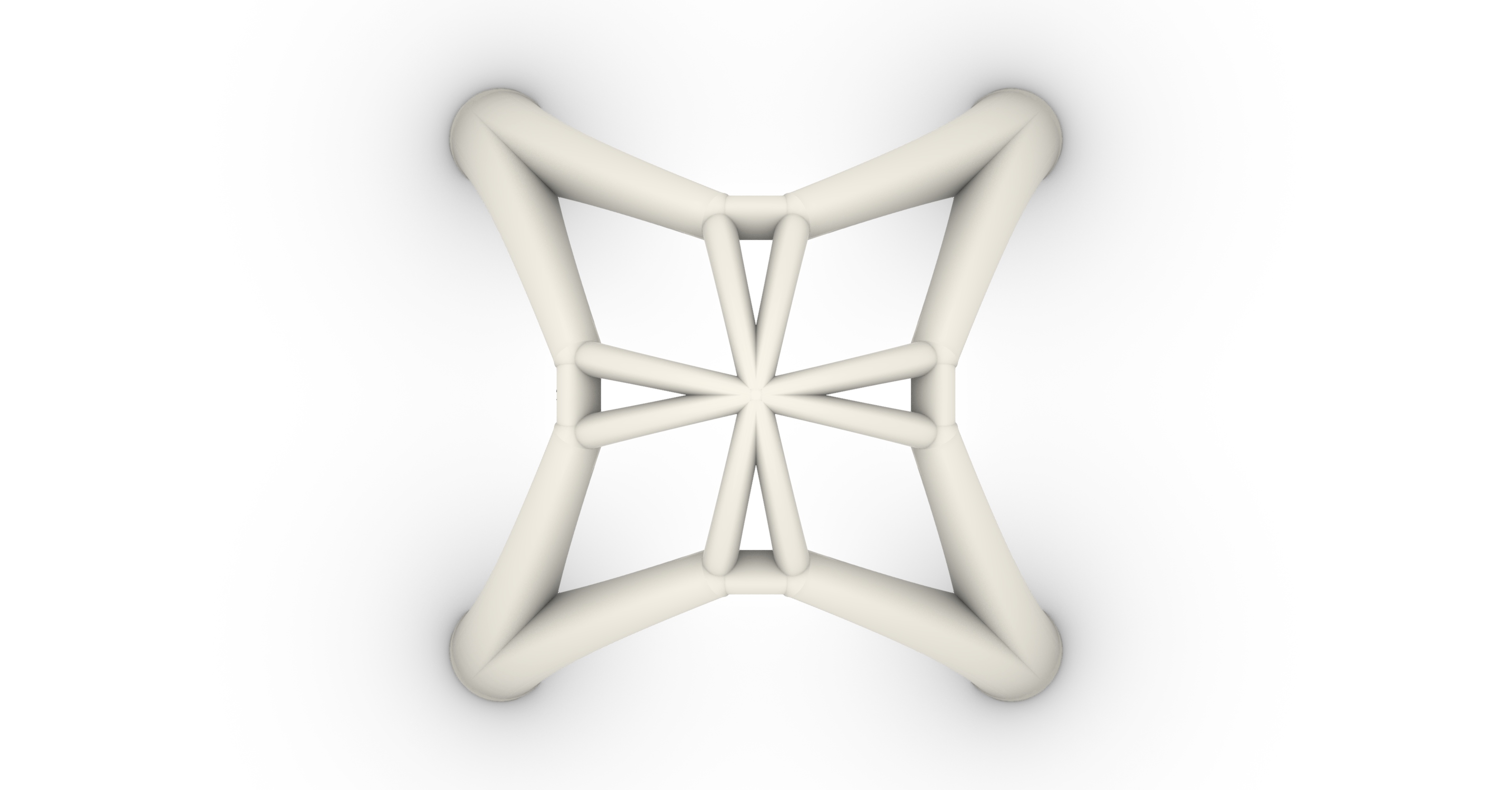} \\
        \raisebox{0.5cm}{\rotatebox{90}{$\rho g = 1.85\frac{\sigma}{L}$}} &
        \includegraphics[width = 0.47\linewidth, trim = 45cm 15cm 33cm 15cm, clip]{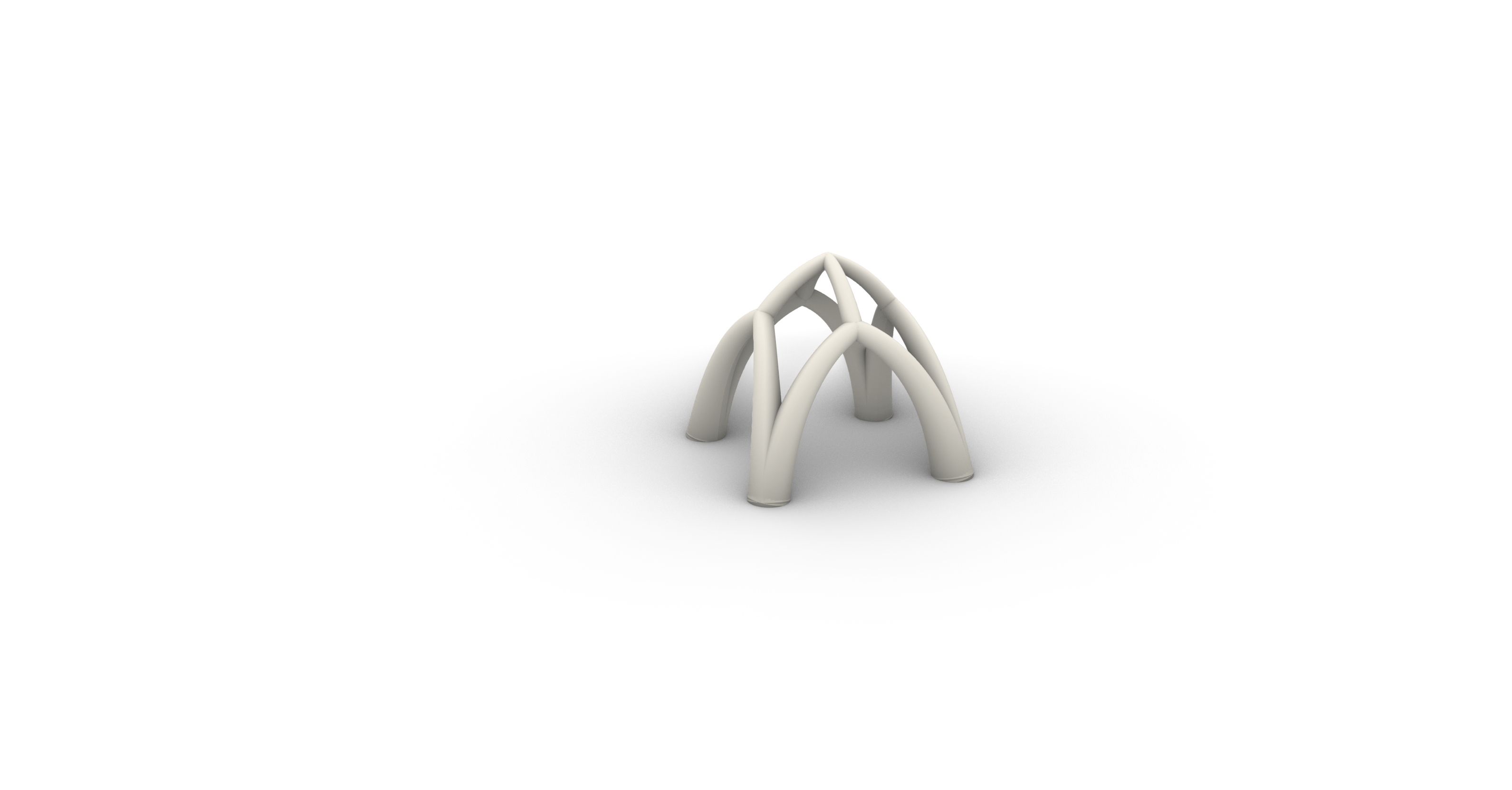} &
        \includegraphics[width = 0.46\linewidth, trim = 20cm 0 20cm 0, clip]{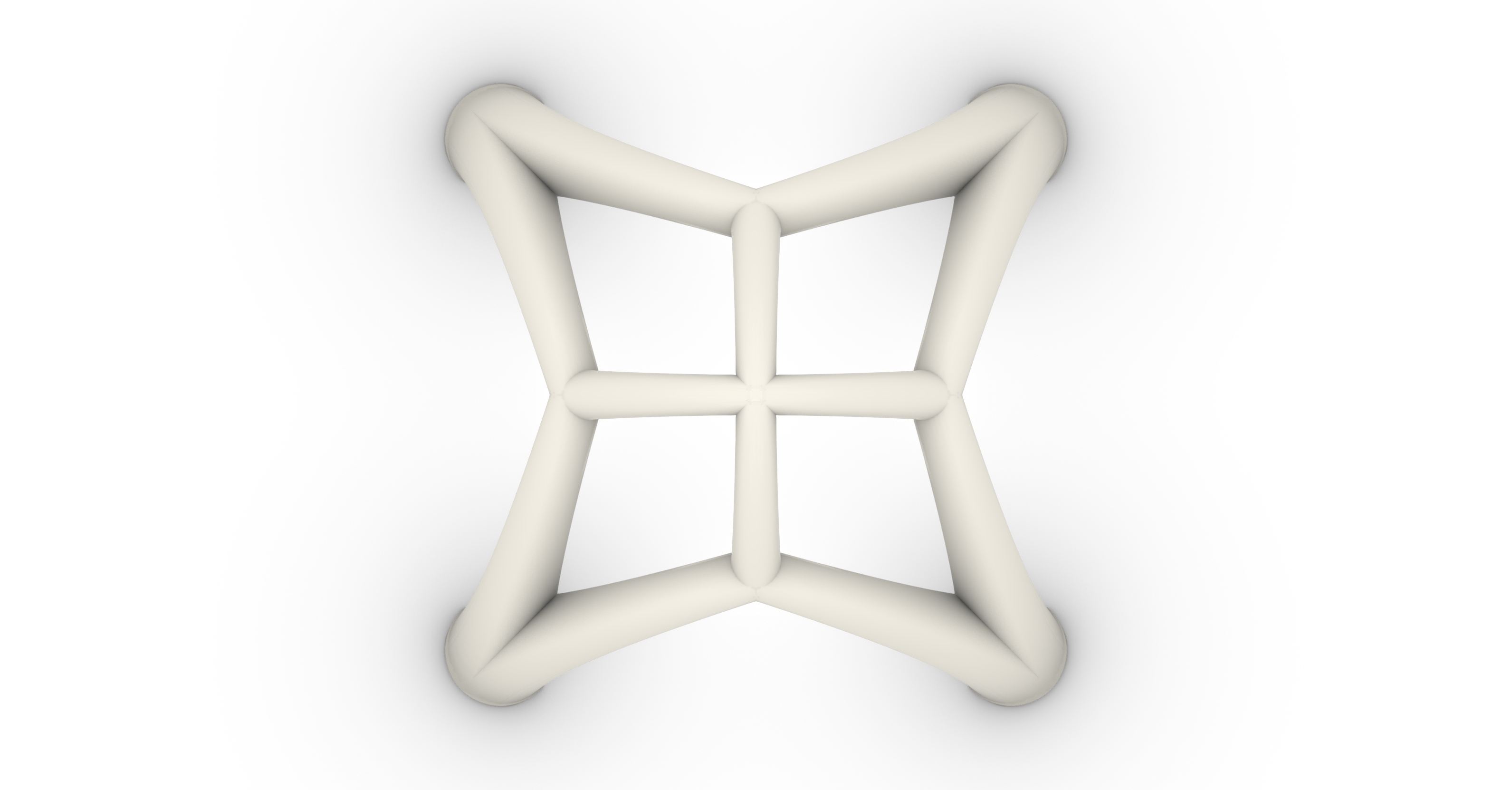} \\
        \end{tabular}
        \caption{Square example with point load: Optimal `13-node' solutions (plan and perspective view) for various values of self-weight}
        \label{fig:pointLoadForms}
    \end{figure}

Figure \ref{fig:pointLoadPlotsB}a shows the resulting volumes -- for a given unit weight -- for all possible locations of the additional point. This shows the simple behavior of the problem which exhibits a single optimum location for the point, denoted as $V_\text{13-nodes}$. Thus, using the initial 11$\times$11 node solution as a guide, an optimal node location can be found easily by manually exploring different locations. With each problem taking less than 0.1 seconds to solve, this is easily achieved using parametric modeling software such as Rhino/Grasshopper \cite{grasshopper}. In all cases, this proved to make only a minor improvement to the total material usage (generally around 0.2\,\%, with some cases up to 0.6\,\% where the optimal node position fell directly between points in the original ground structure). The resulting volumes, $V_\text{13-nodes}$, are also shown in Table \ref{tab:pointLoad}. The forms of these solutions are shown in Figure \ref{fig:pointLoadForms}, while Figure \ref{fig:pointLoadPlotsB}b also shows the location of one of the additional nodes for various levels of self-weight.

    From Figures \ref{fig:pointLoadPlotsA} and \ref{fig:pointLoadForms} it can be seen that when self-weight is low ($\rho g \leq 1.65$), the optimal structure employs the most direct load-path from the point load to the support, aligning the structure to the $\sqrt{2}L$ length span between diagonally opposing supports. However, when self-weight is more dominant (e.g. $\rho g \geq 1.85$) it is preferable to locate as much material as possible close to the shorter spans of length $L$ between adjacent supports, with a secondary set of elements spanning from the point load to these outer elements. For intermediate values of self-weight, the transition occurs by a splitting of the original, diagonal elements.

    \subsection{Example with two holes}

    \begin{figure}
    \centering
        \begin{tikzpicture}[scale = 0.1, >=latex]
            \fill [blue!10] (0,0)--(60,0)--(60,80)--(0,80)--cycle;
            \fill [white] (40, 20) circle(10);
            \fill [white] (20, 60) circle(10);

            \draw [black!30, ultra thick] (0,0)--(60,0)--(60,80)--(0,80)--cycle;
            \draw [black!30, ultra thick, dashed] (40, 20) circle(10);

            \draw [<->, xshift = -3cm] (0,0)--(0,80) node [pos= 0.5, fill=white] {$L$};
            \draw [<->, yshift = -4cm] (0,0)--(60,0) node [pos= 0.5, fill=white] {$\frac{3}{4}L$};
            \draw [white, <->, xshift = 3cm] (60,0)--++(0,80) node [pos= 0.5, fill=white] {$L$};

            \draw [<->] (40,20) -- (60,20) node [pos=0.75, fill=blue!10,, inner sep = 1pt] {$\frac{1}{4}L$};
            \draw [<->] (40,20) -- (40,0) node [pos=0.75, fill=blue!10, inner sep = 1pt] {$\frac{1}{4}L$};
            \draw [<->] (40,20) -- ++(135:10) node [pos = 0.5, fill = white, inner sep = 0 pt] {$\frac{1}{8}L$};

            \draw [<->] (20,60) -- (0, 60) node [pos=0.75, fill=blue!10,, inner sep = 1pt] {$\frac{1}{4}L$};
            \draw [<->] (20,60) -- (20,80) node [pos=0.75, fill=blue!10, inner sep = 1pt] {$\frac{1}{4}L$};
            \draw [<->] (20,60) -- ++(-45:10) node [pos = 0.5, fill = white, inner sep = 0 pt] {$\frac{1}{8}L$};
        \end{tikzpicture}

        (a)

        \vspace{1cm}

        \begin{tikzpicture}
            \begin{axis}[xmin = 0, xmax = 3.5, ymin = 0, ymax = 6257.52,
            ytick = {0,2085.84, 4171.68, 6257.52, 8343.36, 10429.2}, yticklabels = {0, 0.5$\frac{FL}{\sigma}$, 1$\frac{FL}{\sigma}$, 1.5$\frac{FL}{\sigma}$, 2$\frac{FL}{\sigma}$, 2.5$\frac{FL}{\sigma}$}, ylabel = {Volume, $V$}, scaled y ticks = false,
            xtick = {0, 0.5, 1, 1.5, 2, 2.5, 3, 3.5}, xticklabels={0, 0.5$\frac{\sigma}{L}$, 1$\frac{\sigma}{L}$, 1.5$\frac{\sigma}{L}$, 2$\frac{\sigma}{L}$,2.5$\frac{\sigma}{L}$, 3$\frac{\sigma}{L}$, 3.5$\frac{\sigma}{L}$}, xlabel = {Unit weight, $\rho g$}, 
            width = 0.85\linewidth, ylabel style={yshift=-5pt, xshift = 10pt}]


                \addplot[forget plot, no marks, black!30, domain = 0.1666:4 , samples = 50] plot (\x, 1042.92/\x) node[pos= 0.58, sloped, fill=white, inner sep = 1pt] {\footnotesize $W=0.25F$};
                \addplot[forget plot, no marks, black!30, domain = 0.333:4 , samples = 50] plot (\x, 2085.84/\x) node[pos= 0.55, sloped, fill=white, inner sep = 1pt] {\footnotesize $W=0.5F$};
                \addplot[forget plot, no marks, black!30, domain = 0.666:4 , samples = 50, thick] plot (\x, 4171.68/\x) node[pos= 0.5, sloped, fill=white, inner sep = 1pt] {\footnotesize $W=F$};
                \addplot[forget plot, no marks, black!30, domain = 1.333:4 , samples = 50] plot (\x, 8343.36/\x) node[pos= 0.5, sloped, fill=white, inner sep = 1pt] {\footnotesize $W=2F$};
                \addplot[forget plot, no marks, black!30, domain = 2.666:4 , samples = 50] plot (\x, 16686.72/\x) node[pos= 0.25, sloped, fill=white, inner sep = 1pt] {\footnotesize $W=4F$};

                \addplot[no marks, thick] table[x=pg,y=Regular_2,col sep=comma] {BritishMuseum/paper_results.csv}; 
            \end{axis}
        \end{tikzpicture}

        (b)
        \caption{Example with two holes: (a) Problem specification, fully pinned ($x$, $y$ and $z$ direction) supports defined around outer edges indicated with solid line, and vertical-only supports defined around one of the holes indicated with dashed line. Loading is uniform across the domain. (b) Optimal structure volume for various levels of self-weight. For context, relationships between the structure weight $W = \rho g V$ and the total external loading $F$ are given.}
        \label{fig:holeSpec}
    \end{figure}

	The example in Figure \ref{fig:holeSpec} will now be considered. To generate a pure compression structure, it is essential that supports on the convex edges\footnote{The term convex edge is used here to denote locations on the domain boundary where the tangent is (locally) on the outside of the domain, or the point is a corner with internal angle less than $\pi$ radians. Conversely, concave edges are those portions of the boundary where the tangent lies (locally) inside the domain or at corners with internal angles greater than $\pi$ radians. Straight edges may also be unsupported, although equilibrium enforces that they will have only a single arch along them, not connected to the interior of the domain.} of the domain support the structure both vertically and horizontally, allowing the required thrust to be generated. However, for supports on the interior of the domain, or on concave edges, it is possible to model the existence of vertical-only (roller pin) supports.  This example has support in all three directions (full pin) at the outer boundary (solid lines in Figure \ref{fig:holeSpec}a), and vertical-only support at one of the inner hole boundaries (dashed lines in Figure \ref{fig:holeSpec}a). The boundary of the other hole is not supported. A regular grid of nodes is used across the rectangle with $\frac{L}{40}$ spacing, nodes within $0.15L$ of the centers of the holes are removed, and 32 equally spaced nodes are added around each hole. Full connectivity is permitted, excluding elements that would overlap the holes; adjacent nodes around the edge of each hole are connected with potential elements, even though these pass slightly within the circles. Co-linear elements within the grid are removed, and the problem therefore contains 1117 nodes and 258,856 potential elements. Around 30-90 seconds is required to solve the problem (\ref{eqn:primal}) and reconstruct the solution, and a further 30 seconds is needed to construct the geometry, nodal grid and ground-structure. 

     \begin{figure*}
		\centering
        \begin{tabular}{ccc}
           \raisebox{1.5cm}{\rotatebox{90}{$\rho g = 0.1\frac{\sigma}{L}$}} &
        \includegraphics[height = 0.3\linewidth, trim = 20cm 5cm 20cm 15cm, clip]{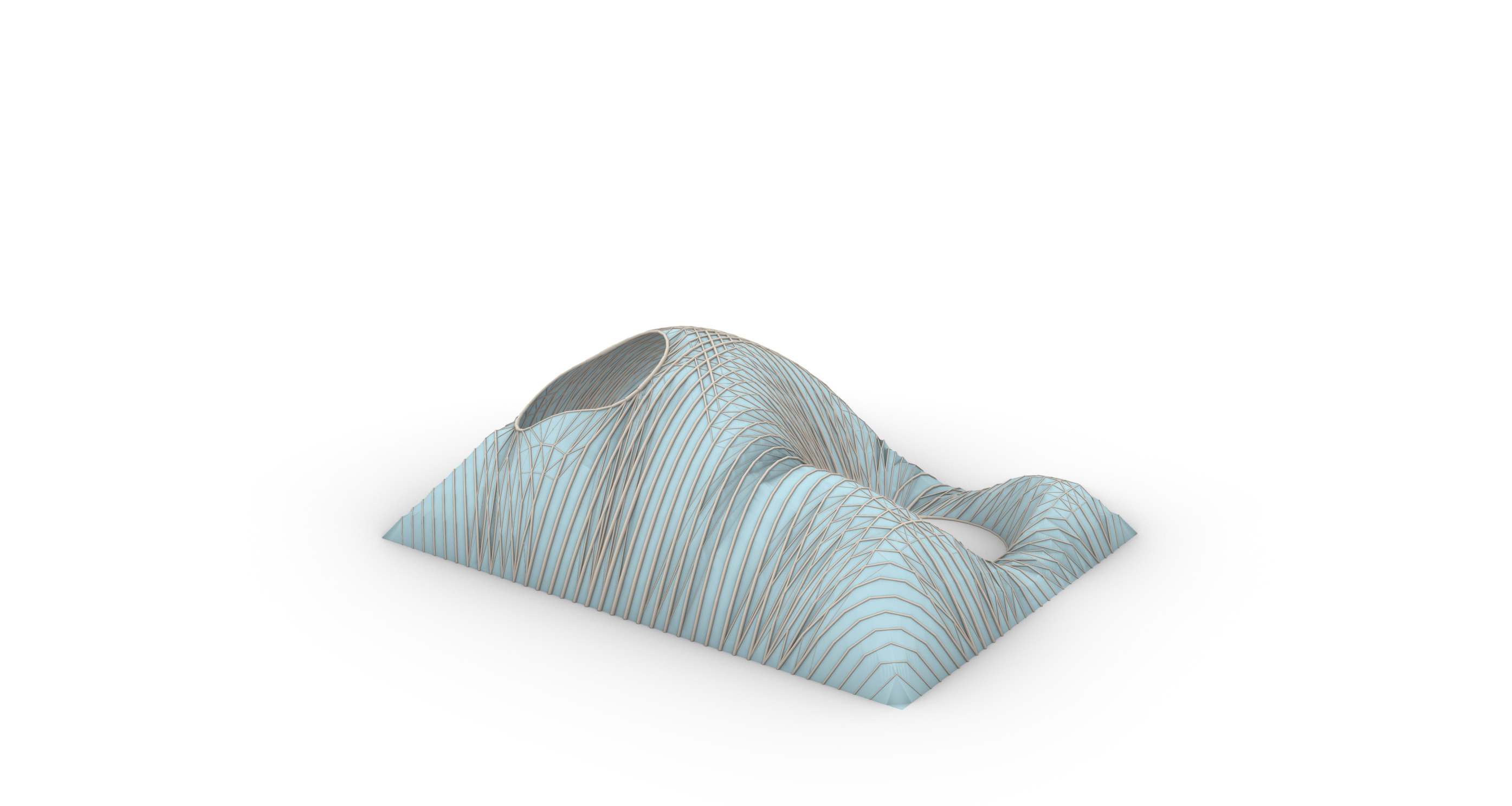} &
        \includegraphics[height = 0.3\linewidth, trim = 31.5cm 3cm 31.5cm 4cm, clip]{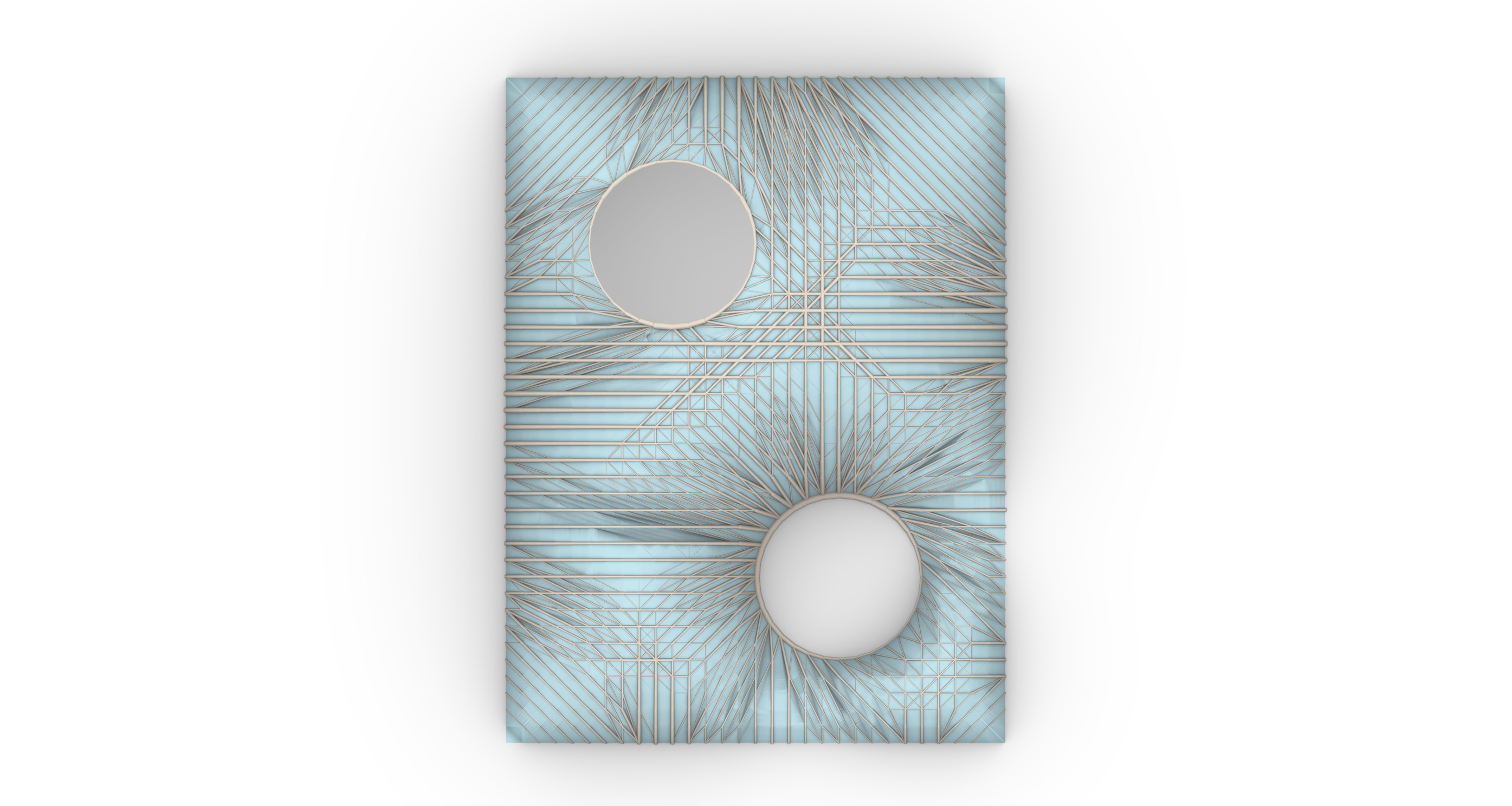} \\  
     
           \raisebox{1.5cm}{\rotatebox{90}{$\rho g = 2\frac{\sigma}{L}$}} &
        \includegraphics[height = 0.3\linewidth, trim = 20cm 5cm 20cm 15cm, clip]{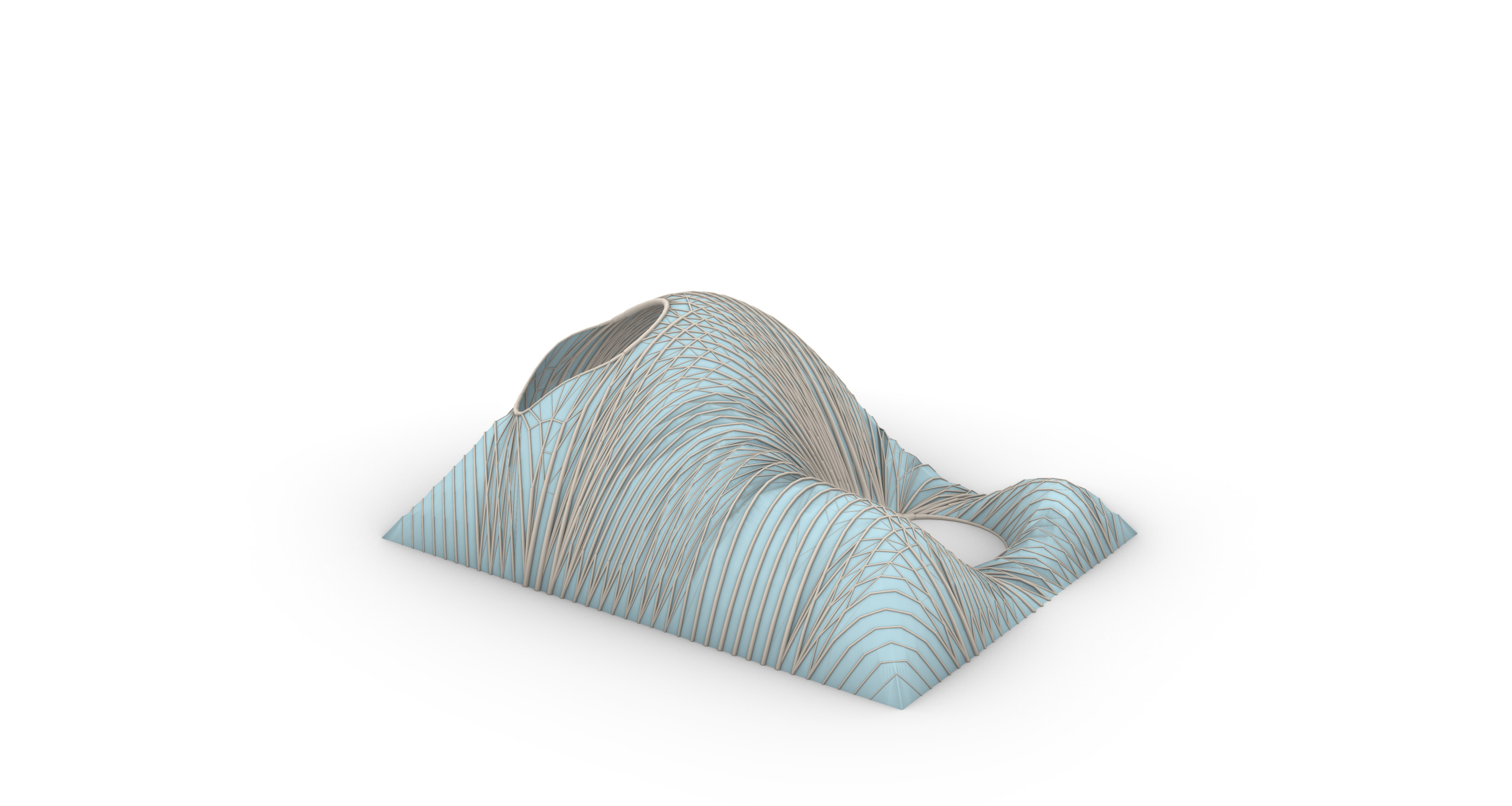} &
        \includegraphics[height = 0.3\linewidth, trim = 31.5cm 3cm 31.5cm 4cm, clip]{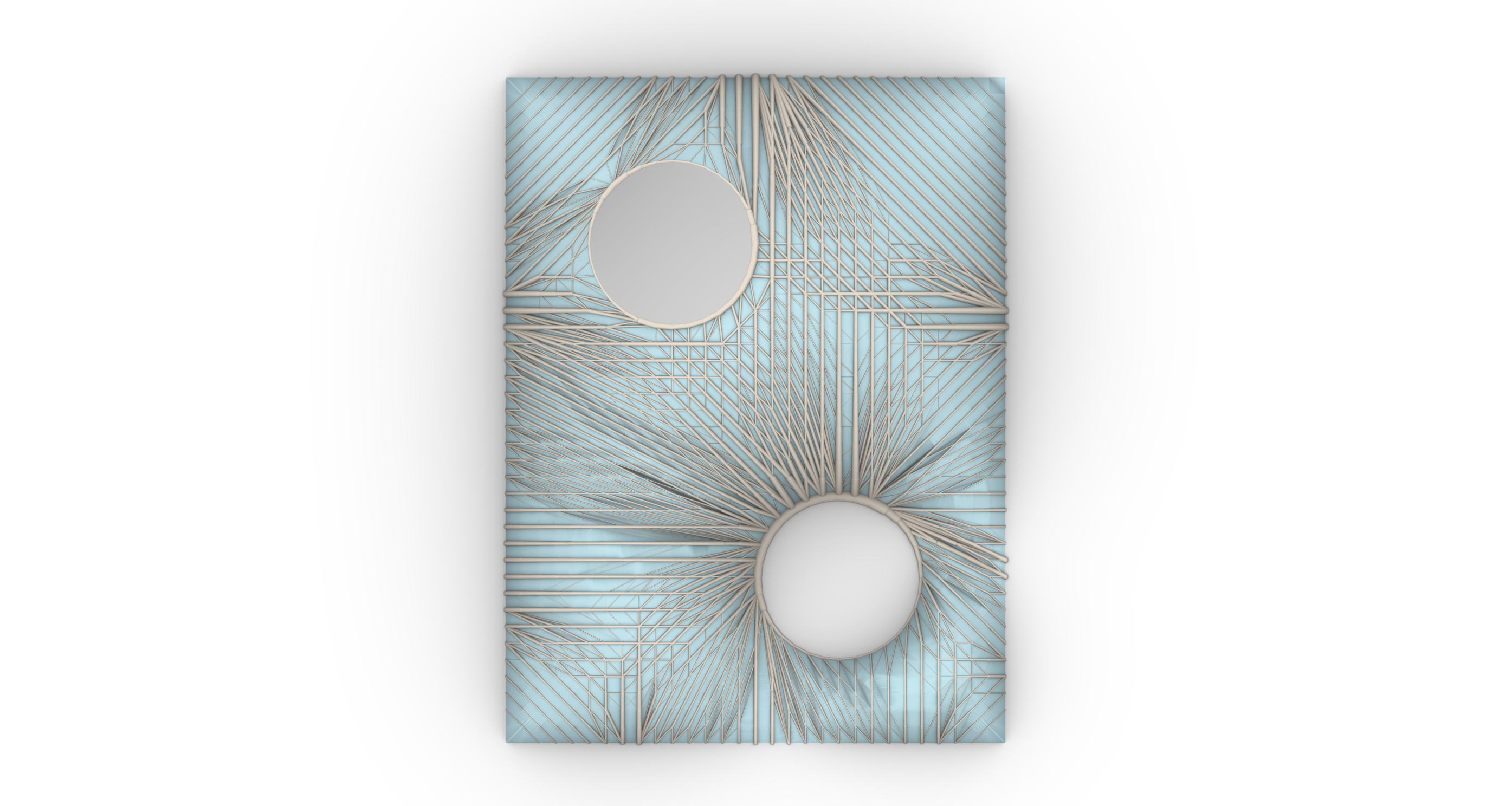} \\  
      
            \raisebox{1.5cm}{\rotatebox{90}{$\rho g = 3.5\frac{\sigma}{L}$}} &
        \includegraphics[height = 0.3\linewidth, trim = 20cm 5cm 20cm 15cm, clip]{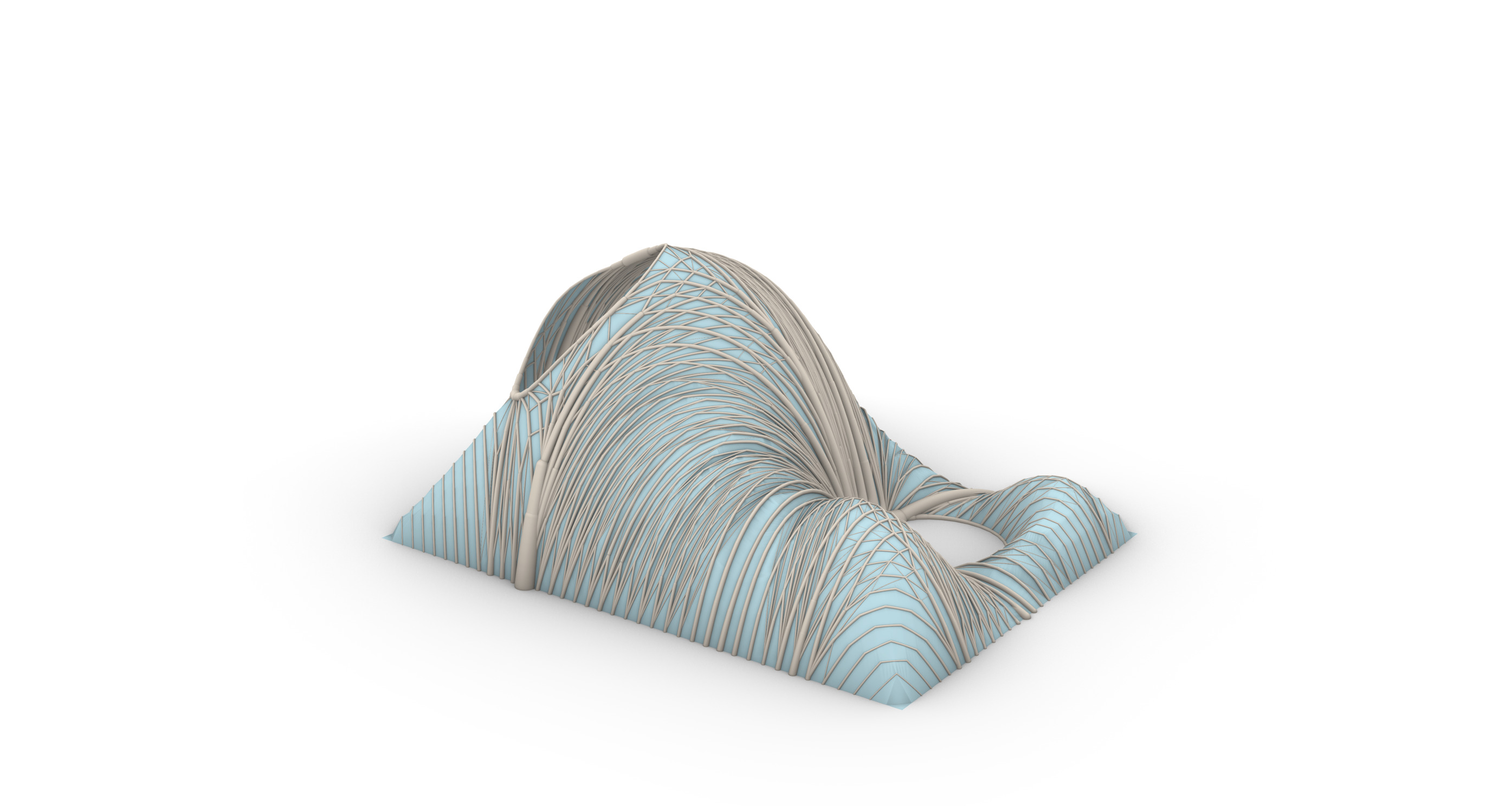} &
        \includegraphics[height = 0.3\linewidth, trim = 31.5cm 3cm 31.5cm 4cm, clip]{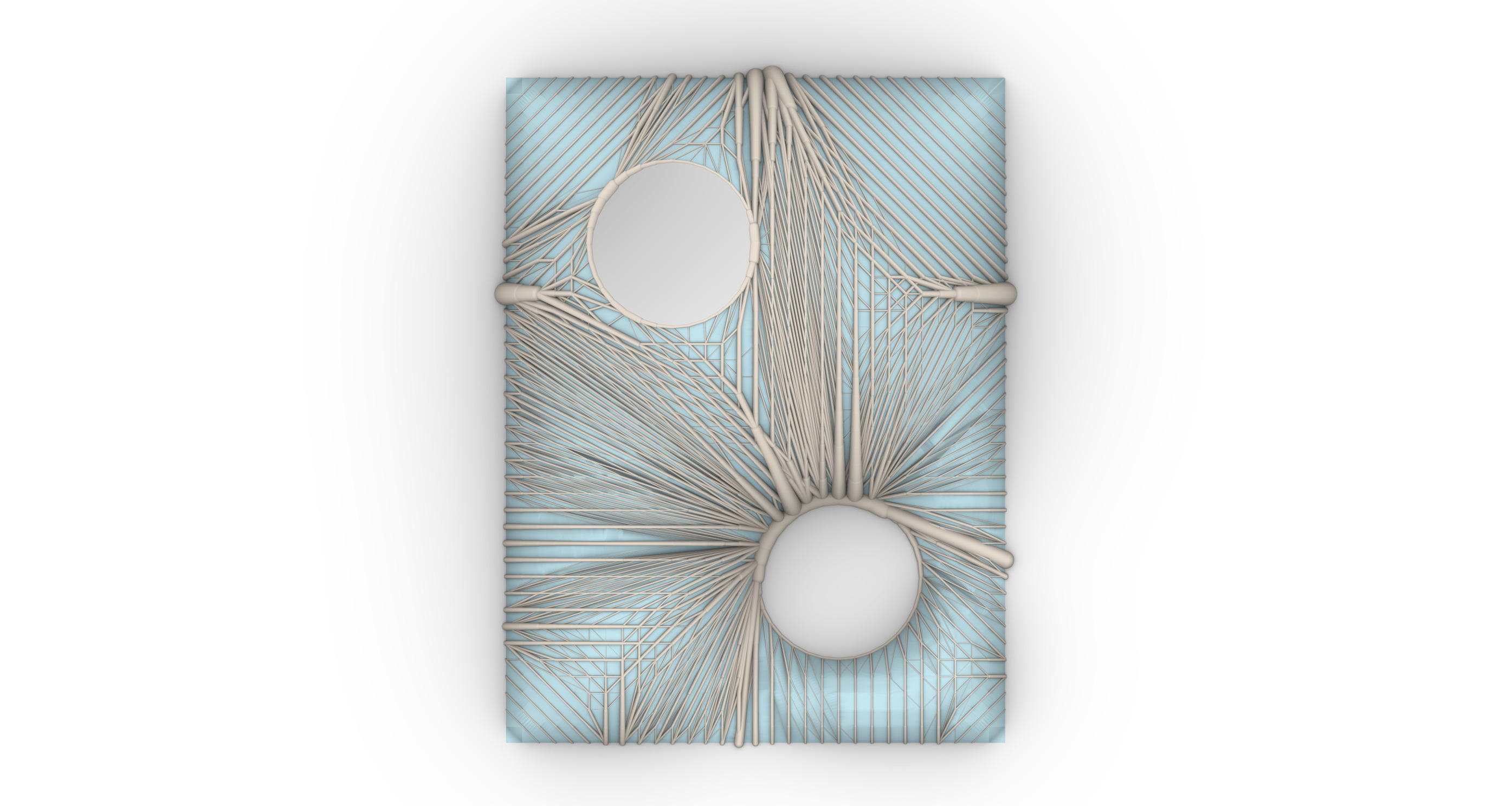} \\        
        \end{tabular}

		\caption{Example with two holes: Optimal results for various values of self-weight, plan and perspective views.} 
		\label{fig:hole}
	\end{figure*}

    This problem has been solved with various levels of self-weight, and the resulting volumes are shown in Figure \ref{fig:holeSpec}b. The optimal forms are shown in Figure \ref{fig:hole} for self-weight values of $\rho g = 0.1 \frac{\sigma}{L}$ (self-weight negligible), $\rho g = 2 \frac{\sigma}{L}$ (total structure weight $\approx$ total external load, see Figure \ref{fig:holeSpec}b)  and $\rho g = 3.5 \frac{\sigma}{L}$ (self-weight more than four times applied load, see Figure \ref{fig:holeSpec}b). The optimal topologies observed are reminiscent of those seen in the simpler square example, see Figure \ref{fig:topology}. When self-weight is low, the optimal topology contains many elements perpendicular to the domain edges, each with similar size; when self-weight is more significant, certain support points appear preferable, with larger elements and more elements fanning out from a single point. In all the results, the vertical-only internal supports result in relatively low rises on the shorter spans, efficiently allowing a larger thrust to be generated to ensure horizontal equilibrium at those points.

\begin{figure}
        \begin{tikzpicture}[scale =0.85]
            \begin{axis}[axis equal, height = 8cm, width = 6cm, scale only axis, xmin = 0, xmax = 60, ymin = 0, ymax = 80,
                ticks = none, 
                colormap={rb}{  rgb(0cm) = (1,0,0);
                                rgb(0.25cm) = (1,1,1); 
                                rgb(0.6cm) = (0,1,1);
                                rgb(1cm) = (0,0,1)}, 
                colorbar left, point meta max = 0.2, point meta min = -0.066666,
                colorbar style={
                ytick={-0.05, 0, 0.05, 0.1, 0.15, 0.2},
                yticklabels = {0.05$L$, 0, 0.05$L$, 0.1$L$, 0.15$L$, 0.2$L$}
                }]
                \draw (20, 60) circle(10);
                \draw (40, 20) circle(10);
                
                \addplot[only marks, thick, scatter, point meta = \thisrow{diff_Z}, colorbar source] table[x=x,y=y,col sep=comma] {BritishMuseum/heightDiffs.csv}; 
            \end{axis}
            \node [below right, inner sep = 0, yshift = -10pt] at (-2.6, 2) {\shortstack[l]{\footnotesize Higher for \\ \footnotesize $\rho g = 0.1 \frac{\sigma}{L}$}}; 
            \node [above right, inner sep = 0, yshift = 10pt] at (-2.6, 2) {\shortstack[l]{\footnotesize Higher for \\ \footnotesize $\rho g = 3.5 \frac{\sigma}{L}$}}; 
            \draw (-2.6cm, 2) -- ++ (1.3,0);
        \end{tikzpicture} 
        \caption{Example with two holes: Elevation difference between optimal structures for $\rho g =0.1 \frac{\sigma}{L}$ and $\rho g = 3.5 \frac{\sigma}{L}$, shown at nodal points. The top hole is the unsupported boundary. For context, the $\rho g = 0.1\frac{\sigma}{L}$ result has a maximum height of approximately $0.32L$}
        \label{fig:holeDiff}
    \end{figure}
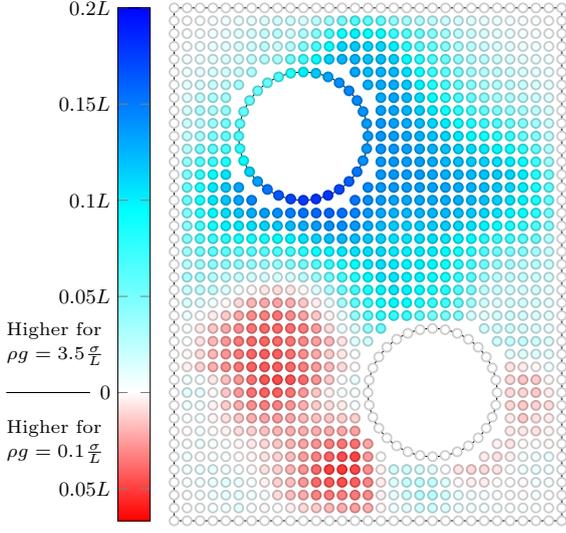

      In the previous examples, it has been seen that the optimal elevation increased across the whole domain when the self-weight increased. However, whilst the maximum elevation increases with self-weight in Figure \ref{fig:hole}, this is not the case across the whole domain. Figure \ref{fig:holeDiff} shows the difference in elevation for each node in the problem. Whilst the largest span regions (around the unsupported hole) show significant increases in elevation, notable reductions in elevations are also seen, particularly around the supported hole. This may be due to the need to generate larger horizontal thrusts to ensure horizontal equilibrium of the supported circle.  

    	\subsection{Self-intersecting example}

\begin{figure}
    \centering
        \begin{tikzpicture}
		\node at (0,0) {\includegraphics[width=6cm]{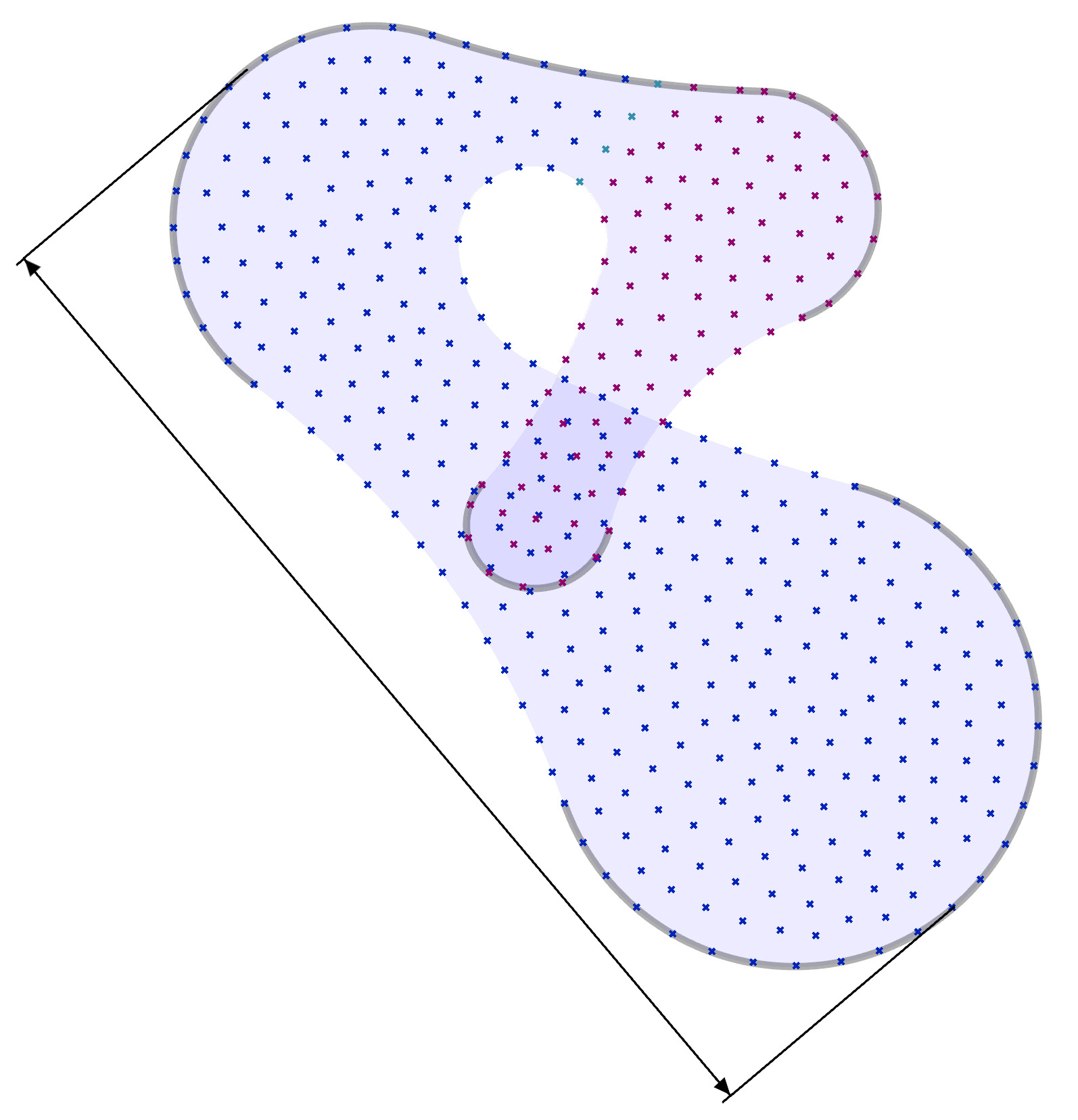}};
        \node[fill=white] at (-0.8cm, -0.8cm) {$L$};
        \draw [cyan!70!black, thick, densely dotted] (0.23,2.1) ++ (-39:-1.9) -- ++ (-39:3.7)
            node [sloped, above right, pos=0, inner sep = 2pt] {\tiny B}
            node [sloped, below right, pos=0, inner sep = 2pt] {\tiny A}
            node [sloped, above left, pos=1, inner sep = 2pt] {\tiny B}
            node [sloped, below left, pos=1, inner sep = 2pt] {\tiny C};
        \end{tikzpicture}

        (a)

        \vspace{1cm}
        
                \begin{tikzpicture}
                \begin{axis}[xmin = 0, xmax = 4, ymin = 0, ymax = 120,
                    ytick = {0, 20, 40, 60, 80, 100, 120}, yticklabels = {0, 0.2$\frac{FL}{\sigma}$, 0.4$\frac{FL}{\sigma}$,0.6$\frac{FL}{\sigma}$, 0.8$\frac{FL}{\sigma}$, 1$\frac{FL}{\sigma}$, 1.2$\frac{FL}{\sigma}$}, ylabel = {Volume, $V$}, scaled y ticks = false, width = 7cm,
            xtick = {0, 1, 2, 3, 4}, xticklabels={0, 1$\frac{\sigma}{L}$, 2$\frac{\sigma}{L}$, 3$\frac{\sigma}{L}$, 4$\frac{\sigma}{L}$}, xlabel = {Unit weight, $\rho g$}]
                    \addplot[forget plot, no marks, black!30, domain = 0.2083333:4 , samples = 50] plot (\x, 25/\x) node[pos= 0.5, sloped, fill=white, inner sep = 1pt] {\footnotesize $W=0.25F$};
                    \addplot[forget plot, no marks, black!30, domain = 0.41666:4 , samples = 50] plot (\x, 50/\x) node[pos= 0.5, sloped, fill=white, inner sep = 1pt] {\footnotesize $W=0.5F$};
                    \addplot[forget plot, no marks, black!30, domain = 0.8333:4 , samples = 50] plot (\x, 100/\x) node[pos= 0.5, sloped, fill=white, inner sep = 1pt] {\footnotesize $W=F$};
                    \addplot[forget plot, no marks, black!30, domain = 1.6666:4 , samples = 50] plot (\x, 200/\x) node[pos= 0.5, sloped, fill=white, inner sep = 1pt] {\footnotesize $W=2F$};
                    \addplot[forget plot, no marks, black!30, domain = 3.3333:4 , samples = 50] plot (\x, 400/\x) node[pos= 0.25,  fill=white, inner sep = 1pt] {\footnotesize $W=4F$};

                    \addplot[no marks, thick] table[x=pg,y=vol,col sep=comma] {selfIntersect/intersectResults.csv}; 
                \end{axis}
            \end{tikzpicture}

        (b)
        
		\caption[Self-intersecting example: Problem specification.]{Self-intersecting example: (a) Problem specification. Grey edges represent locations of supports, note that the self intersecting regions (i.e blue and purple nodal points) do not interact, and the support shown there belongs only to the portion of the domain containing purple node markers. The dashed tangent line and labels A, B, C refer to the regions used to calculate the permissible ground structure. (b)~Optimal volumes for different values of self-weight. For context, relationships between the structure weight $W = \rho g V$ and the total external loading $F$ are given. }
		\label{fig:intersectSpec}
	\end{figure}

	It is interesting to note that, although the problem is specified in 2D, it is not essential that the domain's outer boundary is a simple closed curve. Specifically, it is possible to consider self-intersecting domains, whilst defining that the intersecting parts are disconnected. To demonstrate this, the problem shown in Figure \ref{fig:intersectSpec}a will be considered. The boundary is a planar self-intersecting curve, formed from a series of arcs. Supports are available along the boundary sections shown with grey lines in Figure \ref{fig:intersectSpec}a, and the maximum straight line distance between two points in the domain is denoted as $L$.

	\begin{figure*}[t!]
		\centering
        \begin{tabular}{ccc}
            \raisebox{1.5cm}{\rotatebox{90}{$\rho g = 1\frac{\sigma}{L}$}} &
        \includegraphics[trim = 20cm 20cm 20cm 0, clip, height = 0.3\linewidth]{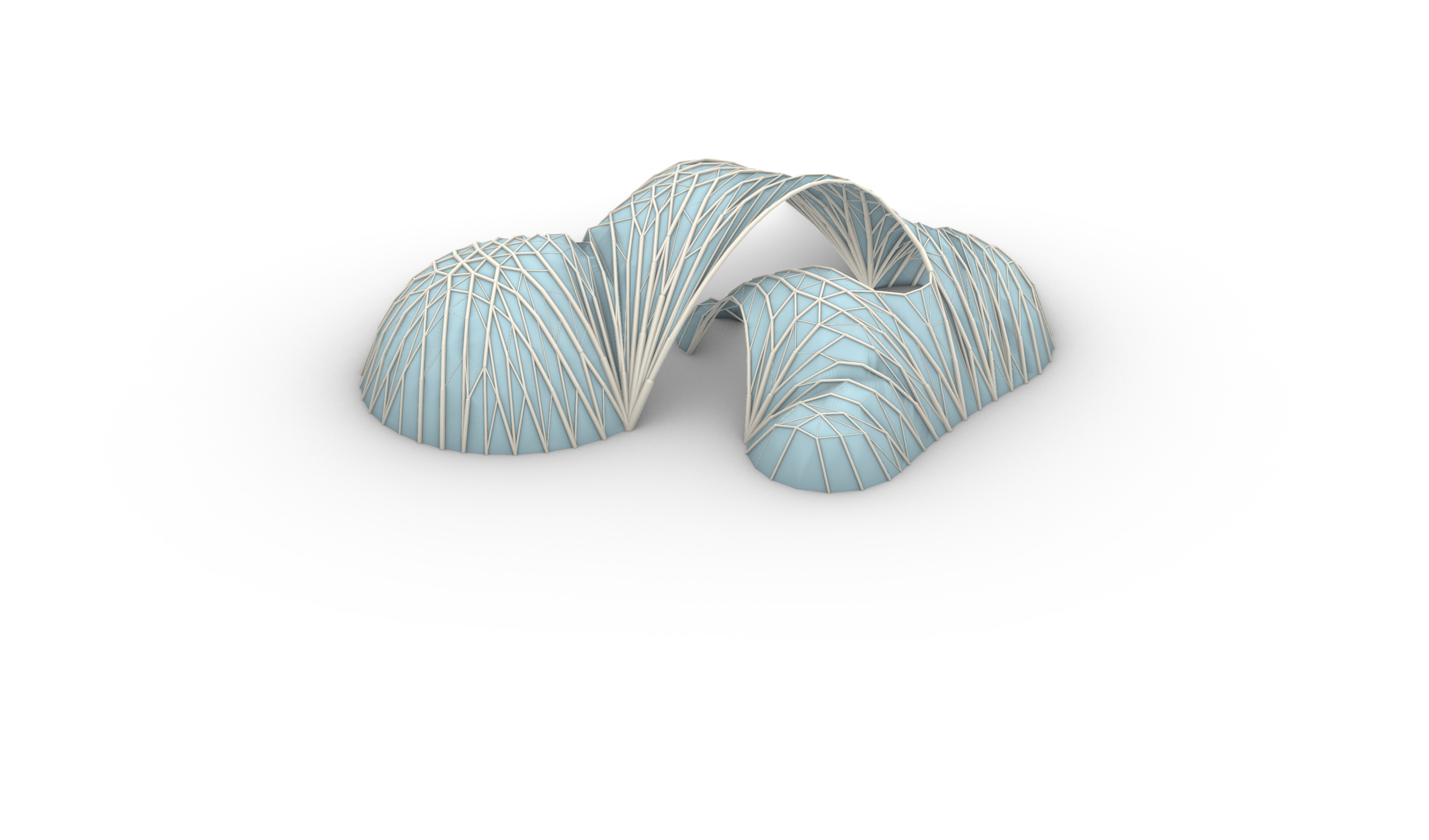} &
        \includegraphics[trim = 15cm 0 15cm 0, clip, height = 0.3\linewidth]{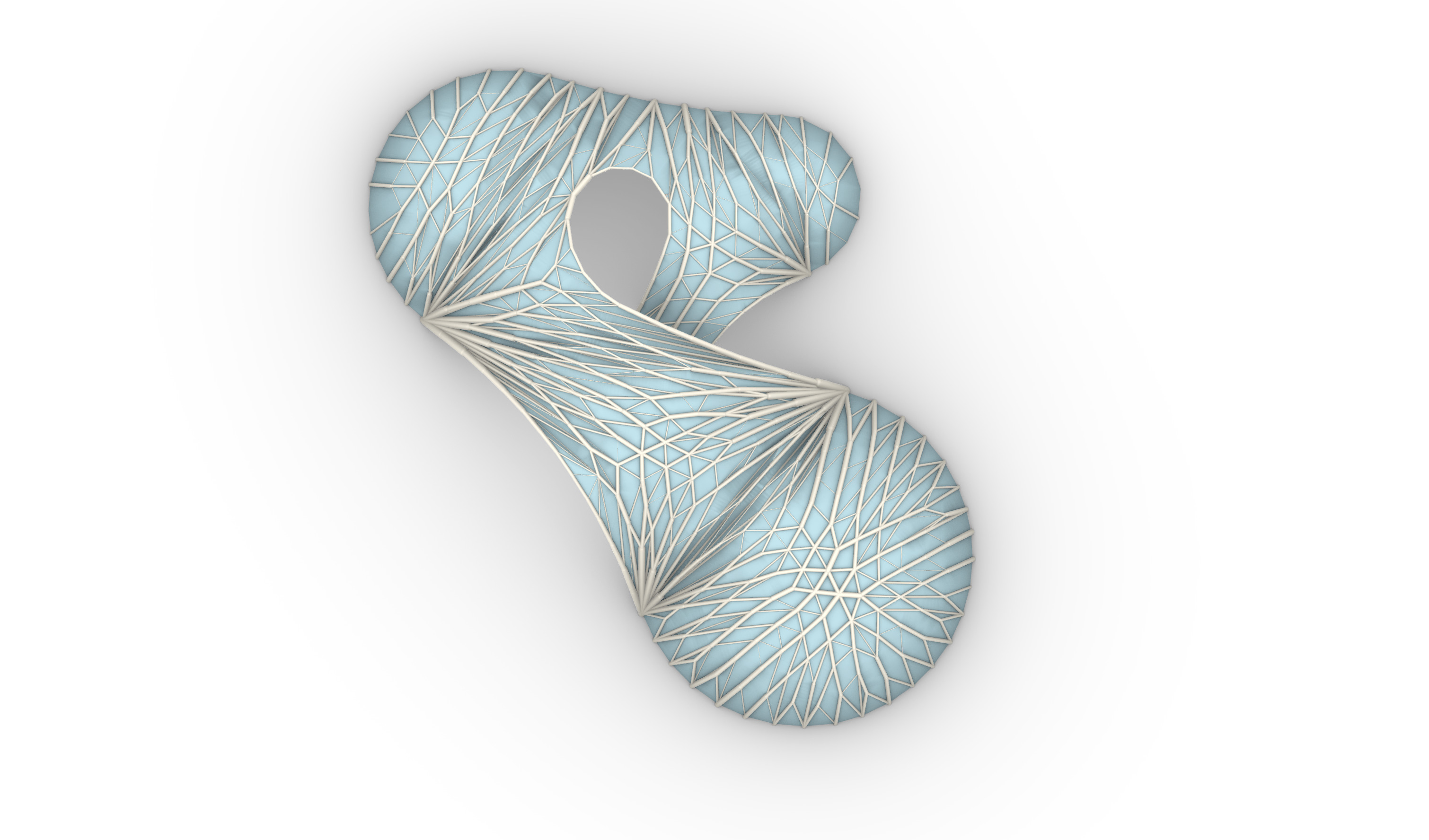}\\
             \raisebox{1.5cm}{\rotatebox{90}{$\rho g = 4\frac{\sigma}{L}$}} &
        \includegraphics[trim = 20cm 20cm 20cm 0, clip, height = 0.3\linewidth]{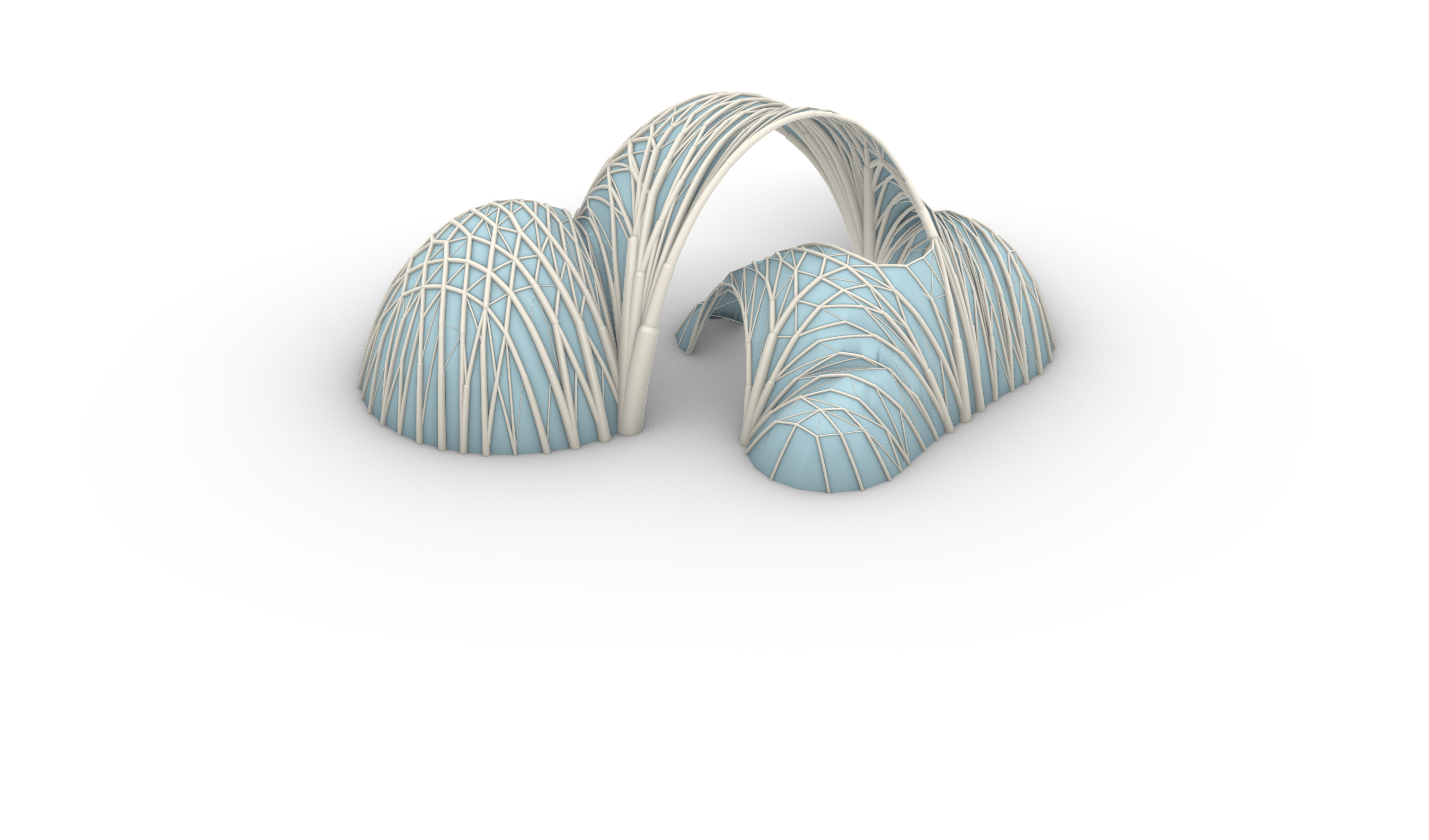} &
        \includegraphics[trim = 15cm 0 15cm 0, clip, height = 0.3\linewidth]{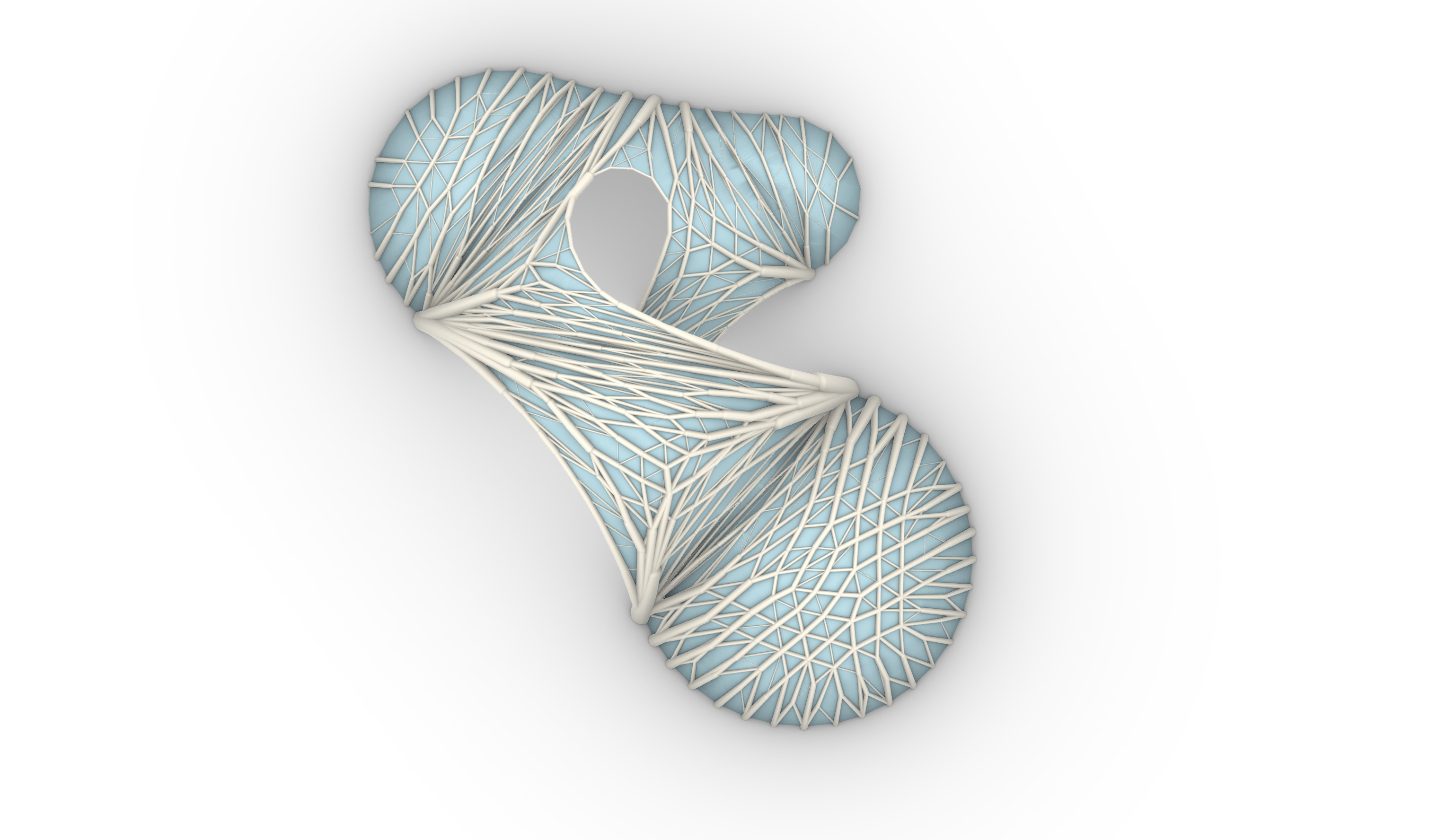}\\
        \end{tabular}
		
		\caption{Self-intersecting example: Results, plan and perspective view.}
		\label{fig:intersecting}
	\end{figure*}	

    For such a case, no special changes need be made to the method described. However, great care must be taken in constructing the nodal grid and connectivity of the ground structure, as standard algorithms will not be able to distinguish between the overlapping regions. Visual modelling environments such as Rhino/Grasshopper \cite{grasshopper} can be invaluable for this to evaluate the problem-specific logic at each stage.

    Here, the approach used was to select a point on the interior hole, and from there take a normal and tangent line to split up the domain. Nodal points were defined separately in the regions each side of the normal line (purple/blue points in Figure \ref{fig:intersectSpec}), with evenly spaced points pre-defined along the normal line. The Grasshopper Triremesh component was used to generate approximately equally spaced points across the irregular shaped regions, and a Voronoi diagram was used to assign forces to each point representing a uniformly distributed imposed load. The potential element list was then generated by considering three regions (A, B, C) of the domain defined by the tangent line (shown as a dotted line in Figure \ref{fig:intersectSpec}). Note that any element connecting section A to section C would not be permitted, regardless of the proximity of the points in actual 2D space. By separately considering the union of A and B and the union of B and C, potential elements can be checked using standard approaches, and finally the ground structure is combined for solving. The complete list of nodal coordinates and potential elements can be found in the supplementary information. 

    The results shown in Figure \ref{fig:intersectSpec}b required between 2 and 5 seconds to solve the optimization problem for each value of self-weight. Selected optimal forms are shown in Figure \ref{fig:intersecting}. From Figure \ref{fig:intersectSpec}b, it can be seen that the two cases shown in Figure \ref{fig:intersecting} correspond to the cases where the self-weight is just over one quarter of the external load, and where self-weight is just over four times the external load, respectively. 
    
    For self-intersecting examples such as this one, it is not possible to specify which of the overlapping regions should occur on top, or indeed to prevent the two areas from crossing over. Nonetheless, when the overlapping regions have very different spans, it will generally be the case that the optimal elevations will differ sufficiently for useful results to be obtained.

    \section{Concluding remarks}
    \label{sec:conclusions}
    A methodology has been presented to obtain the optimal topology and elevation profile for a pure compression grid-shell structure carrying a predefined external loading plus the self-weight of the structure itself. By solving a convex conic programming problem, globally optimal solutions (for the given ground structure) can be obtained rapidly. The method provides speed-up of several orders of magnitude compared to a 3D truss optimization approach, whilst simultaneously increasing accuracy and increasing the clarity of the solution by ensuring a single-layer structure is returned. The method could equivalently be used to design pure tension structures by following the logic herein with signs reversed. 

    Through the use of this approach, many features of optimal grid-shell structures under self-weight loading have been discovered. It is observed that as self-weight becomes more significant, the optimal peak height of the structure typically increases. For the weightless case, the maximum elevation cannot be more than the radius of the domain's circumcircle, a limit which the solutions with self-weight frequently violate. This means that these optimal solutions could not be obtained by approximating the self-weight as an external load, even if an iterative approach was used to re-distribute the load in the most appropriate proportion. 

    Whilst the overall increase in optimal height with increasing influence of self-weight has been previously noted for simpler structures, the numerical method presented here allows more complex scenarios which show that it is not necessarily true that elevation should increase at \textit{all} points in a structure. Particularly when vertical-only supports are present, reducing the rise in some areas can be beneficial in generating required horizontal thrusts.

    Furthermore, it has been shown that the optimal topology can change markedly when self-weight becomes significant. The use of an incorrect topology in either direction (i.e. the optimal topology found for the weightless case being used for a problem where self-weight is significant, or vice versa) is shown to increase the material demands of the structure. Nonetheless, it is generally observed that the penalty for an inefficient topology increases as self-weight becomes more significant, echoing previous findings in optimization of other structure types.  

    Finally, it has been shown that when self-weight loading is considered, it is possible for the optimal structure to require joints at points which are not the locations of external loads or supports. This increases the need to optimize the topology in these scenarios, in conjunction with the elevation function. Furthermore, it can be seen from the results that the chosen topology and the vertical shape are closely linked, highlighting the importance of the coupled optimization method presented here. 

    Overall, the method presented here provides a powerful tool for investigating this class of material-efficient structures. This study has shown that these structures have many interesting properties, many of which demonstrate significant challenges to purely intuitive design approaches, or to existing fixed-topology methods. By overcoming these issues, this method can facilitate the adoption of such structural forms, and Python scripts and Grasshopper files implementing examples from this paper have been made available \cite{codes} to encourage this.

    	\subsection*{Declaration of Competing Interest}
	
	The authors declare that they have no known competing financial	interests or personal relationships that could have appeared to influence	the work reported in this paper.

    \subsection*{CRediT authorship contribution statement}
    \textbf{Helen Fairclough: }Conceptualization, Formal analysis, Investigation, Methodology, Software, Visualization, Writing - original draft. \textbf{Karol Bo{\l}botowski: }Formal Analysis, Methodology, Writing - original draft. \textbf{Linwei He:} Software, Writing - review and editing. \textbf{Andrew Liew:} Writing - review and editing. \textbf{Matthew Gilbert: }Conceptualization, Writing - review and editing.

    \subsection*{}
	\bibliographystyle{elsarticle-num-names}
	\bibliography{VaultCat}
	
	\appendix	
 \section{Lumped mass approach}
\label{sec:lumped}
 Within truss optimization, it is common to consider the effects of self-weight using a `lumped mass' approach, where it is assumed that the weight of each element can be modeled as two equal point loads acting directly on the end-points of the element. This approach neglects the bending within a straight element which would be required to transmit the self-weight load to the nodes. However the simplicity of the model means it remains popular. This section will discuss the seemingly promising approach of combining this self-weight model with the vault design formulation. However, it will be shown that this does not produce valid solutions. 

 From the weightless vault formulation (\ref{eqn:vault}), it can be seen that the volume of an element $i$ is given by $\frac{l_i}{\sigma}\left(s_i + \frac{q_i^2}{s_i}\right)$. Using the material's unit weight $\rho g$ and the definition of the variable $r_i$ given in (\ref{eqn:rVar}), then the lumped mass model in this case would involve the imposing of self-weight forces with magnitude $0.5 \frac{\rho g}{\sigma} l_i \left(s_i + r_i\right)$ at each end of element $i$. Applying these forces involves adding these terms to the vertical equilibrium constraints, to give the optimization problem:
 \begin{subequations} \label{eqn:lumped}
	\begin{align}
		\min \sum_{i\in M} \frac{l_i}{\sigma_i} \left(s_i + r_i\right), \label{eqn:lumpedObj}\\ 
		\text{s.t. } \mathbf{Bs} &= \mathbf{f}_{xy}, \label{eqn:lumpedInplane} \\
		\mathbf{Dq} + \mathbf{Zs} + \mathbf{Zr}  &= \mathbf{f}_z, \label{eqn:lumpedOutplane}\\
        \big(r_i &\geq \frac{q_i^2}{s_i} \big)_{\forall i} \\
		\mathbf{s} & \geq 0, \label{eqn:lumpedComp}
	\end{align}
	\end{subequations}
    where the new $\mathbf{Z}$ matrix contains zeros everywhere except at entry $i,j$ when element $i$ is connected to node $j$, at which point it contains the value $Z_i = -0.5 l_i \frac{\rho g }{\sigma}$.

     \begin{figure}[t!]
    \centering
        \includegraphics[width = \linewidth]{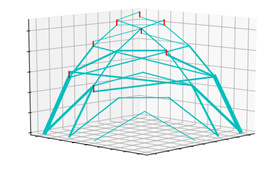}

        (a)

        \includegraphics[width = \linewidth]{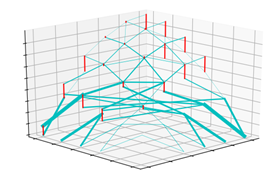}

        (b)

        \caption{Lumped mass formulation: Results showing inconsistency in nodal elevation calculations. Vertical red lines link elements ends which should connect at a single point, highlighting the inconsistency }
        \label{fig:lumped}
    \end{figure}

    Problem (\ref{eqn:lumped}) can be easily set up and solved for a given scenario. However, interpreting the optimal solution uncovers the problems with this model. The variables $q$ and $s$ describe the vertical and horizontal components of the force in an element, and thus define the inclination of its centre-line\footnote{Note that here we cannot assume $q$ to include an additional mass contribution as used in Figure \ref{fig:lumps}. In this formulation $q$ defines two equal and opposite forces, so any variation in $q$ would be imposed downward at one end of the element and upward at the other.}. From this, the elevation difference between the two ends can be calculated. It is then possible to work through the structure, starting from the supported points with 0 elevation, and thereby establish heights for all of the nodes. (Note: this procedure was suggested and successfully used in the weightless case by \cite{he2025minimum} as a conceptually simpler alternative to the use of the dual variables and also as a way to validate the solutions.)

    To demonstrate the issue, (\ref{eqn:lumped}) has been solved for problems representing a quarter of the square-based domain problem, with distributed loads. When the procedure is followed, it is found that different routes through the structure result in different elevations of the same node. These errors are represented in Figure \ref{fig:lumped} by the vertical red lines. This therefore shows that the solutions of (\ref{eqn:lumped}) do not represent physically meaningful solutions. 

    Note that within the catenary formulation presented in this paper, the equivalent process of validation via primal variables is the calculation of (non-negative) lumped masses from the relaxed geometrical coupling condition described in \ref{sec:geometrical}. This has been successfully carried out for all examples herein, validating the analytical proof of  correctness of the catenary approach.

\section{Analysis of the optimal primal and dual solutions}

\label{sec:analysis}

     Below the properties (I)-(III) of the optimal primal and dual solutions $\mathbf{s},\mathbf{q}_A, \mathbf{q}_B$ and $\mathbf{u},\mathbf{w},\mathbf{g}_1,\mathbf{g}_2,\mathbf{g}_3$ are derived, see Section \ref{sec:properties}.
    It must be emphasized that the criteria from Section \ref{sec:assumptions} are assumed to hold throughout. In particular, the strict inequality below holds for every node,
    \begin{equation}
        \mathbf{w} < \frac{1}{\rho g}
        \label{eqn:w_bound_again}
    \end{equation}

    \smallskip
    
    \noindent \textit{Proof of properties (I), (II):} These properties will be shown for every element for which $s >0$ at optimality.

    First, from the complementary slackness for the primal constraint $s \geq 0$, it immediately follows that the dual linear constraint \eqref{eqn:dualS} is an equality for such an element. This allows to explicitly express all $g_j$ that enter the dual conic constraint $2 g_1 g_2 \geq g_3^2$,
    \begin{align}
        g_1 &= \tfrac{1}{\sin \bar{l}}\big(\tfrac{1}{\rho g} - w_A\big), \quad g_2 = \tfrac{1}{\sin \bar{l}}\big(\tfrac{1}{\rho g} - w_B\big), \\
        g_3 &= - \sqrt{2}\,\Big(\tfrac{1}{2}\, \Delta u + \tfrac{\cos \bar{l}}{\sin \bar{l}} \,\big(\tfrac{1}{\rho g} - \tfrac{w_A + w_B}{2}\big) \Big),
    \end{align}
    where $\Delta u$ represents the horizontal extension of the element, calculated as the relevant entry of $\mathbf{B}^\T \mathbf{u}$.
    On the other hand, the primal conic constraint \eqref{eqn:primalCone} can be expressed as $2 t_1 t_2 \geq t_3^2$, where
    \begin{align}
        t_1 &= \sin \bar{l}\, q_A + \cos \bar{l} \,s , \\
        t_2 &= \sin \bar{l}\, q_B + \cos \bar{l} \,s, \qquad t_3 = \sqrt{2} s.
    \end{align}
    It is well established that at optimality there holds a complementary slackness condition associated to the conic constraints, see Section 2.5 in \cite{ben2001},
    \begin{equation}
        \label{eqn:slackness}
        t_1 g_1 + t_2 g_2 + t_3 g_3 = 0.
    \end{equation}
    Since $s>0$, there is $t_3 >0$ and thus also $t_1,t_2>0$ due to the conic constraint. In addition, $g_1,g_2 >0$, due to the assumption \eqref{eqn:w_bound_again}. As result, the condition \eqref{eqn:slackness} implies that $g_3 < 0$. Starting from the simple inequality $a^2 + b^2 \geq 2ab$, the chain below follows,
    \begin{align*}
        t_1 g_1 + t_2 g_2 &\geq 2 \sqrt{t_1 g_1} \sqrt{t_2 g_2}\\
        &= \sqrt{2 t_1 t_2} \sqrt{2 g_1 g_2} \geq |t_3| |g_3| = -t_3 g_3, 
    \end{align*}
    where the second inequality simply combines the two conic constraints. By virtue of the condition \eqref{eqn:slackness}, this chain must be a chain of equalities only. The first equality implies that $t_1g_1 =t_2 g_2$. Indeed, for positive numbers, $a^2 + b^2 = 2ab$ ensures that $a=b$. Up to multiplying by $\sin \bar{l}$ on each side, this is exactly the statement (III). Next, the equality previous to last ensures that both conic constraints are satisfied as equalities. In particular, $2t_1 t_2 = t_3^2$, which furnishes the statement (II) in the case $s>0$. For $s=0$ it will follow from the assertion (III), proved below.

    Note that \eqref{eqn:w_bound_again} was fundamental, since, otherwise, the complementary slackness condition \eqref{eqn:slackness} could have been trivially satisfied for $g_1 = g_2 = g_3 =0$. It was thus necessary to make sure that $\mathbf{w} < \frac{1}{\rho g}$.

    \smallskip

    \noindent\textit{Proof of property (III):} For an element with $s=0$, there must also be $t_3 = 0$ and so the complementary slackness condition becomes 
    \begin{equation}
        t_1g_1 + t_2g_2 + 0 = 0.
    \end{equation} 
    The definition of the primal cone requires $t_1, t_2 \geq 0$, while property (I) furnishes $g_1, g_2 > 0$. Hence both remaining terms are $\geq 0$, and so both must be equal to 0. This implies $t_1 = t_2 = 0$, which when combined with the initial definition $s=0$ proves that $q_A = q_B = 0$.

\end{document}